\newif\ifanon

\newif\ifwhitepaper

\newif\iflncs

\newif\ifshorten

\iflncs
  \whitepaperfalse
\fi

\iflncs
  \documentclass[runningheads]{llncs}
\else
  \documentclass[twocolumn]{article}
\fi

\usepackage{xspace}
\usepackage{graphicx}
\usepackage{tablefootnote}
\usepackage{array}
\usepackage{pythonhighlight}
\usepackage{amsmath}
\iflncs\else\usepackage{amsthm}\fi 
\usepackage{thmtools}

\usepackage{xurl}
\usepackage[most]{tcolorbox}
\usetikzlibrary{tikzmark, calc, fit}
\usepackage{listings}
\usepackage{xcolor}
\usepackage{cite}
\usepackage{amsmath,amssymb,amsfonts}
\usepackage{algorithm}
\usepackage{algpseudocode}
\algnewcommand{\LineComment}[1]{\State $\triangleright$ #1} 
\usepackage{placeins} 
\usepackage{graphicx}
\usepackage{adjustbox} 
\usepackage{textcomp}
\usepackage{xcolor}
\iflncs\else\usepackage{geometry}\fi 
\usepackage{cuted}

\usepackage[font=small,labelfont=bf,labelsep=period]{caption}
\usepackage[font=small,labelfont=rm,labelsep=space]{subcaption}
\usepackage[breaklinks=true]{hyperref}

\usepackage{cleveref}
\makeatletter
\AtBeginDocument{\def\refstepcounter{\@ifnextchar[\refstepcounter@optarg\refstepcounter@noarg}}
\makeatother

\usepackage{etoolbox}
\makeatletter
\newcounter{algorithmicH}
\newcounter{ALGlinecont}
\AtBeginEnvironment{algorithmic}{\stepcounter{algorithmicH}}
\expandafter\apptocmd\csname\string\algorithmic\endcsname{\setcounter{ALG@line}{\value{ALGlinecont}}}{}{\PackageWarning{macros}{could not patch algorithmic}}
\AtEndEnvironment{algorithmic}{\setcounter{ALGlinecont}{\value{ALG@line}}}
\def\theHALG@line{ALG@line.\thealgorithmicH.\arabic{ALG@line}}
\makeatother
\usepackage[inline]{enumitem}
\setlist{itemsep=2pt, topsep=4pt, parsep=0pt, leftmargin=10pt}

\makeatletter
\renewcommand\@makefntext[1]{\noindent\@makefnmark\ #1}
\makeatother

\newcommand{\BZ}{\ensuremath{\mathbb{Z}}\xspace}
\newcommand{\BN}{\ensuremath{\mathbb{N}}\xspace}

\newcommand{\CR}{\ensuremath{\mathcal{R}}\xspace}

\newcommand{\false}{\ensuremath{\mathsf{false}}\xspace}

\newcommand{\getsrandomly}{\overset{\$}{\gets}}

\newcommand{\dotparagraph}[1]{\paragraph*{\textbf{#1:}}}

\newcommand{\concat}{\,\|\,\allowbreak}
\newcommand{\eg}{\emph{e.g., }}
\newcommand{\Wlog}{{w.l.o.g.}\xspace}

\newcommand{\omitit}[1]{}

\usepackage[normalem]{ulem}
\AtBeginDocument{}

\makeatletter
\newcommand{\ipc@xspace}{\futurelet\@let@token\ipc@xspace@check}
\newcommand{\ipc@xspace@check}{\ifx\@let@token(\else\expandafter\xspace\fi}
\DeclareRobustCommand{\ipcCommand}[1]{\texttt{#1}\ipc@xspace} 
\makeatother

\newcommand{\createCommand}{\ipcCommand{create}}
\newcommand{\joinCommand}{\ipcCommand{join}}
\newcommand{\leaveCommand}{\ipcCommand{leave}}
\newcommand{\stakeCommand}{\ipcCommand{stake}}

\newcommand{\withdrawCommand}{\ipcCommand{withdraw}}
\newcommand{\depositCommand}{\ipcCommand{deposit}}
\newcommand{\transferCommand}{\ipcCommand{transfer}}
\newcommand{\checkpointCommand}{\ipcCommand{checkpoint}}
\newcommand{\killCommand}{\ipcCommand{kill}}

\newcommand{\createKeyword}{\ipcCommand{CRT}}
\newcommand{\checkpointKeyword}{\ipcCommand{CPT}}
\newcommand{\transferKeyword}{\ipcCommand{TFR}}
\newcommand{\slashKeyword}{\ipcCommand{SLH}}
\newcommand{\joinKeyword}{\ipcCommand{JOI}}
\newcommand{\stakeKeyword}{\ipcCommand{STK}}
\newcommand{\depositKeyword}{\ipcCommand{FND}}

\newcommand{\script}[1]{\big\langle\, #1 \,\big\rangle}
\newcommand{\scriptSep}{;\ }
\newcommand{\opCode}{\ensuremath{\textsc{Op}}\xspace}
\newcommand{\opReturn}{\ensuremath{\textsc{Op\_Return}}\xspace}
\newcommand{\opPush}{\ensuremath{\textsc{Op\_Push}}\xspace}
\newcommand{\opDrop}{\ensuremath{\textsc{Op\_Drop}}\xspace}
\newcommand{\opTrue}{\ensuremath{\textsc{Op\_True}}\xspace}
\newcommand{\opCheckMultiSig}{\ensuremath{\textsc{Op\_CheckMultiSig}}\xspace}

\newcommand{\opPushBytes}{\ensuremath{\textsc{Op\_PushBytes}}\xspace}
\newcommand{\opPushDataOne}{\ensuremath{\textsc{Op\_PushData1}}\xspace}
\newcommand{\opPushDataTwo}{\ensuremath{\textsc{Op\_PushData2}}\xspace}

\newcommand{\sighashAll}{\texttt{SIGHASH\_ALL}\xspace}
\newcommand{\sighashSingle}{\texttt{SIGHASH\_SINGLE}\xspace}
\newcommand{\sighashACP}{\texttt{ANYONECANPAY}\xspace}
\newcommand{\sighashSingleACP}{\texttt{SIGHASH\_SINGLE|ANYONECANPAY}\xspace}

\newcommand{\createTxOne}{\ensuremath{\mathsf{tx_{crt}^1}}\xspace}
\newcommand{\createTxTwo}{\ensuremath{\mathsf{tx_{crt}^2}}\xspace}
\newcommand{\createTx}{\ensuremath{[\mathsf{tx_{crt}^1}, \mathsf{tx_{crt}^2}]}\xspace}
\newcommand{\joinTxOne}{\ensuremath{\mathsf{tx_{join}^1}}\xspace}
\newcommand{\joinTxTwo}{\ensuremath{\mathsf{tx_{join}^2}}\xspace}
\newcommand{\joinTx}{\ensuremath{[\mathsf{tx_{join}^1}, \mathsf{tx_{join}^2}]}\xspace}
\newcommand{\stakeTx}{\ensuremath{\mathsf{tx_{stake}}}\xspace}
\newcommand{\depositTx}{\ensuremath{\mathsf{tx_{dep}}}\xspace}
\newcommand{\stakeTxPrev}{\ensuremath{\mathsf{tx_{stake}'}}\xspace}
\newcommand{\createTxTwoPrime}{\ensuremath{\mathsf{tx_{crt}^{2*}}}\xspace}
\newcommand{\joinTxTwoPrime}{\ensuremath{\mathsf{tx_{join}^{2*}}}\xspace}

\newcommand{\op}[1]{\textsl{#1}}
\newcommand{\estimatesmartfee}{\op{estimatesmartfee}\xspace}

\newcommand{\computeStateCom}[1]{\ipcCommand{computeCom}(#1)\xspace}
\makeatletter
\newcommand{\retrieveStateCom}{\texttt{getSubnet}\@ifnextchar\bgroup\retrieveStateCom@arg\xspace}
\newcommand{\retrieveStateCom@arg}[1]{\ifx\relax#1\relax\else(#1)\fi\xspace}
\makeatother

\ExplSyntaxOn
\tl_new:N \l__bsl_txtable_in_tl
\tl_new:N \l__bsl_txtable_out_tl
\tl_new:N \l__bsl_txtable_wit_tl
\tl_new:N \l__bsl_txtable_tmp_tl
\tl_new:N \l__bsl_txtable_pre_tl
\int_new:N \l__bsl_txtable_idx_int
\box_new:N \l__bsl_txtable_box
\dim_new:N \l__bsl_txtable_wd_dim
\bool_new:N \l__bsl_txtable_opt_bool
\bool_new:N \l__bsl_txtable_pad_bool  
\tl_new:N \l__bsl_txtable_kind_tl
\tl_new:N \txMarkPrefix
\NewDocumentCommand \txMarks { m } { \tl_set:Nn \txMarkPrefix {#1} }
\cs_new:Npn \__bsl_txtable_mark:n #1
 {
  \tl_if_empty:NF \txMarkPrefix
   {
    \exp_not:N \tikzmark
     { \txMarkPrefix - \l__bsl_txtable_kind_tl - \int_use:N \l__bsl_txtable_idx_int - #1 }
   }
 }
\cs_new_protected:Npn \__bsl_txtable_strip:w ( #1 ) \q_stop
 { \tl_set:Nn \l__bsl_txtable_tmp_tl {#1} }
\cs_new_protected:Npn \__bsl_txtable_qstrip:w ? #1 \q_stop
 { \tl_set:Nn \l__bsl_txtable_tmp_tl {#1} }
\cs_new_protected:Npn \__bsl_txtable_plainpre:
 {
  \bool_if:NTF \l__bsl_txtable_pad_bool
   { \tl_set:Nx \l__bsl_txtable_pre_tl
      { \exp_not:N \phantom{(} \int_use:N \l__bsl_txtable_idx_int \exp_not:N \phantom{)} } }
   { \tl_set:Nx \l__bsl_txtable_pre_tl { \int_use:N \l__bsl_txtable_idx_int } }
 }
\cs_new_protected:Npn \__bsl_txtable_cell:Nn #1#2
 {
  \tl_clear:N #1
  \dim_zero:N \l__bsl_txtable_wd_dim
  \int_zero:N \l__bsl_txtable_idx_int
  \tl_map_inline:nn {#2}
   {
    \tl_set:Nn \l__bsl_txtable_tmp_tl {##1}
    \str_if_eq:eeTF { \tl_item:Nn \l__bsl_txtable_tmp_tl { 1 } } { ? }
     { \tl_set:Nx \l__bsl_txtable_pre_tl { ( \int_use:N \l__bsl_txtable_idx_int ) } }
     { \__bsl_txtable_plainpre: }
    \hbox_set:Nn \l__bsl_txtable_box
     { \scriptsize $ \tl_use:N \l__bsl_txtable_pre_tl {:} \; $ }
    \dim_set:Nn \l__bsl_txtable_wd_dim
     { \dim_max:nn { \l__bsl_txtable_wd_dim } { \box_wd:N \l__bsl_txtable_box } }
    \int_incr:N \l__bsl_txtable_idx_int
   }
  \int_zero:N \l__bsl_txtable_idx_int
  \tl_map_inline:nn {#2}
   {
    \tl_set:Nn \l__bsl_txtable_tmp_tl {##1}
    \bool_set_false:N \l__bsl_txtable_opt_bool
    \str_if_eq:eeT { \tl_item:Nn \l__bsl_txtable_tmp_tl { 1 } } { ? }
     {
      \bool_set_true:N \l__bsl_txtable_opt_bool
      \exp_after:wN \__bsl_txtable_qstrip:w \l__bsl_txtable_tmp_tl \q_stop
     }
    \bool_lazy_and:nnT
     { \str_if_eq_p:ee { \tl_item:Nn \l__bsl_txtable_tmp_tl { 1 } } { ( } }
     { \str_if_eq_p:ee { \tl_item:Nn \l__bsl_txtable_tmp_tl { -1 } } { ) } }
     { \exp_after:wN \__bsl_txtable_strip:w \l__bsl_txtable_tmp_tl \q_stop }
    \tl_replace_all:Nnn \l__bsl_txtable_tmp_tl { , } { , \allowbreak }
    \bool_if:NTF \l__bsl_txtable_opt_bool
     { \tl_set:Nx \l__bsl_txtable_pre_tl { ( \int_use:N \l__bsl_txtable_idx_int ) } }
     { \__bsl_txtable_plainpre: }
    \tl_if_empty:NF #1 { \tl_put_right:Nn #1 { \par } }
    \tl_put_right:Nx #1
     {
      \exp_not:n { \rightskip 0pt~plus~1fil \relax }
      \exp_not:N \hangindent \dim_use:N \l__bsl_txtable_wd_dim
      \exp_not:N \hangafter 1
      \exp_not:N \noindent
      \__bsl_txtable_mark:n { a }
      \tl_if_empty:NF \txMarkPrefix
       { \exp_not:N \rlap { \exp_not:n { \hskip \hsize } \__bsl_txtable_mark:n { r } } }
      \exp_not:N \hbox_to_wd:nn { \dim_use:N \l__bsl_txtable_wd_dim }
       { \exp_not:N \hfil $ \exp_not:V \l__bsl_txtable_pre_tl
         \exp_not:n { {:} } \exp_not:N \; $ }
      $ \exp_not:V \l__bsl_txtable_tmp_tl $
      \__bsl_txtable_mark:n { b }
     }
    \int_incr:N \l__bsl_txtable_idx_int
   }
 }
\bool_new:N \l__bsl_txtable_winline_bool
\NewDocumentCommand \txWitnessInline { } { \bool_set_true:N \l__bsl_txtable_winline_bool }
\cs_new_protected:Npn \__bsl_txtable_icell:Nn #1#2
 {
  \tl_clear:N #1
  \int_zero:N \l__bsl_txtable_idx_int
  \tl_map_inline:nn {#2}
   {
    \tl_set:Nn \l__bsl_txtable_tmp_tl {##1}
    \bool_set_false:N \l__bsl_txtable_opt_bool
    \str_if_eq:eeT { \tl_item:Nn \l__bsl_txtable_tmp_tl { 1 } } { ? }
     {
      \bool_set_true:N \l__bsl_txtable_opt_bool
      \exp_after:wN \__bsl_txtable_qstrip:w \l__bsl_txtable_tmp_tl \q_stop
     }
    \bool_if:NTF \l__bsl_txtable_opt_bool
     { \tl_set:Nx \l__bsl_txtable_pre_tl { ( \int_use:N \l__bsl_txtable_idx_int ) } }
     { \tl_set:Nx \l__bsl_txtable_pre_tl { \int_use:N \l__bsl_txtable_idx_int } }
    \tl_if_empty:NF #1 { \tl_put_right:Nn #1 { \hskip 0.5em plus 0.5em \relax } }
    \tl_put_right:Nx #1
     { \exp_not:N \mbox { $ \exp_not:V \l__bsl_txtable_pre_tl \exp_not:n { {:} } \exp_not:N \, \exp_not:V \l__bsl_txtable_tmp_tl $ } }
    \int_incr:N \l__bsl_txtable_idx_int
   }
 }
\bool_new:N \l__bsl_txtable_linline_bool
\NewDocumentCommand \txLabelsInline { } { \bool_set_true:N \l__bsl_txtable_linline_bool }
\cs_new_protected:Npn \__bsl_txtable_label:Nn #1#2
 {
  \bool_if:NTF \l__bsl_txtable_linline_bool
   { \tl_put_left:Nn #1 { \noindent \textit{#2} \quad } }
   { \tl_put_left:Nn #1 { { \centering \textit{#2} \par } \vspace{0.7ex} } }
 }
\NewDocumentCommand \txTable { O{0.5} m m m m }
 {
  \tl_set:Nn \l__bsl_txtable_kind_tl { in }
  \__bsl_txtable_cell:Nn \l__bsl_txtable_in_tl  {#3}
  \tl_set:Nn \l__bsl_txtable_kind_tl { out }
  \__bsl_txtable_cell:Nn \l__bsl_txtable_out_tl {#4}
  \tl_set:Nn \l__bsl_txtable_kind_tl { wit }
  \bool_if:NTF \l__bsl_txtable_winline_bool
   { \__bsl_txtable_icell:Nn \l__bsl_txtable_wit_tl {#5} }
   { \__bsl_txtable_cell:Nn \l__bsl_txtable_wit_tl {#5} }
  \__bsl_txtable_label:Nn \l__bsl_txtable_in_tl { Inputs }
  \__bsl_txtable_label:Nn \l__bsl_txtable_out_tl { Outputs }
  \bool_if:NTF \l__bsl_txtable_linline_bool
   { \tl_put_left:Nn \l__bsl_txtable_wit_tl { \textit{Witness} \quad } }
   { \tl_put_left:Nn \l__bsl_txtable_wit_tl { \textit{Witness} \par \vspace{-0.3ex} } }
  \par \addvspace { \medskipamount }
  \begingroup \centering \small
  \begin{tabular}
   { | p{ \dimexpr #1\dimexpr ( \linewidth - 4\tabcolsep - 3\arrayrulewidth ) \relax \relax }
     | p{ \dimexpr ( \linewidth - 4\tabcolsep - 3\arrayrulewidth ) - #1\dimexpr ( \linewidth - 4\tabcolsep - 3\arrayrulewidth ) \relax \relax } | }
  \multicolumn{2}{c}{#2} \\
  \hline
  \scriptsize \tl_use:N \l__bsl_txtable_in_tl & \scriptsize \tl_use:N \l__bsl_txtable_out_tl \\
  \hline
  \multicolumn{2}
   { | >{ \centering \arraybackslash } p{ \dimexpr \linewidth - 2\tabcolsep - 2\arrayrulewidth \relax } | }
   { \scriptsize \tl_use:N \l__bsl_txtable_wit_tl } \\
  \hline
  \end{tabular}
  \par \endgroup
 }
\NewDocumentCommand \txTableV { O{0.2} m m m m }
 {
  \bool_set_true:N \l__bsl_txtable_pad_bool
  \tl_set:Nn \l__bsl_txtable_kind_tl { in }
  \__bsl_txtable_cell:Nn \l__bsl_txtable_in_tl  {#3}
  \tl_set:Nn \l__bsl_txtable_kind_tl { out }
  \__bsl_txtable_cell:Nn \l__bsl_txtable_out_tl {#4}
  \tl_set:Nn \l__bsl_txtable_kind_tl { wit }
  \__bsl_txtable_cell:Nn \l__bsl_txtable_wit_tl {#5}
  \par \addvspace { \medskipamount }
  \begingroup \centering \small
  \begin{tabular}
   { | p{ \dimexpr #1\linewidth - 2\tabcolsep \relax }
       p{ \dimexpr \linewidth - #1\linewidth - 2\tabcolsep - 2\arrayrulewidth \relax } | }
  \multicolumn{2}{c}{#2} \\
  \hline
  \scriptsize \textit{Inputs}  & \scriptsize \tl_use:N \l__bsl_txtable_in_tl  \\
  \hline
  \scriptsize \textit{Outputs} & \scriptsize \tl_use:N \l__bsl_txtable_out_tl \\
  \hline
  \scriptsize \textit{Witness} & \scriptsize \tl_use:N \l__bsl_txtable_wit_tl \\
  \hline
  \end{tabular}
  \par \endgroup
  \bool_set_false:N \l__bsl_txtable_pad_bool
 }
\ExplSyntaxOff

\newcommand{\pseudocodeWidth}{0.8\textwidth}
\NewDocumentEnvironment{pseudocodefloat}{}{%
    \iflncs
        \begin{algorithm*}[ht!]%
    \else
        \begin{figure*}[tp]\centering
        \begin{minipage}{\pseudocodeWidth}%
        \begin{algorithm}[H]%
    \fi
}{%
    \iflncs
        \end{algorithm*}%
    \else
        \end{algorithm}%
        \end{minipage}%
        \end{figure*}%
    \fi
}

\NewDocumentEnvironment{txfigure}{O{black} m m}
 {\begin{figure}[!htp]\color{#1}\centering}
 {\caption{#2}\label{#3}\end{figure}}
\NewDocumentEnvironment{txfigure*}{O{black} m m}
 {\begin{figure*}[p]\color{#1}\centering}
 {\caption{#2}\label{#3}\end{figure*}}

\makeatletter
\newdimen\bsl@xmin \newdimen\bsl@xmax \newdimen\bsl@ymin \newdimen\bsl@ymax
\newdimen\bsl@ax \newdimen\bsl@ay \newdimen\bsl@bx \newdimen\bsl@by
\newcommand{\txBoxPadTop}{7.5pt}    
\newcommand{\txBoxPadBottom}{7.5pt}  
\newcommand{\txBoxPadLeft}{1pt}
\newcommand{\txBoxPadRight}{3pt}
\def\bsl@mark#1{\tikz@scan@one@point\pgfutil@firstofone(pic cs:\txMarkPrefix-#1)\relax}
\def\bsl@entry#1{%
  \bsl@mark{#1-a}\bsl@ax=\pgf@x \bsl@ay=\pgf@y
  \bsl@mark{#1-b}\bsl@bx=\pgf@x \bsl@by=\pgf@y
  \ifdim\dimexpr\bsl@ay-\bsl@by\relax>1pt \bsl@mark{#1-r}\bsl@bx=\pgf@x \fi 
  \ifdim\bsl@ax<\bsl@xmin \bsl@xmin=\bsl@ax \fi
  \ifdim\bsl@bx>\bsl@xmax \bsl@xmax=\bsl@bx \fi}
\NewDocumentCommand{\txBox}{O{} m O{}}{%
  \bsl@xmin=\maxdimen \bsl@xmax=-\maxdimen \bsl@ymin=\maxdimen \bsl@ymax=-\maxdimen
  \edef\bsl@list{\zap@space#2 \@empty}%
  \@for\bsl@item:=\bsl@list\do{%
    \bsl@entry{\bsl@item}%
    \ifdim\bsl@ay>\bsl@ymax \bsl@ymax=\bsl@ay \fi
    \ifdim\bsl@by<\bsl@ymin \bsl@ymin=\bsl@by \fi}%
  \edef\bsl@list{\zap@space#3 \@empty}%
  \@for\bsl@item:=\bsl@list\do{\bsl@entry{\bsl@item}}%
  \advance\bsl@xmin-\txBoxPadLeft\relax \advance\bsl@xmax\txBoxPadRight\relax
  \advance\bsl@ymin-\txBoxPadBottom\relax \advance\bsl@ymax\txBoxPadTop\relax
  \node[draw, dashed, thin, rectangle, inner sep=0pt,
        fit={(\the\bsl@xmin,\the\bsl@ymin)(\the\bsl@xmax,\the\bsl@ymax)}, #1] {};}
\makeatother

\ifwhitepaper
  \newcommand{\operand}[1]{\textsl{#1}}
  \newcommand{\tx}{\operand{tx}\xspace}

  \newcommand{\data}{\operand{data}\xspace}
  
  \newcommand{\fee}{\operand{fee}\xspace}
  
  \newcommand{\feeRate}{\operand{feeRate}\xspace}
  \newcommand{\subnetData}{\operand{subnetData}\xspace}
  \newcommand{\joinData}{\operand{joinData}\xspace}
  \newcommand{\stakeData}{\operand{stakeData}\xspace}

  \newcommand{\userAddress}{\operand{userAddress}\xspace}
  
  \newcommand{\amountShort}{\ensuremath{v}\xspace}

  \newcommand{\transferData}{\operand{transferData}\xspace}
  \newcommand{\slashData}{\operand{slashData}\xspace}
  \newcommand{\depositData}{\operand{depositData}\xspace}
  \newcommand{\subnetId}{\operand{subnetId}\xspace}
  \newcommand{\whitelist}{\operand{initialConf}\xspace}
  \newcommand{\whitelistThreshold}{\operand{min\_validators}\xspace}
  
  \newcommand{\checkpointInterval}{\operand{checkpoint-period}\xspace}
  
  \newcommand{\killVotePeriod}{\operand{kill-proposal-period}\xspace}
  \newcommand{\killDelayPeriod}{\operand{kill-delay-period}\xspace}
  \newcommand{\epochLength}{\ensuremath{\mathcal{K}}\xspace}
  
  \newcommand{\colDuration}{\ensuremath{\mathcal{T}}\xspace}
  \newcommand{\subnetHeight}{\ensuremath{h}\xspace}
  \newcommand{\height}{\ensuremath{H}\xspace}
  \newcommand{\comBlock}{\ensuremath{c_\text{bl}}\xspace}
  
  \newcommand{\heightActive}{\ensuremath{H_0}\xspace}
  \newcommand{\subnetConf}{\ensuremath{\mathcal{C}}\xspace}
  \newcommand{\minTimeout}{\ensuremath{s_{min}}\xspace}
  \newcommand{\maxTimeout}{\ensuremath{s_{max}}\xspace}
  \newcommand{\coopExit}{\operand{cooperative-exit}\xspace}
  \newcommand{\uniExit}{\operand{unilateral-exit}\xspace}
  \newcommand{\uniUnbond}{\operand{unilateral-unbonding}\xspace}
  
  \newcommand{\permissionPolicy}{\textsc{PermissionPolicy}\xspace}

  \newcommand{\minCollateral}{\operand{collateral}\xspace}

  \newcommand{\checkpointTxOne}{\ensuremath{\mathsf{tx_{cpt}^1}}\xspace}
  \newcommand{\checkpointTxTwo}{\ensuremath{\mathsf{tx_{cpt}^2}}\xspace}
  \newcommand{\checkpointTx}{\ensuremath{\mathsf{tx_{cpt}^1}, \mathsf{tx_{cpt}^2}}\xspace}

  \newcommand{\colS}{\ensuremath{\mathsf{colS}\xspace}}
  \newcommand{\economicalMode}{\textsc{Economical}\xspace}
  \newcommand{\conservativeMode}{\textsc{Conserative}\xspace}
  \newcommand{\activeSubnet}{\emph{{active}}\xspace}
  \newcommand{\initializedSubnet}{\emph{{initialized}}\xspace}
\else
  \newcommand{\operand}[1]{\ensuremath{#1}}
  \newcommand{\tx}{\operand{tx}\xspace}

  \newcommand{\data}{\operand{D}\xspace}
  
  \newcommand{\fee}{\operand{f}\xspace}
  
  \newcommand{\feeRate}{\operand{feeRate}\xspace}
  \newcommand{\subnetData}{\operand{D_\text{crt}}\xspace}
  \newcommand{\joinData}{\operand{D_\text{joi}}\xspace}
  \newcommand{\stakeData}{\operand{D_\text{stk}}\xspace}

  \newcommand{\userAddress}{\operand{u}\xspace}
  
  \newcommand{\amountShort}{\ensuremath{v}\xspace}

  \newcommand{\transferData}{\operand{D_\text{tfr}}\xspace}
  \newcommand{\slashData}{\operand{D_\text{slh}}\xspace}
  \newcommand{\depositData}{\operand{D_\text{dep}}\xspace}
  \newcommand{\subnetId}{\ensuremath{\mathsf{sid}}\xspace}
  \newcommand{\whitelist}{\operand{C_\text{init}}\xspace}
  \newcommand{\whitelistThreshold}{\ensuremath{n_{\text{init}}}\xspace}
  
  \newcommand{\checkpointInterval}{\operand{\ell_\text{cpt}}\xspace}
  
  \newcommand{\killVotePeriod}{\operand{\ell_\text{kill}}\xspace}
  \newcommand{\killDelayPeriod}{\operand{d_\text{kill}}\xspace}
  \newcommand{\epochLength}{\ensuremath{\mathcal{K}}\xspace}
  
  \newcommand{\colDuration}{\ensuremath{\mathcal{T}}\xspace}
  \newcommand{\subnetHeight}{\ensuremath{h}\xspace}
  \newcommand{\comBlock}{\ensuremath{c_\text{bl}}\xspace}
  
  \newcommand{\height}{\ensuremath{H}\xspace}
  \newcommand{\heightActive}{\ensuremath{H_0}\xspace}
  \newcommand{\subnetConf}{\ensuremath{\mathcal{C}}\xspace}
  \newcommand{\minTimeout}{\ensuremath{s_{min}}\xspace}
  \newcommand{\maxTimeout}{\ensuremath{s_{max}}\xspace}
  \newcommand{\coopExit}{{CE}\xspace}
  \newcommand{\uniExit}{{UE}\xspace}
  \newcommand{\uniUnbond}{{UU}\xspace}
  
  \newcommand{\permissionPolicy}{\textsc{MembPolicy}\xspace}

  \newcommand{\minCollateral}{\ensuremath{v_{\text{min}}}\xspace}

  \newcommand{\checkpointTxOne}{\ensuremath{\mathsf{tx_{cpt}^1}}\xspace}
  \newcommand{\checkpointTxTwo}{\ensuremath{\mathsf{tx_{cpt}^2}}\xspace}
  \newcommand{\checkpointTx}{\ensuremath{\mathsf{tx_{cpt}^1}, \mathsf{tx_{cpt}^2}}\xspace}

  \newcommand{\colS}{\mathsf{colS}\xspace}
  \newcommand{\economicalMode}{\textsc{Economical}\xspace}
  \newcommand{\conservativeMode}{\textsc{Conserative}\xspace}
  \newcommand{\activeSubnet}{\emph{{active}}\xspace}
  \newcommand{\initializedSubnet}{\emph{{initialized}}\xspace}
\fi

\newcommand{\tokenName}[1]{\operand{#1}\xspace}
\newcommand{\BTC}{\tokenName{BTC}}
\newcommand{\wBTC}{\tokenName{wBTC}}
\newcommand{\IPCBTC}{\tokenName{IPC-BTC}}

\newcommand{\ipcAwareNode}{\ifanon{ScaleBTC node}\else{\emph{IPC-aware} node}\fi\xspace}
\newcommand{\ipcAwareNodes}{\ifanon{ScaleBTC nodes}\else{\emph{IPC-aware} nodes}\fi\xspace}
\newcommand{\anIpcAwareNode}{\ifanon{a ScaleBTC node}\else{an \emph{IPC-aware} node}\fi\xspace}
\newcommand{\AnIpcAwareNode}{\ifanon{A ScaleBTC node}\else{An \emph{IPC-aware} node}\fi\xspace}
\newcommand{\btcProvider}{\ipcAwareNode}
\newcommand{\btcMonitor}{\ipcAwareNode}

\newcommand{\bitcoinIpc}{\ifanon{ScaleBTC}\else{\emph{Bitcoin-IPC}}\fi\xspace}
\newcommand{\versionNumber}{\emph{v0.4.0}\xspace}

\newcommand{\BslBitcoinIpcRepo}{\ifanon\emph{Anonymized}\else\texttt{BitcoinScalingLabs/bitcoin-ipc}\fi\xspace}

\newcommand{\return}{\State{\textbf{return}}\xspace}
\algrenewcommand\textproc[1]{#1}
\algblockdefx[On]{On}{EndOn}[1]{\textbf{upon} #1 \textbf{do}}{\textbf{end upon}}
\algblockdefx[Operation]{Operation}{EndOperation}[1]{\textbf{operation} #1}{\textbf{end operation}}
\newcommand{\Let}[2]{\State{$#1 \gets #2$}}
\newcommand{\pcop}[2]{\ensuremath{\textsl{#1}(#2)}}
\newcommand{\opInit}[1]{\pcop{init}{#1}}
\newcommand{\opSign}[1]{\pcop{sign}{#1}}
\newcommand{\opVerify}[1]{\pcop{verifySig}{#1}}
\newcommand{\opTxid}[1]{\pcop{txid}{#1}}
\newcommand{\opMerkleRoot}[1]{\pcop{merkleRoot}{#1}}
\newcommand{\subnetParams}{\ensuremath{P}\xspace}
\newcommand{\nodeParams}{\ensuremath{\mathit{params}}\xspace}
\newcommand{\nodeStatus}{\ensuremath{\mathit{status}}\xspace}
\newcommand{\nodeEpoch}{\ensuremath{\mathit{epoch}}\xspace}
\newcommand{\nodeConf}{\ensuremath{\mathit{conf}}\xspace}

\newcommand{\nodeCollaterals}{\ensuremath{\mathit{collaterals}}\xspace}
\newcommand{\nodeCheckpoints}{\ensuremath{\mathit{checkpoints}}\xspace}
\newcommand{\nodeSlashable}{\ensuremath{\mathit{slashable}}\xspace}
\newcommand{\nodeActHeight}{\ensuremath{\mathit{actHeight}}\xspace}
\newcommand{\nodeValidators}{\ensuremath{\mathit{validators}}\xspace}
\newcommand{\nodeNonces}{\ensuremath{\mathit{nonces}}\xspace}
\newcommand{\nodeEotsSk}{\ensuremath{\mathit{eotsSk}}\xspace}
\newcommand{\nodeSecNonces}{\ensuremath{\mathit{secNonces}}\xspace}
\newcommand{\statusInitialized}{\ensuremath{\textsc{initialized}}\xspace}
\newcommand{\statusActive}{\ensuremath{\textsc{active}}\xspace}
\newcommand{\statusToBeKilled}{\ensuremath{\textsc{to-be-killed}}\xspace}
\newcommand{\statusKilled}{\ensuremath{\textsc{killed}}\xspace}
\newcommand{\opProcessCreate}[1]{\pcop{processCreate}{#1}}
\newcommand{\opProcessJoin}[1]{\pcop{processJoin}{#1}}

\newcommand{\opBuildCollateralScript}[1]{\pcop{buildColScript}{#1}}
\newcommand{\opInitEotsKeys}[1]{\pcop{initEotsKeys}{#1}}
\newcommand{\opGenEotsNonce}[1]{\pcop{genEotsNonce}{#1}}
\newcommand{\opEotsSign}[1]{\pcop{eotsSign}{#1}}
\newcommand{\opEotsVerify}[1]{\pcop{eotsVerify}{#1}}
\newcommand{\opEotsExtract}[1]{\pcop{eotsExtract}{#1}}
\newcommand{\opGetConfiguration}[1]{\pcop{getConfiguration}{#1}}
\newcommand{\opValidateJoinParams}[1]{\pcop{validateColParams}{#1}}
\newcommand{\opProcessStake}[1]{\pcop{processStake}{#1}}
\newcommand{\opCheckNonceCom}[1]{\pcop{checkNonceCom}{#1}}
\newcommand{\opValidCreateParams}[1]{\pcop{validateCreateParams}{#1}}
\newcommand{\opDeriveSubnetId}[1]{\pcop{deriveSubnetId}{#1}}
\newcommand{\opCreateCheckpointTxs}[1]{\pcop{createCheckpointTxs}{#1}}
\newcommand{\opProcessCheckpoint}[1]{\pcop{processCheckpoint}{#1}}
\newcommand{\opUnspent}[1]{\pcop{listUnspent}{#1}}
\newcommand{\opBuildUnbondingScript}[1]{\pcop{buildUnbS}{#1}}
\newcommand{\opSubnetOf}[1]{\pcop{subnetOf}{#1}}
\newcommand{\opBuildSlashTx}[1]{\pcop{buildSlashTx}{#1}}

\newcommand{\andl}{\textbf{and}\xspace}

\newcommand{\pk}{\mathsf{pk}}
\newcommand{\sk}{\mathsf{sk}}
\newcommand{\pkExit}[1][V]{\ensuremath{\pk_{#1}^{\mathsf{L1}}}\xspace}

\newcommand{\pkCons}[1][V]{\ensuremath{\pk_{#1}^{\mathsf{cons}}}\xspace}

\newcommand{\pkEots}[1][V]{\ensuremath{\pk_{#1}^{\mathsf{eots}}}\xspace}
\newcommand{\skEots}[1][V]{\ensuremath{\sk_{#1}^{\mathsf{eots}}}\xspace}

\newcommand{\slashTx}{\ensuremath{\mathsf{tx_{slash}}}\xspace}
\newcommand{\unbS}{\mathsf{unbS}\xspace}

\newcommand{\nonceCom}{\ensuremath{\Gamma}\xspace}
\newcommand{\nonceComSig}{\ensuremath{\sigma_{\nonceCom}}\xspace}
\newcommand{\nonceStart}{\ensuremath{h_{\mathsf{start}}}\xspace}
\newcommand{\nonceCount}{\ensuremath{\ell_{\mathsf{nonce}}}\xspace}
\newcommand{\nonceRoot}{\ensuremath{\rho}\xspace}
\newcommand{\nonce}[1]{\ensuremath{N_{#1}}}
\newcommand{\secNonce}[1]{\ensuremath{k_{#1}}}
\newcommand{\slashFee}{\ensuremath{\amountShort_\text{fee}}\xspace}
\newcommand{\eotsSig}{\ensuremath{\sigma_V^{\mathsf{eots}}}\xspace}

\newcommand{\depth}{\ensuremath{k}\xspace}
\newcommand{\inclDelay}{\ensuremath{\Delta_{\mathsf{incl}}}\xspace}
\newcommand{\slashDelay}{\ensuremath{\Delta_{\mathsf{slash}}}\xspace}
\newcommand{\adoptDelay}{\ensuremath{\Delta_{\mathsf{adopt}}}\xspace}
\newcommand{\obsDelay}{\ensuremath{\Delta_{\mathsf{obs}}}\xspace}

\iflncs\else\newcounter{example}\fi 
\makeatletter
\newif\if@resuming@example
\@resuming@examplefalse

\iflncs 
  \let\example\@undefined
  \let\endexample\@undefined
\fi
\NewTColorBox{example}{O{}m}{
  before title={%
    \def\@example@label{#1}%
    \def\@example@title{#2}%
    \ifx\@example@label\@empty
      \@resuming@examplefalse
      \stepcounter{example}%
    \else
      \@ifundefined{@exnum@\@example@label}{%
        \@resuming@examplefalse
        \stepcounter{example}%
        \expandafter\gdef\csname @exnum@\@example@label\endcsname{\number\value{example}}%
      }{%
        \@resuming@exampletrue
        \setcounter{example}{\csname @exnum@\@example@label\endcsname}%
      }%
    \fi
    \ifx\@example@title\@empty
      \def\@example@title@final{Example \theexample}%
    \else
      \def\@example@title@final{Example \theexample: \@example@title}%
    \fi
  },
  before upper={%
    \def\@example@label{#1}%
    \ifx\@example@label\@empty
      \addtocounter{example}{-1}%
      \refstepcounter{example}%
    \else
      \@ifundefined{@exnum@\@example@label}{%
        \addtocounter{example}{-1}%
        \refstepcounter{example}%
      }{%
        \addtocounter{example}{-1}%
        \refstepcounter{example}%
      }%
    \fi
  },
  after upper={%
    \if@resuming@example
      \@resuming@examplefalse
    \fi
  },
  colback=gray!5,
  colframe=gray!50!black,
  title=\@example@title@final,
  left=6pt,
  right=6pt,
  top=6pt,
  bottom=6pt,
  boxrule=0.5pt,
  breakable
}
\makeatother

\crefname{example}{Example}{Examples}
\crefname{case}{case}{cases}
\Crefname{case}{Case}{Cases}
\Crefname{example}{Example}{Examples}
\creflabelformat{example}{#2#1#3}

\iflncs\else
\theoremstyle{definition}
\newtheorem{definition}{Definition}

\newtheorem{lemma}{Lemma}
\newtheorem{remark}{Remark}
\fi

\newcommand{\anonurl}[1]{\ifanon{URL omitted for anonymization reasons.}\else\url{#1}\fi}

\iflncs
\makeatletter
\renewcommand\subsubsection{\@startsection{subsubsection}{3}{\z@}%
        {-12\p@ \@plus -4\p@ \@minus -4\p@}%
        {0.1\p@}%
        {\normalfont\small\bfseries\boldmath}}
\makeatother
\fi

\iflncs\else
\fi
\begin{document}
\title{\bitcoinIpc: Scaling Bitcoin\\with a Network of Proof-of-Stake Subnets}

\iflncs
  \titlerunning{\bitcoinIpc: Scaling Bitcoin with a Network of Proof-of-Stake Subnets}
  \ifanon
    \author{}
    \institute{}
  \else
    \authorrunning{O. Alpos et al.}
    \author{Orestis Alpos\inst{1} \and
    Jakov Mitrovski\inst{2} \and
    Themis Papameletiou\inst{2} \and
    Nikola Risti\'c\inst{2} \and
    Dionysis Zindros\inst{2} \and
    Marko Vukoli\'c\inst{1}}
    \institute{Bitcoin Scaling Labs \and Common Prefix}
  \fi
\else 
\ifanon
    \author{}
  \else
    \author{
    Orestis Alpos\\ Bitcoin Scaling Labs \and
    Jakov Mitrovski\\ Common Prefix \and
    Themis Papameletiou\\Common Prefix \and
    Nikola Risti\'c\\Common Prefix  \and
    Dionysis Zindros\\Common Prefix \and
    Marko Vukoli\'c\\ Bitcoin Scaling Labs}
  \fi
\fi

\date{}
\maketitle

\begin{abstract}
Despite holding over $\$1.7T$ in value, Bitcoin scales poorly: Layer-1 (L1) processes only a few transactions per second, and Layer-2 (L2) solutions suffer from operational and liquidity fragmentation that requires external bridges for cross-chain connectivity.

We present \bitcoinIpc, a protocol that horizontally scales Bitcoin via permissionless, interconnected, programmable Proof-of-Stake L2 subnets staked in L1 BTC. Subnets leverage Bitcoin L1 for interconnectivity, settlement, and security, including Sybil protection, equivocating validator slashing, unilateral exit for honest stake, long-range attack protection, and compromised subnet firewalling. \bitcoinIpc cross-subnet transfers involve only source/destination  subnet validators and L1, with no external bridges or per-path liquidity lock-up. Embedded and batched in Bitcoin's witness mechanism, cross-subnet settlements can approach 6 vB per transfer, 23x less than a native L1 payment. Filling Bitcoin blocks with such batches corresponds to L1 monetary throughput of about 273 transfers per second, without modifying Bitcoin.




\end{abstract}

\section{Introduction} \label{sec:intro}

The Bitcoin \cite{Nakamoto:Bitcoin} community has accepted relatively low throughput on Bitcoin mainnet (L1), preferring addressing scalability on so-called Layer 2 (L2) to increasing Bitcoin block size \cite{BlocksizeWar}.
The commonly cited 7~transactions per second (tps) L1 throughput \cite{Vukolic21}  corresponds to an average transaction (tx) size of about 238~virtual-bytes (vB) and roughly matches Bitcoin average throughput we  observed over weeks/months of operation. In principle, the achievable throughput on Bitcoin L1 is somewhat higher: with 141~vB/tx (size of a baseline ``send + change'' tx produced by modern wallets), 1M~vB per block -- corresponding to a full Bitcoin block, and one block every 10 minutes, \emph{baseline throughput} is 11.8 tps, still insufficient for global use of Bitcoin. 

To  address  this low L1 throughput, the main L2 scalability approach today remains the Lightning Network (LN)~\cite{poon2016bitcoinlightning}, 
where a \emph{pair} of Bitcoiners record payments between them in a dedicated off-chain channel and connect such 2-party channels into a payment network. LN's mixed success \cite{waugh2020empirical,beres2020cryptoeconomic,divakaruni2022lightning} helped spawn a large wave of Bitcoin L2 projects (hundreds of proposals with fewer live chains \cite{qi2025sok}), many adding smart-contract support on a multi-party ledger. However, most of them are isolated from one another: value locked in one L2 cannot be used in another L2 without an \emph{external} bridge or a swap~\cite{tang2025bitcoinbridge}, leaving this L2 ecosystem  fragmented~\cite{hassan2025sok,qi2025sok}.


\bitcoinIpc horizontally scales Bitcoin, avoiding aforementioned limitations. It is a \emph{network} of programmable L2 chains, called \emph{subnets}, interconnected by Bitcoin itself. 
The network is \emph{open}: any \BTC holder can instantiate a subnet, which may be permissioned or permissionless. To become a subnet validator, \BTC holders \emph{stake} their \BTC as \emph{collateral}; subnet validators run a BFT consensus protocol weighted by their stake (e.g., \cite{cometbft}). Importantly, subnets allow \emph{unilateral exit} to honest validators, that can withdraw their collateral without cooperation from other validators. The value is transfered across subnets without per-path or per-subnet liquidity lock-ups. Transfers are executed by validators of communicating  subnets and routed through Bitcoin, without relying on third-party bridges. 

Transfers are \emph{batched}, packing many cross-subnet transfers into Bitcoin transaction witness data. Thus, \bitcoinIpc amortizes the size and the cost for monetary transfers on Bitcoin and reduces them up to \emph{23x} relative to a typical native Bitcoin payment (6 vB/transfer in large-batch limit vs. 141 vB/tx).  Filling L1 blocks with such batches corresponds to about 273 monetary transfers per second, 23x the baseline throughput. Intra-subnet transfers remain on separate paths, which horizontally scale, independently of L1 throughput. At the same time, each subnet runs its own programmable engine (\eg EVM) for smart-contract programmability. \bitcoinIpc leverages Bitcoin for security in multiple aspects. It uses \BTC as collateral for sybil-attack protection, slashing collateral of equivocating validators, periodical anchoring to Bitcoin, persisting a tamper-evident history that bounds PoS long-range attacks \cite{steinhoff2021bms,DBLP:conf/consensusday/AzouviV22}. Moreover,  cross-subnet interoperability inherits a \emph{firewall property}~\cite{DBLP:conf/sp/GaziKZ19} from UTXO conservation: a compromised subnet cannot extract more \BTC than was deposited into it. \bitcoinIpc provides these features using only existing Bitcoin script capabilities, requiring no consensus update, and is open source available.\footnote{\anonurl{https://github.com/bitcoinscalinglabs/bitcoin-ipc}} 
\smallskip

\noindent\textbf{Related work.} 
Horizontal scaling (HS) via PoS sidechains has been proposed and implemented for contract-capable L1s  \cite{DBLP:conf/sp/GaziKZ19,DBLP:conf/icdcs/RochaKSV22}. 
The novelty of \bitcoinIpc lies in the instantiation of this paradigm on Bitcoin, which offers no general computation.  Subnet creation, dynamic validator membership, anchoring and cross-subnet transfers are encoded as ordinary Bitcoin transactions.
Among proposed Bitcoin scaling techniques, only the Lightning Network~\cite{poon2016bitcoinlightning} offers HS; it is a network of two-party payment channels without general-purpose programmability.


While both \bitcoinIpc and Babylon~\cite{BabylonTR,DBLP:journals/corr/abs-2408-01896} enable non-custodial Bitcoin staking and equivocation slashing, their goals and architectures differ fundamentally. Babylon proxies information from foreign PoS chains (called BSNs) via a dedicated Bitcoin L2 (\emph{Babylon Genesis}) that BSNs must trust, making them sort of Bitcoin L3s. Babylon does not aim at scaling Bitcoin, lacking cross-BSN interoperability and requiring manual BSN bootstrap. In contrast, \bitcoinIpc subnets act as independent L2s, trusting \emph{only} Bitcoin, can be created permissionlessly, and feature built-in cross-subnet interoperability via Bitcoin L1.

Stacks~\cite{stacks-nakamoto-wp,ali2020pox} 
offers a \emph{single} L2 chain, anchored to Bitcoin for reorg protection. Its security rests on its own STX token:
consensus is signed by STX holders staking on the L2. \bitcoinIpc, instead, supports a dynamic network of interoperable subnets, uses \BTC for both voting weight and slashing upon equivocation, and treats any staking reward as optional rather than a credential for joining. 

Protocols such as BRC-20, Taproot Assets, and Runes~\cite{brc20-docs,taproot-assets-docs,runes-docs} allow creation of custom assets with limited logic on Bitcoin. 
RGB~\cite{rgb-specs} supports richer logic, and BRC2.0~\cite{brc20-programmable-docs}
aims at full EVM support. However, they work as overlays over Bitcoin, with 
and \emph{indexer} reading smart contracts and their inputs from Bitcoin and executing them off-chain. Users only interact with the programs through Bitcoin, and the indexer is fully trusted to provide the correct result. 
In contrast, \bitcoinIpc enables autonomous, permissionlessly-created, fault-tolerant chains.




\smallskip

\noindent\textbf{Organization.}
\iflncs
    \Cref{sec:tx-notation} contains preliminary definitions, \bitcoinIpc interface and properties. \Cref{sec:operations} describes the core construction and \Cref{sec:benchmarks} the most interesting benchmarks. Appendices contain protocol pseudocode, complete formalization and proofs, additional benchmarks, incentivization and discussion.
\else
    \Cref{sec:tx-notation} contains preliminary definitions, \bitcoinIpc interface and properties. \Cref{sec:operations} describes the core construction, \Cref{app:staking} the optional staking incentives, and \Cref{sec:benchmarks} the benchmarks. \Cref{sec:discussion} concludes with a discussion and future work. Appendices contain protocol pseudocode, complete formalization and proofs, and implementation notes.
\fi

\section{Preliminaries}
\label{sec:preliminaries}
\label{sec:tx-notation}
\label{sec:interface}
\label{sec:bitcoin-definition}

\begin{description}[leftmargin=0pt, font=\normalfont\bfseries]
\item[Bitcoin transaction model{.}]
In Bitcoin, value resides in transaction outputs (UTXOs), each with a  \emph{value} (in BTC) and a \emph{locking script}. A UTXO can only be spent once by providing a \emph{witness} that satisfies its locking script. A transaction consumes existing UTXOs as \emph{inputs} and creates new UTXOs at \emph{outputs}. We allow for 0-value outputs, \Wlog
Locking scripts $S$ follow the grammar:
{\small\begin{center}\binoppenalty=10000 \interlinepenalty=10000
$S ::= \opCode \mid \mathsf{CheckSig}_{\mathit{pk}} \mid \mathsf{CheckMultiSig}_{C}\footnote{In Tapscript, the script language used in script trees, \opCheckMultiSig, can be implemented from other opcodes: \url{https://en.bitcoin.it/wiki/BIP_0342}.} \mid \mathsf{AbsTimelock}(H) \mid \mathsf{RelTimelock}(\delta) \mid S \wedge S \mid S \vee S$
\end{center}}where \opCode are Bitcoin \emph{opcodes} with their standard semantics (we will introduce them as needed), $C$ is a set of weighted public keys and $\mathsf{CheckMultiSig}_{C}$ is satisfied with signatures from more than $2/3$ of the total weight.
A \emph{script tree} $T = \langle P_{\mathit{int}}, \mathcal{S}\rangle$ pairs an \emph{internal} public key $P_{\mathit{int}}$ with a set of \emph{leaf} scripts $\mathcal{S}$. Spending a UTXO under $T$ requires either a valid signature under $P_{\mathit{int}}$ or a triplet $(S_k, \pi_k, w_k)$, where $w_k$ is a witness that satisfies $S_k$, and $\pi_k$ proves $S_k \in T$. Here, unless stated otherwise, $P_{\mathit{int}}$ is a NUMS point~\cite{bip341}, hence a script tree can only be spent through a leaf script. Each signature in the witness (of some input $k$) carries a \emph{sighash} type, determining what it commits to; \sighashAll: every input and output; \sighashSingle: every input and the output at index $k$; modifier \sighashACP (in conjunction with a type): keep only input $k$. Transaction fields, such as version and locktime, are always included.

A transaction $\mathsf{tx}$ consists of inputs $((\mathsf{tx_{prev}}, n), S)$, outputs $(v, S')$, and per-input witnesses, where $(\mathsf{tx_{prev}}, n)$ is the spent outpoint, $S$ the unlocking script, $v$ is value, and $S'$ the locking script. 
Wildcards $*$ mark logic-irrelevant components, and parenthesized indices indicate optional fields.
A transaction is \emph{valid} if: (1) every input spends an existing, unspent UTXO; (2) every input witness satisfies its script $S \in T$ and respects all timelocks; and (3) total output value does not exceed total input value ($\sum v_{\mathit{out}} \le \sum v_{\mathit{in}}$), the difference being \emph{transaction fees}.

\item[Blockchain, safety and liveness.] 
    An \emph{honest} party follows the prescribed protocol, while a \emph{Byzantine} party may arbitrarily deviate. A \emph{blockchain} totally-orders, finalizes, and executes transactions batched into \emph{blocks}, indexed by \emph{block height}.
    Following ~\cite{DBLP:conf/crypto/GarayKL17}, a blockchain is \emph{safe} if honest users agree on transaction order and state, and \emph{live} if transactions by honest users are finalized. Moreover, we call a blockchain \emph{correct} if it is safe and live, and \emph{compromised} otherwise.

    \item[Bitcoin.] We model Bitcoin network as \emph{synchronous}, that is, all messages arrive within a known delay $\Delta$, and the adversary may arbitrarily reorder network messages. We assume Bitcoin is correct and we parse and process Bitcoin transactions after six confirming blocks, when we consider them finalized~\cite{DBLP:conf/crypto/GarayKL17}.

    \item[Parties.]  The parties relevant to our protocol are \emph{\ipcAwareNodes}, \emph{subnet validators} and \emph{users}. Subnet validators are also \ipcAwareNodes, and a party may have all three roles. \AnIpcAwareNode is the core party in our protocol. It runs Bitcoin full node, parses finalized Bitcoin blocks for \bitcoinIpc transactions, maintains the \bitcoinIpc state, and constructs Bitcoin transactions implementing \bitcoinIpc operations.
     Optionally, it may join one or more subnets as a \emph{validator}, in which case it also runs the subnet-specific consensus and execution protocols, augmented with protocol hooks for creating and signing \bitcoinIpc-specific transactions 
     and relaying them to Bitcoin. \emph{Users} interact with subnets by submitting transactions or querying state. Any party can be Byzantine. 

\item[\bitcoinIpc subnets.] 
    A \emph{subnet} is an independent blockchain with PoS consensus \cite{PoSbitcointalk}, where each validator's weight is determined by its stake in \BTC. Its \emph{configuration} $\subnetConf_e$ (dynamic across epochs) is the set of validator-stake pairs in \emph{epoch} $e$, where a validator $V$ is identified by its Bitcoin key \pkExit. Epochs are parameterized per subnet by length $\epochLength$ (in Bitcoin blocks). Epoch $e \ge 1$ spans Bitcoin heights $h \in (\height_{e-1}, \height_e]$, bounded by \emph{cutoff heights} $\height_e = \height_{e-1} + \epochLength$. Epoch~$0$ is an exception: it spans from the finalization of the subnet's \createCommand command on Bitcoin to its \emph{activation} at height $\heightActive$ (\Cref{sec:create-subnet}).
    

\item[Interface and properties of \bitcoinIpc.] 
    \AnIpcAwareNode can perform the following \emph{operations}:
    \begin{enumerate*}[label=\textbf{(\arabic*)}]
    \item \createCommand: Create a subnet with permissionless or permissioned membership policy, persisting its parameters on Bitcoin; 
    \item \joinCommand and \leaveCommand: Become a validator in a subnet (if its permission policy allows it) or stop being one, by locking and retrieving collateral;
    \item \stakeCommand: Add or remove collateral on permissionless subnets;
    \item \killCommand\iflncs\footnote{Due to space limitations, \killCommand will be presented in \Cref{sec:protocol-details}.}\fi: Initiate termination of a subnet;
    \item \retrieveStateCom: Retrieve a subnet's current state (including joined validators and anchored blocks), as persisted on Bitcoin.
    Moreover, users can perform:
    \item \depositCommand: Lock \BTC on Bitcoin to mint an equal amount of its representation (\wBTC) on a chosen subnet;
    \item \withdrawCommand: Burn \wBTC on a subnet to release an equal amount of \BTC on Bitcoin; \item \transferCommand: Send \wBTC across/within a  subnet -- transfers across subnets are routed via Bitcoin.
    \end{enumerate*}
    Operations (1)-(6)
    interact directly with Bitcoin, while (7)-(8) 
    are submitted to a subnet, which then interacts with Bitcoin. Operations are implemented as standard Bitcoin transactions, requiring no Bitcoin upgrade; standard Bitcoin nodes may simply ignore \bitcoinIpc transactions.
    
    \bitcoinIpc creates three UTXO types.
    \emph{Collateral UTXOs} hold validator stake and are locked with a validator-defined timeout, supporting \emph{unilateral exit (\uniExit)} and \emph{cooperative exit (\coopExit)}. 
    \emph{Fund UTXOs} hold user deposits and incoming transfers, \emph{Fee UTXOs} hold reserves that pay Bitcoin fees for subnet checkpoints, both locked by \opCheckMultiSig scripts.
 
\end{description}

\dotparagraph{Security properties} 
    We make \emph{no assumptions} on the number of Byzantine validators within each subnet. 
    However, for each individual subnet, assuming that fraction $F_e$ of the total stake within configuration $\subnetConf_e$ of epoch $e$ is controlled by Byzantine validators, \bitcoinIpc achieves:
    \textbf{(i)} $F_e \le \frac{1}{3}$ for every $e$: The subnet is correct, and we call it \emph{$1/3$-bounded};
    \textbf{(ii)} $\frac{1}{3} < F_e \le \frac{2}{3}$ for some $e$: Safety violations may happen during epoch $e$ but imply validator equivocation, which is attributable and \emph{slashed}.
    Liveness violations may happen (e.g., stalling \joinCommand of new validators and checkpoints), but funds are never stolen: spending them requires more than $2/3$ of the stake.
    \textbf{(iii)} $F_e > \frac{2}{3}$ for some $e$: The subnet is compromised, with no guarantees to its users.
    Moreover, the security of a subnet does \emph{not} depend on validators that have withdrawn their collateral. This model --- usual in PoS constructions~\cite{steinhoff2021bms,DBLP:conf/consensusday/AzouviV22,DBLP:conf/icdcs/RochaKSV22} --- gives rise to \emph{Long-Range Attacks (LRA)}~\cite{DeirmentzoglouP19}: Not being at risk of slashing, exited validators can extend an old block (where their collateral was still active) and fork-out legitimate transactions. \bitcoinIpc relies on Bitcoin to mitigate LRA.
    
To further protect honest validators from Byzantine validators and users of correct subnets from other compromized subnets, \bitcoinIpc always provides the following properties (formalized in \Cref{sec:model-full}).
    
    \noindent\textbf{P1: Firewall \cite{DBLP:conf/sp/GaziKZ19,DBLP:conf/icdcs/RochaKSV22}.}
    Compromised subnets cannot send out (\withdrawCommand or \transferCommand) more funds that what has actually be sent into them,
    thereby protecting correct subnets from compromized ones.
    
    \noindent\textbf{P2: Long-Range Attack (LRA) protection.} \bitcoinIpc mitigates LRA through a Bitcoin-anchored history, using the commitments returned by \retrieveStateCom.
    

    \noindent\textbf{P3: Unilateral exit.}
    An honest validator can always retrieve its collateral.

    \noindent\textbf{P4: No external bridges.} In contrast to prior inter-blockchain protocols~\cite{DBLP:journals/iacr/AumayrALMPSZ24,DBLP:conf/sp/GaziKZ19,DBLP:conf/fc/ZamyatinAZKMKK21} that rely on additional parties running a dedicated \emph{bridge} protocol, \bitcoinIpc uses only the validators of the communicating subnets and Bitcoin for \depositCommand, \transferCommand, and \withdrawCommand, and thus assumes no trust in any auxiliary parties.


\ifwhitepaper
    \input{sections/subnet-lifecycle.tex}
    \input{sections/subnet-addressing.tex}
    \input{sections/components.tex}
\fi

\section{\bitcoinIpc design and implementation}\label{sec:operations}

\subsection{\bitcoinIpc data in Bitcoin transactions} \label{sec:bitcoin-scripting}\label{sec:write-arbitrary}\label{sec:data-script}

\bitcoinIpc persists data on Bitcoin (subnet parameters, block commitments, transfer details) in a way verifiable by all parties, using two mechanisms: (1) an $\opReturn(\data)$ output, an opcode that is never satisfied but carries data \data, which we limit to 80 bytes per transaction, and (2) the \emph{commit-reveal technique}, which carries larger payloads (up to Bitcoin's maximum standard transaction size of $100$K virtual bytes) in a transaction witness, enjoying 
the $4\times$ discounted fee rate.
It uses the \emph{data script}
{\small$\mathsf{dataS}(\data) \;=\; \script{\opPush(\data) \scriptSep \opDrop \scriptSep \opTrue}$}, which pushes and drops \data (in chunks of at most 520 bytes) and then pushes $\opTrue$, so it is satisfied by any witness.
A first transaction creates an output locked by the script tree $\langle P_{\mathit{int}}, {\mathsf{dataS}(\data)}\rangle$, which commits to $\data$ without revealing it; a second transaction spends it through its single leaf and its witness must exhibit the script, disclosing \data on-chain.

\iflncs\else
    \Cref{table:ipc-data-techniques} summarizes the used techniques.
    \begin{table}[h]
        \centering
        \resizebox{\columnwidth}{!}{%
        \begin{tabular}{|c|c|c|}
            \hline
            \textbf{Command} & \textbf{Data attached} & \textbf{Technique}\\ \hline
            \createCommand~(\ref{sec:create-subnet}) & subnet parameters & commit-reveal \\ \hline
            \joinCommand~(\ref{sec:join-subnet}) & validator details  & commit-reveal \\ \hline
            \stakeCommand~(add/remove stake)~(\ref{sec:dynamic-participation}) & validator, amount & \opReturn \\ \hline
            persist state~(\ref{sec:persist-state}) & subnet state  & \opReturn \\ \hline
            \depositCommand~(\ref{sec:deposit}) & recipient address  & \opReturn \\ \hline
            \transferCommand~(\ref{sec:transfer}) & recipients, amounts  & commit-reveal \\ \hline
        \end{tabular}}
        \caption{Data attachment techniques used by \bitcoinIpc commands. Commands whose metadata fits in 80 bytes use \opReturn for simplicity; commands carrying larger payloads use the commit-reveal technique, taking advantage of witness discount and the higher capacity.}
        \label{table:ipc-data-techniques}
    \end{table}
\fi
\iflncs
    \subsection{Subnet lifecycle: \createCommand, \joinCommand, and \leaveCommand}
\else
    \subsection{Subnet lifecycle: \createCommand, \joinCommand, \leaveCommand, and \killCommand}
\fi
\label{sec:create-subnet}\label{sec:join-subnet}

\iflncs\vspace{-3pt}\fi
\dotparagraph{The \createCommand operation} 
Any \ipcAwareNode can permissionlessly create a new subnet on Bitcoin, parameterizing \createCommand as follows:
\begin{enumerate*}[label=\textbf{(\arabic*)}]
    \item \minCollateral: minimum collateral required per validator.
    \item \checkpointInterval: number of subnet blocks between successive checkpoints.
    \item \emph{epoch length} \epochLength:
    number of Bitcoin blocks between successive cutoffs.
    \item $\minTimeout, \maxTimeout$: the minimum and maximum number of epochs for which a collateral can give its owner a seat in the subnet, used in \joinCommand operations.
    \item $\permissionPolicy$: It defines the policy for accepting new validators. Arbitrary policies can be defined, enforced by each validator when signing a checkpoint message.
    \item \killVotePeriod, \killDelayPeriod: parameters related to the \killCommand operation.
    \item \whitelist and \whitelistThreshold: They define the \emph{initial configuration} of the subnet, explained below.
\end{enumerate*}

The creator encodes tag \createKeyword and the parameters in a string \subnetData and commits it to Bitcoin through transactions \createTx (shown in \iflncs\Cref{sec:tx-details}\else{\Cref{fig:tx-all}}\fi):
output~0 of \createTxOne commits to \subnetData, \createTxTwo spends that output revealing \subnetData in the witness.

\iflncs\vspace{-3pt}\fi
\dotparagraph{Subnet lifecycle and the \whitelist parameter}
When the \createCommand transactions finalize on Bitcoin, the subnet is \initializedSubnet. It has an identifier \subnetId and parameters but no validators, so users cannot yet interact with it. Once at least \whitelistThreshold validators join, they post the first checkpoint (\Cref{sec:checkpoint}) at height $\height_0$. This makes the subnet \activeSubnet, opening epoch~1 with live configuration $\subnetConf_1$, determined by the validators that joined (subset of $\whitelist$) and their collaterals, and enabling user operations (\Cref{sec:subnet-interact}).

The \emph{initial configuration} \whitelist is a bootstrap mechanism. Slash transactions and collateral UTXOs (both presented below) require a script naming a concrete validator set, and pre-activation joins cannot construct this script without an existing configuration. The \whitelist solves this by defining an initial bootstrap configuration, a set of Bitcoin keys with equal weight. 
All subsequent validator joins embed their epoch's current configuration $\subnetConf_e$ and follow \permissionPolicy.

\ifwhitepaper
\begin{figure}[h]
    \centering
    \includegraphics[width=\linewidth]{images/create-subnet-example.png}
    \caption{Create-subnet example. The commit transaction locks a Taproot output with the hash of a script containing the subnet parameters. At this point only the script hash is visible on Bitcoin. The reveal transaction spends this output by including the full script in the witness, making the subnet parameters publicly readable. For simplicity, the figure shows both transactions paying the same fee.}
    \label{fig:create-subnet-example}
\end{figure}
\fi

\ifwhitepaper
    \begin{example}[create-subnet-example]{Create subnet} \label{ex:create-subnet}
    Let's visualize how the commit-reveal technique works through an example.
   \footnote{The corresponding coding for writing arbitrary data to Bitcoin can be found in \url{https://github.com/bitcoinscalinglabs/bitcoin-ipc/blob/baab505db8dfaf32917f2fd5a8da852b1c7d8348/src/bitcoin_utils.rs\#L209}.}
    Assume we want to create a subnet with 4 validators in the whitelist, with required collateral 1 BTC.
    The \subnetData string looks like this:
    \begin{verbatim}
min_collateral="1", min_validators="4",
checkpoint_period="100",
active_validators_limit="100",
min_cross_msg_fee="1",
whitelist="e327a66b169732bde49d827d90781
327af558fc12d5cd2d5004e7551ec00c662,0a2f
008d849514ed8d9388cc0b1cc52fb281c7510c7d
57abcdb0e4eb5438bbdc,a0f982a368e43a47ae6
2e2f3806536404a11e6991f4f6a43e75b3b7a40c
6df29,71ac1eb874233999e11cd050f388f1dd6d
a9b446180fdf9b06740419cc487b6f"
    \end{verbatim}

    Some parameters, such as \textsl{active\_validators\_limit}, are subnet specific, and \whitelist is a comma-separated list of Bitcoin public keys.
    The \subnetData string is then encoded in binary format (so as to compress the data), to obtain \textsl{binary-subnetData}. The Bitcoin script contains the data ``\createKeyword\textsl{binary-subnetData}'', where \createKeyword is a keyword that identifies the transaction as an IPC-related \createCommand operation.

    \medskip
    First the \textbf{commit transaction} is constructed:
    \begin{itemize}
    \item The input is a UTXO of value at least $2\cdot\fee$ (this will be used for the fees of the two transactions), spendable by the user that creates the subnet.
    \item The output is a UTXO locked with a P2TR script, i.e., the script serves as a scriptPubKey. The value of this UTXO must be at least \fee.
    \end{itemize}

    Observe that, at this point, only the hash of the script is visible on the Bitcoin network. The commit transaction is signed (using the wallet of the user that creates the subnet) and submitted to Bitcoin.

    \medskip
    Then, the \textbf{reveal transaction} is constructed, with one input and one output UTXO:
    \begin{itemize}
    \item The input is the UTXO that we previously locked with the script.
    \item The output is a change UTXO that returns the whole value of the input minus \fee to the wallet that initiated the \createCommand operation (hence, \fee is the fee for the reveal transaction).
    \end{itemize}

    For the reveal transaction to be valid, it must contain a witness that unlocks the input UTXO. The witness contains the full \subnetData string.
    The two transactions are also shown in \Cref{fig:create-subnet-example}.
    In this example we have assumed for simplicity that both transactions pay the same fee.
    \end{example}
\fi

\iflncs\vspace{-3pt}\fi
\dotparagraph{The \joinCommand operation and collateral script}
To join a subnet, \anIpcAwareNode locks its collateral under the non-custodial \emph{collateral script tree} $\colS$ (shown in \Cref{tab:collateral-script}), which holds three leaves:
\begin{enumerate*}[label=\textbf{(\arabic*)}]
    \item \emph{Unilateral-exit (\uniExit) leaf}: After block height $\colDuration$, the owner alone can unconditionally spend the UTXO, guaranteeing \emph{unilateral exit} for honest validators.
    \item \emph{Slash leaf}: Allows slashing upon provable L2 equivocation.
    \item \emph{Cooperative-exit (\coopExit) leaf}: Allows early exit, stake modifications, and collateral refresh (\Cref{sec:checkpoint}) prior to maturity $\colDuration$, requiring signatures from both the owner $\pkExit$ and  more than $2/3$ of the stake of configuration $\subnetConf$.
\end{enumerate*}

\sloppy{
The \ipcAwareNode $V$ creates commit-reveal transactions $\joinTx$, which create the collateral UTXO under $\colS$ and reveal the data string $\joinData = (\joinKeyword, \subnetId, \pkCons, \pkExit, \colDuration, \pkEots, \nonceCom, \nonceComSig)$, where \joinKeyword is a tag, $\pkCons$ is the L2 consensus key, \pkExit the Bitcoin key of $V$, \pkEots is a Schnorr public key, $\nonceCom=(\nonceStart, \nonceCount, \nonceRoot)$ is a nonce commitment, and \nonceComSig is a signature under \pkEots on all preceding fields of \joinData.
Further, \nonceRoot is a Merkle commitment to \emph{public nonces} $\nonce{h}$, one per height $h \in [\nonceStart, \nonceStart + \nonceCount)$, for $\nonceStart,\nonceCount >0$. These will be used for voting in the subnet consensus, explained below.
All transactions are shown in \iflncs\Cref{sec:tx-details}\else\Cref{fig:tx-all}\fi.
}

\begin{table}[t]
\centering
\small
\begin{tabular}{l|c}
 \multicolumn{2}{c}{$\colS(\pkExit, \pkEots, \colDuration, \subnetConf)$} \\
\hline
\emph{Internal key} & (disabled) \\
\hline
\emph{Leaf 0 (unil.\ exit, \uniExit)} & $\mathsf{CheckSig}_{\pkExit} \wedge \mathsf{AbsTimelock}(\colDuration)$ \\
\emph{Leaf 1 (slash leaf)} & $\mathsf{CheckSig}_{\pkEots} \wedge \mathsf{CheckMultiSig}_{\subnetConf}$ \\
\emph{Leaf 2 (coop.\ exit, \coopExit)} & $\mathsf{CheckSig}_{\pkExit} \wedge \mathsf{CheckMultiSig}_{\subnetConf}$ \\
\hline
\end{tabular}
\setlength{\belowcaptionskip}{0pt}%
\caption{The collateral script tree $\colS(\pkExit, \pkEots, \colDuration, \subnetConf)$, locking the collateral of validator $V$ (public key $\pkExit$) until Bitcoin height \colDuration. The slash leaf uses the key \pkEots declared in \joinCommand. Validator $V$ can unlock it through leaf~0 after block $T$; the other leaves use configuration \subnetConf.}
\label{tab:collateral-script}
\end{table}

\iflncs\vspace{-3pt}\fi
\dotparagraph{Updating collateral and \leaveCommand operation} \label{sec:dynamic-participation}\label{sec:coop-exit}
Once a validator $V$ has joined, collateral updates (increase, reactivation, or \leaveCommand) are handled with a single transaction \stakeTx.
As with \joinCommand, new collateral is locked under the collateral script $\colS$. As $V$ is already registered, \stakeTx only needs to reference the subnet (by \subnetId) and that identity (by $\pkCons$), plus the new timeout $\colDuration$ (for stake renewal). An \opReturn is enough for this, hence \stakeTx is a single transaction.
Transaction \stakeTx also allows $V$ to create more commitments to subnet nonces, carrying commitment \nonceCom and signature \nonceComSig. 

Once the collateral matures, a validator can spend it using the UE leaf of $\colS$.
Moreover, a validator can exit before $\colDuration$ with the consent of the subnet's validators. The collateral is not sent directly to the validator but it gets locked by the \emph{unbonding script tree} $\unbS$, shown in \Cref{tab:unbonding-script}, which holds two leaves:
\begin{enumerate*}[label=\textbf{(\arabic*)}]
    \item \emph{Unilateral-unbonding (\uniUnbond) leaf:} It is a \emph{relative} timelock of three epochs ($3\epochLength$ blocks). The owner remains exposed to slashing through the slash script during this period.
    \item The same \emph{slash script} used in the collateral script tree (\Cref{tab:collateral-script}), with \subnetConf the configuration that signs the checkpoint creating the UTXO.
\end{enumerate*}
Cooperative exits are signed by the subnet's validators in checkpoints.

\begin{table}[t]
\centering
\small
\begin{tabular}{l|c}
 \multicolumn{2}{c}{$\unbS(\pkExit, \pkEots, \epochLength, \subnetConf)$} \\
\hline
\emph{Internal key} & (disabled) \\
\hline
\emph{Leaf 0 (unil. unbond)} & $\mathsf{CheckSig}_{\pkExit} \wedge \mathsf{RelTimelock}(3\epochLength)$ \\
\emph{Leaf 1 (slash leaf)} & $\mathsf{CheckSig}_{\pkEots} \wedge \mathsf{CheckMultiSig}_{\subnetConf}$ \\
\hline
\end{tabular}
\setlength{\belowcaptionskip}{0pt}%
\caption{The unbonding script tree $\unbS(\pkExit, \pkEots, \epochLength, \subnetConf)$. Validator $V$ can unlock it through leaf~0 after $3\epochLength$ blocks, unless slashed. The slash leaf uses the key \pkEots of $V$ and the configuration \subnetConf that created the UTXO.}
\label{tab:unbonding-script}
\end{table}

\iflncs
    \dotparagraph{Details on Bitcoin transactions}
    In \Cref{fig:tx-join,fig:tx-stake,fig:tx-create,fig:tx-deposit} show the transactions submitted to Bitcoin for the \joinCommand, \stakeCommand, \createCommand, and \depositCommand operations, respectively (\Cref{sec:create-subnet,sec:deposit}).
\fi
\label{sec:tx-details}

\begin{figure*}[t]
\begin{subfigure}[t]{0.48\linewidth}\vspace{0pt}\centering
\txTableV{\joinTxOne}{{(*)}}{{(f, \mathsf{dataS}(\joinData))}
 {(\amountShort, \colS(\pkExit, \pkEots, \colDuration, \subnetConf))}
 {?(*)}}{{*}}
\txTableV{\joinTxTwo}{{((\joinTxOne, 0), \mathsf{dataS}(\joinData))}}{{(*)}}{{\joinData = (\joinKeyword, \subnetId, \pkCons, \pkExit, \colDuration, \pkEots, \nonceCom, \nonceComSig)}}
\caption{The \joinCommand transactions \joinTx. The string \joinData carries the \joinKeyword tag, the subnet, the keys and timeout of the joining validator, its nonce commitment \nonceCom and a signature \nonceComSig on it (details in \Cref{sec:create-subnet}). In \joinTxOne, input $0$ spends arbitrary UTXOs of the funder, output $0$ is the commitment output, carrying the fee for \joinTxTwo, output $1$ locks the collateral $\amountShort$ under the collateral script tree $\colS$ (see \Cref{tab:collateral-script}), and output $2$ is optional change. Transaction \joinTxTwo spends the commitment output of \joinTxOne and reveals \joinData.}
\label{fig:tx-join}
\end{subfigure}\hfill
\begin{subfigure}[t]{0.48\linewidth}\vspace{0pt}\centering
\txTableV{\stakeTx}{{?((\stakeTxPrev, 1), \mathsf{CheckSig}_{\pkExit} \wedge \mathsf{AbsTimelock}(\colDuration'))}
 {?(*)}}{{?(0, \opReturn(\stakeData))}
 {?(\amountShort, \colS(\pkExit, \pkEots, \colDuration, \subnetConf))}
 {?( *)}}{{\sigma_V^{\mathsf{L1}}}
 {*}}
\caption{The \stakeCommand transaction \stakeTx.
Input $0$ spends a matured collateral (timeout $\colDuration'$ smaller than current height) ofs $V$ through its \uniExit leaf, while input $1$ adds fresh funds.
Either input may be absent.
Output $0$ carries $\stakeData = (\stakeKeyword, \subnetId, \pkCons, \colDuration)$, identifying the operation, the subnet, and the validator, and declaring the new timeout $\colDuration$, while output~1 contains the new collateral UTXO under $\colS$ (see \Cref{tab:collateral-script}). Output $2$ is an optional payout to the owner. A \stakeTx carrying a nonce commitments $\nonceCom, \nonceComSig$ is instead implemented as a commit-reveal pair, similar to \joinCommand (not shown for simplicity).}
\label{fig:tx-stake}
\end{subfigure}\\[2ex]
\begin{subfigure}[t]{0.48\linewidth}\vspace{0pt}\centering
\txTableV{\createTxOne}{{(*)}}{{(f, \mathsf{dataS}(\subnetData))}
 {?(*)}}{{*}}
\txTableV{\createTxTwo}{{((\createTxOne, 0), \mathsf{dataS}(\subnetData))}}{{(*)}}{{\subnetData = (\createKeyword, \minCollateral, \whitelistThreshold, \checkpointInterval, \epochLength, \minTimeout, \maxTimeout, \permissionPolicy, \killVotePeriod, \killDelayPeriod, \whitelist)}}
\caption{The \createCommand transactions \createTx. The string \subnetData carries the \createKeyword tag, which identifies the transaction, and all required subnet parameters (explained in \Cref{sec:create-subnet}). Output~0 of \createTxOne commits to data script $\mathsf{dataS}(\subnetData) = \script{\opPush(\subnetData) \scriptSep \opDrop \scriptSep \opTrue}$ and output~$1$ is optional change; \createTxTwo reveals \subnetData in the witness.}
\label{fig:tx-create}
\end{subfigure}\hfill
\begin{subfigure}[t]{0.48\linewidth}\vspace{0pt}\centering
\txTableV{\depositTx}{{(*)}
 {?(*)}}{{(\amountShort, \mathsf{CheckMultiSig}_{\subnetConf})}
 {(0, \opReturn(\depositData))}
 {?(*)}}{{*}
 {*}}
\caption{The \depositCommand transaction \depositTx. Its inputs spend UTXOs of the user's wallet. Output $0$ locks the deposited amount \amountShort under the multisig of the current configuration \subnetConf of the subnet, output $1$ carries $\depositData = (\depositKeyword, \subnetId, \userAddress)$, identifying the operation, the subnet, and the address \userAddress on the subnet that receives the minted \wBTC, and output $2$ is optional change.}
\label{fig:tx-deposit}
\end{subfigure}
\caption{The Bitcoin transactions of the \joinCommand, \stakeCommand, \createCommand, and \depositCommand operations.}\label{fig:tx-all}
\end{figure*}

\iflncs\vspace{-3pt}\fi
\dotparagraph{Validating the operations on Bitcoin}\label{sec:detecting}
Each \ipcAwareNode parses Bitcoin transactions in the same, deterministic way after they are finalized.
When the \createKeyword tag is detected in the witness of a transaction, it parses it as \createTxTwo and identifies \createTxOne through its output that \createTxTwo spends, and validates the subnet-creation parameters.
If everything checks out, the \ipcAwareNode deterministically derives the subnet identifier \emph{\subnetId} of the new subnet from the txid\footnote{\url{https://github.com/bitcoin/bips/blob/master/bip-0141.mediawiki}} of \createTxTwo.

Similarly, \joinCommand operations are detected through the \joinKeyword tag. The \ipcAwareNode locates \joinTxOne through the spent outpoint and builds the expected script tree $\colS$ from the declared parameters, using the current \emph{and} the previous configurations, $\subnetConf_e, \subnetConf_{e-1}$, of the subnet (\bitcoinIpc allows \emph{delayed arrivals} of transactions for up to one epoch). It accepts the \joinCommand only if output~$1$ of \joinTxOne is unspent, it is locked under one of the two possible scripts ($\subnetConf_e$ or $\subnetConf_{e-1}$), carries at least the required collateral, and the timeout respects the subnet's bounds: $e + \minTimeout + 3 \leq E_{\colDuration} \leq e + \maxTimeout + 3$, where $e$ is the current epoch of the subnet and $E_{\colDuration}$ is the epoch that contains height \colDuration. The latter ensures that the collateral gives a seat for at least \minTimeout and at most \maxTimeout epochs.
It further requires that \nonceComSig verifies under \pkEots{} and that \pkEots has not been used before.
The validation of \stakeCommand transactions is analogous, verifying that $\nonceStart \geq \nonceStart' + \nonceCount'$ for every accepted nonce commitment $(\nonceStart', \nonceCount', \cdot)$ of the same \pkEots, ensuring that no subnet height is covered by different nonces.

Notice that subnets admit validators according to their \permissionPolicy, which ranges from open (permissionless) to restricted policies requiring explicit approval by the active configuration via checkpoint signatures.
\AnIpcAwareNode accepts a \joinCommand or \stakeCommand only if all conditions above hold \emph{and} \permissionPolicy allows it. 
Before activation, \ipcAwareNodes accept a \joinCommand only from \whitelist keys.

\iflncs\else
    \dotparagraph{Remarks on \createCommand and \joinCommand}
Transactions \createTxTwo and \joinTxTwo can be front-run: once broadcast in the mempool, the contents of the script $\mathsf{dataS}$ are public, so anyone can create transaction $\createTxTwoPrime$ or $\joinTxTwoPrime$ spending the commitment output of \createTxOne or \joinTxOne, respectively. This poses no danger, as \createTxTwo and \joinTxTwo transfer no value, and the witness of $\createTxTwoPrime$ or $\joinTxTwoPrime$ will necessarily disclose the correct parameters (those committed to by $\mathsf{dataS}$).
Moreover, third-party stakes are explicitly permitted: the funder of a \joinCommand need not be the holder of the secret key behind $\pkCons$ or $\pkExit$.

\dotparagraph{The \killCommand operation}\label{sec:killing-a-subnet}

Any validator of a subnet can propose the termination of that subnet, i.e., to \emph{kill} the subnet,
by submitting a dedicated kill-proposal transaction to the subnet.
The proposal expires after \killVotePeriod subnet blocks, a parameter set at subnet creation.
The validators vote on the proposal in the subnet, and the proposal is accepted when validators holding more than $2/3$ of the stake of the current configuration vote for it before it expires. The acceptance is recorded in the next checkpoint, upon which every \ipcAwareNode marks the subnet as \statusToBeKilled.

For the next \killDelayPeriod epochs, the subnet remains fully operational: it issues checkpoints, rotates its configuration, and accepts \joinCommand and \stakeCommand operations. Users are expected to withdraw their funds within this period.
In its last rotation checkpoint, the configuration cooperatively exits every collateral UTXO, and every \ipcAwareNode marks the subnet as \statusKilled; the subnet issues no further checkpoints. A collateral that is not exited in this way is spent by its owner through the \uniExit leaf once it matures.
\fi

\subsection{Slashing validators for equivocation}\label{sec:slashing}

In \bitcoinIpc subnets, slashing punishes validators who commit equivocation (signing two conflicting blocks at the same height). We adopt Babylon’s EOTS primitive~\cite{DBLP:journals/corr/abs-2408-01896} for non-custodial Bitcoin slashing, adapting the signing orchestration to a multi-subnet, dynamic setting. Namely, unlike Babylon, which proxies security information through a dedicated chain, \bitcoinIpc has no central L2. Validator keys, nonce commitments, and collateral are self-declared in Bitcoin transactions that Bitcoin cannot validate; hence, the protocol must enforce that a \joinCommand is consistent with the locked collateral and cannot be copied or duplicated.
The slashing design consists of the following steps.

\begin{figure}[t]
\iflncs\begin{minipage}[c]{0.48\linewidth}\fi
\txTableV{\slashTx}
{{((\mathsf{tx}, n), \mathsf{CheckSig}_{\pkEots} \wedge \mathsf{CheckMultiSig}_{\subnetConf})}}
{{(\amountShort - \slashFee, \opReturn)}
 {(\slashFee, \opTrue)}}
{{\eotsSig, \{\sigma_{U}\}^{> 2/3}_{U \in \subnetConf}}}
\iflncs
\end{minipage}\hfill
\begin{minipage}[c]{0.49\linewidth}
\vspace{5pt}
\fi
\setlength{\belowcaptionskip}{0pt}%
\caption{The slashing transaction $\slashTx$ for a collateral output $o = (\mathsf{tx}, n)$ of value \amountShort, spent through the slash leaf of $o$. The witness carries a signature under \pkEots and signatures from validators holding more than $2/3$ of the stake of \subnetConf.}
\label{fig:tx-slash}
\iflncs\end{minipage}\fi
\end{figure}

\begin{description}[leftmargin=0pt, font=\normalfont\bfseries]
    \item[Voting on the subnet:] During consensus finalization, $V$ signs block $b$ at height $h$ using \pkEots and the secret nonce behind the committed $\nonce{h}$, published in \joinCommand and \stakeCommand transactions. It attaches $\nonce{h}$ and its Merkle proof to the signature, and other validators accept the signature only if this equals the committed value. Notice that $V$ can only produce valid signatures for subnet heights in the committed ranges $[\nonceStart, \nonceStart + \nonceCount)$. A validator uses a single EOTS key in a subnet, hence an equivocation exposes all of its collateral. 
    \item[Secret key extraction:] If $V$ signs two different blocks $(\subnetId, h, b)$ and $(\subnetId, h, b')$, it reuses nonce $\nonce{h}$. By Schnorr signature properties, combining these two signatures mathematically reveals $V$'s secret key $\skEots$. 
    \item[Pre-signing slashing transaction:] When a collateral UTXO $o$ is created under script tree $\colS$ (or during unbonding under $\unbS$), the active configuration \subnetConf pre-signs a slashing transaction $\slashTx$, shown in \Cref{fig:tx-slash}. This can be deterministically created given $o$, and spends $o$ through leaf~1 and pays $\amountShort - \slashFee$ to an unspendable output and \slashFee, a protocol constant, to an output anyone can spend.
    The pre-signed $\slashTx$ is published on Bitcoin through checkpointing.
    \item[Slash on Bitcoin:] The slash leaf of $\colS$ requires both a signature under $\pkEots$ and a multisig from $\subnetConf$. Because $\slashTx$ is already pre-signed, anyone holding the extracted $\skEots$ can fulfill the first requirement, sign $\slashTx$ and broadcast it to burn $V$'s collateral on Bitcoin.
\end{description}

\subsection{Subnet consensus and the checkpoint mechanism} \label{sec:checkpoint}\label{sec:persist-state}

\dotparagraph{Determining the configuration with \retrieveStateCom}\label{sec:configuration}
Operation \retrieveStateCom allows \anIpcAwareNode to read the configuration $\subnetConf_e$ of a subnet for epoch $e$, as a function of finalized Bitcoin data:
it consists of every validator that at cutoff height $\height_{e-1}$ holds unspent collateral UTXOs, accepted by the \joinCommand or \stakeCommand implementation or refreshed by a checkpoint, that are locked under $\subnetConf_{e-1}$ and whose timeout satisfies $\colDuration > \height_{e+2}$ (i.e., the collateral matures \emph{after epoch} $e+2$), weighted by their total value. A collateral UTXO counts only if a checkpoint carrying the signatures on its slashing transaction is finalized by that cutoff.

\Cref{tab:stake-lifecycle} depicts this. Denoting by $E_{\colDuration}$ the epoch that contains height $\colDuration$, the collateral holds a seat in the subnet's configuration in every epoch up to and including $E_{\colDuration}-3$, hence a ``seated collateral'' is always at least three epochs away from being spendable. It remains \emph{bonded} through epochs $E_{\colDuration}-2$ to $E_{\colDuration}$, giving the subnet time to slash it for misbehaviour committed while seated, and ensuring that validators that are no longer bonded have no authority over the collateral of seated validators, as the slash leaf of a seated collateral names the previous configuration. Moreover, the validator remains online for signing checkpoint messages during epoch $E_{\colDuration}-2$, until the rotation checkpoint that closes that epoch is signed.

\begin{table}[t]
\iflncs\begin{minipage}[c]{0.52\linewidth}\fi
\centering
\small
\begin{tabular}{l l l}
\hline
\emph{Epochs} & \emph{Seat} & \emph{Collateral} \\
\hline
$\leq E_{\colDuration}-3$  & yes (active) & locked \\
$E_{\colDuration}-2$ to $E_{\colDuration}$  & no (bonded) & locked until $\colDuration$ \\
$\geq E_{\colDuration}+1$ & no (matured) & spendable \\
\hline
\end{tabular}
\iflncs
\end{minipage}\hfill
\begin{minipage}[c]{0.44\linewidth}
\fi
\setlength{\belowcaptionskip}{0pt}%
\caption{Lifecycle of a stake with timeout $\colDuration$. $E_{\colDuration}$ is the epoch containing $\colDuration$. From $E_{\colDuration}-2$ to $E_{\colDuration}$ it assigns no seat in the subnet's configuration, but remains exposed for slashing.}
\label{tab:stake-lifecycle}
\iflncs\end{minipage}\fi
\end{table}

\dotparagraph{Subnet consensus}\label{sec:subnet-consensus}
\bitcoinIpc supports any proof-of-stake consensus protocol with deterministic finality. In epoch $e$, the validators of a subnet are the members of $\subnetConf_e$, weighted by their stake, as returned by \retrieveStateCom. A block is finalized once validators holding more than $2/3$ of the stake have signed it with their EOTS keys.
Checkpointed subnet blocks are treated as final: a block that does not extend the latest checkpoint is invalid, and an honest validator never votes for such a block. 
A validator that joins or resynchronizes the subnet starts from the latest checkpoint. These rules protect the subnet from long-range attacks~\cite{DeirmentzoglouP19}, as no honest party accepts chains incompatible with checkpointed blocks.

\dotparagraph{Checkpointing}
Each subnet independently issues a checkpoint every \checkpointInterval subnet blocks (fixed at subnet creation). The first checkpoint of an epoch is a \emph{rotation} checkpoint; the rest are \emph{regular}. A checkpoint consists of two Bitcoin transactions \checkpointTx, which the subnet validators produce via consensus on the latest finalized Bitcoin and subnet blocks, then sign and submit to Bitcoin. 
\iflncs\else\newline\fi
Checkpoints periodically write a block commitment, committee rotation, and batch outgoing \transferCommand and \withdrawCommand operations, as follows:

\begin{enumerate}[wide=0pt, font=\normalfont\bfseries]
    \item\textbf{Committing to the latest block:}
    Transaction \checkpointTxOne carries an \opReturn output with the \checkpointKeyword tag, the height \subnetHeight of the subnet block at which the checkpoint was created, and a commitment \comBlock to that block. In \Cref{fig:tx-checkpoint}, this is output $7$.
    From this point on, honest validators only accept blocks that extend \comBlock.
\ifwhitepaper
\begin{figure}[h]
    \centering
    \includegraphics[width=\linewidth]{images/transfer-example.png}
    \caption{Example showing the inputs and outputs added to the \checkpointTxOne and \checkpointTxTwo transactions to implement the \transferCommand operation. Four transfers with two different target subnets are batched together in a pair of Bitcoin transactions.
    The \checkpointTxOne contains one input with value equal to the sum of all transfers, and three outputs: one with value equal to the sum of the transfers to subnet $B$, one with value equal to the sum of the transfers to subnet $C$, and one locked with the hash of a script. The script contains the details of each transfer, and is revealed by the \checkpointTxTwo transaction.
    For simplicity, the figure shows both transactions paying the same fee.}
    \label{fig:transfer-example}
\end{figure}

    \begin{example}[transfer-example]{Batching transfers}
    Let's look at an example that involves three subnets, $A$, $B$, and $C$, with the following accounts:
    \begin{itemize}
        \item Accounts on subnet $A$: \emph{u,v,w,x}.
        \item Accounts on subnet $B$: \emph{d,e}.
        \item Accounts on subnet $C$: \emph{j,k}.
    \end{itemize}
    At checkpoint time, the following outgoing transfers have been submitted on subnet $A$:
    \begin{itemize}
        \item $A.u \rightarrow B.d$ with amount $v_1$
        \item $A.v \rightarrow B.e$ with amount $v_2$
        \item $A.w \rightarrow C.j$ with amount $v_3$
        \item $A.x \rightarrow C.k$ with amount $v_4$
    \end{itemize}
    These transfers will be batched together and submitted to the Bitcoin network with two transactions. Transaction \checkpointTxOne will contain one input, with value $v_1 + v_2 + v_3 + v_4$ (plus the necessary fees), and three outputs, the first with value $v_1 + v_2$, locked with the address of subnet $B$, and the second with value $v_3 + v_4$, locked with the address of subnet $C$, and the third locked with the hash of a script. The script contains the details of each transfer, and is revealed by the \checkpointTxTwo transaction.
    \Cref{fig:transfer-example} depicts this example.
    \end{example}
\fi

    \item\textbf{Pre-sign \slashTx transactions:}
    A checkpoint (regular or rotation) contains \slashTx pre-signatures for each collateral UTXO created since the last checkpoint (either by new \joinTxOne transactions or by collateral refresh). Hence, validators in $\subnetConf_e$ (and $\subnetConf_{e-1}$ for late \joinCommand arrivals) create and sign the slashing transaction \slashTx.
    A \slashTx can be created locally and deterministically from leaf~1 of the collateral script $\colS$.
    They are published in string \slashData in the witness of \checkpointTxTwo.

    \item\textbf{Handle outgoing \transferCommand and \withdrawCommand operations:}
    As we detail in \Cref{sec:subnet-interact}, these operations are initiated by the user in a subnet, which then handles them at the next checkpoint. In \Cref{fig:tx-checkpoint}, box~(ii) contains batched \transferCommand-related and box~(iii) \withdrawCommand-related UTXOs. 

    \item\textbf{Configuration update in rotation checkpoints:}\label{sec:rotation}
    Each subnet creates a rotation checkpoint when a Bitcoin block at cutoff height $\height_e$, $e>0$, becomes finalized, which indicates a new configuration $\subnetConf_{e+1}$.
    All fee and fund UTXOs locked under $\subnetConf_e$ are spent and locked under $\subnetConf_{e+1}$, and so are the collateral UTXOs whose owners have signed.
\end{enumerate}

\noindent
A rotation checkpoint sweeps UTXO types as follows:
\iflncs\vspace{-3pt}\fi
\begin{description}[leftmargin=0pt, font=\normalfont\bfseries]
    \item[Sweeping fee and fund UTXOs:]
    These UTXOs are only locked by a multisig script, hence we need signatures from validators holding more than $2/3$ of the stake of the configuration that locks them.
    However, that configuration is not necessarily the exiting configuration $\subnetConf_e$: a transaction \depositTx (\Cref{sec:deposit}) uses the configuration of the subnet that is current when \depositTx is created, but it may be confirmed on Bitcoin later, in a following epoch.
    \bitcoinIpc supports \emph{delayed arrivals} of \emph{one} epoch: the rotation checkpoint at the end of epoch $e$ must rotate the fund UTXOs that are locked under the multisig of configuration $\subnetConf_{e-1}$ as well. This is possible because all validators in a \bitcoinIpc subnet remain active for one epoch after they leave a subnet.

    \item[Sweeping collateral UTXOs:]\label{sec:coop-refresh}
    Collateral UTXOs need to be rotated to the new configuration, so that no seated validator has its collateral locked under a configuration that has exited.
    This requires a signature from the collateral owner (key $\pkExit)$ and from the subnet's validators.
    Note that at the end of epoch $e$, these UTXOs can be locked under $\subnetConf_e$ or under $\subnetConf_{e-1}$ --- due to delayed arrivals, same as fund UTXOs, and due to the following situation: $\subnetConf_{e+1}$ becomes known only \emph{after} the Bitcoin block at height $\height_e$ is finalized, and creating \checkpointTxOne requires $\subnetConf_{e+1}$ to be known, implying that \checkpointTxOne can only be signed after Bitcoin height $\height_e$ is finalized.
    This leaves a very short interval for collateral owners to sign,
    and we do not want to make the assumption that collateral owners are always online.
    
    To rectify this, we allow a \emph{delay of one epoch} for the rotation of collateral UTXOs.
    Namely, a collateral UTXO can be spent in the following ways:
    \begin{enumerate*}[label=\textbf{(\roman*)}]
    \item\label[case]{item:refresh-1} If locked under $\subnetConf_e$ and the owner provides its signature: it is rotated to $\subnetConf_{e+1}$ by any checkpoint of epoch $e+1$.
    \item\label[case]{item:refresh-2} If locked under $\subnetConf_{e-1}$ and the owner provides its signature: it is rotated to $\subnetConf_{e+1}$ by the rotation checkpoint, as the validators of $\subnetConf_{e-1}$ are still online at the end of epoch $e$ to provide theirs.
    We call these two cases a \emph{refresh}: the collateral UTXO is recreated with the same value, the same owner keys, and the same timeout $\colDuration$.
    \item\label[case]{item:coop-exit} Instead of a refresh, the collateral may be exited into an unbonding UTXO (\Cref{sec:coop-exit}): as this requires no knowledge of the next configuration, the owner can sign throughout the whole epoch $e$.
    \end{enumerate*}

    A checkpoint carries one input and one output with the same value per collateral UTXO it spends. In \Cref{fig:tx-checkpoint}, inputs/output pairs with index $0$, $1$, and $2$ are a refresh of \cref{item:refresh-1}, \cref{item:refresh-2}, and a cooperative exit of \cref{item:coop-exit}, respectively.
\end{description}

\begin{txfigure*}{Example checkpoint transactions \checkpointTx of a subnet $A$ with configuration $\subnetConf^A_e$, taken at Bitcoin cutoff $\height_e$ and subnet height \subnetHeight.
\textbf{(i) Collateral operations:} Input/output pairs $0$--$2$ encode operations for collateral UTXOs. They always occur in pairs: here, index $0$ refreshes the collateral with value $\amountShort_0$ (created in previous transaction $\mathsf{tx}_0$ at output index $n_0$) of a validator $V_0$, locking it under the next configuration $\subnetConf^A_{e+1}$; similarly, index $1$ refreshes collateral $\amountShort_1$ of a validator $V_1$, which was still locked under the previous configuration $\subnetConf^A_{e-1}$; index $2$ unbonds the collateral with value $\amountShort_2$ of a validator $V_2$.\textbf{(ii) Transfers:} Outputs $3-5$ encode outgoing transfers to subnets $B$ and $C$, each with value the sum of all outgoing transfers to recipients in that subnet, $\amountShort_B$ and $\amountShort_C$, respectively. $\subnetConf^B_{e'}$ and $\subnetConf^C_{e''}$ are the configurations of $B$ and $C$, as seen by subnet $A$ at Bitcoin height $\height_e$, with $e'$ and $e''$ their own current epoch numbers. Output $5$ commits to \transferData, the description of all outgoing transfers, and to \slashData, the pre-signatures on slashing transactions of UTXOs created by earlier checkpoints, and carries the fee $f'$ of \checkpointTxTwo, which spends it and reveals \transferData and \slashData in its witness.
\textbf{(iii) Withdrawals:} Output $6$ withdraws $\amountShort_u$ to user $u$. All outgoing transfers and withdrawals can be funded by any fund UTXO locked under the subnet multisig (such as a \depositCommand or an incoming \transferCommand) -- here input $3$, the output of some previous transaction $\mathsf{tx}'$ with value $\amountShort$, such that $\amountShort \geq \amountShort_B+\amountShort_C+\amountShort_u$.
\textbf{(iv) Anchoring:} Output $7$ contains the checkpoint metadata: identifier \checkpointKeyword, subnet height \subnetHeight, commitment \comBlock to the block at that height.
Finally, output $8$ is the change, of value $\amountShort' = \amountShort - \amountShort_B - \amountShort_C - \amountShort_u$, in a rotation checkpoint locked under the next configuration $\subnetConf^A_{e+1}$, as in this example, and under $\subnetConf^A_e$ otherwise, and input $4$ spends fee UTXOs (which are always locked under the subnet's multisig) of value $f+f'$ to pay the fees for \checkpointTxOne and \checkpointTxTwo, respectively.
\textbf{Witness:} $\{\sigma_{V}\}^{> 2/3}_{V \in \subnetConf}$ denotes signatures from validators of some configuration $\subnetConf$, collectively holding more than 2/3 of the stake in that configuration.
}{fig:tx-checkpoint}
\txWitnessInline
\txMarks{cpt}
\txTable[0.5]{\checkpointTxOne}
{{?((\mathsf{tx}_0, n_0), \mathsf{CheckSig}_{\pk_{V_0}^{\mathsf{L1}}} \wedge \mathsf{CheckMultiSig}_{\subnetConf^A_e}\vspace{8pt})}
 {?((\mathsf{tx}_1, n_1), \mathsf{CheckSig}_{\pk_{V_1}^{\mathsf{L1}}} \wedge \mathsf{CheckMultiSig}_{\subnetConf^A_{e-1}}\vspace{8pt})}
 {?((\mathsf{tx}_2, n_2), \mathsf{CheckSig}_{\pk_{V_2}^{\mathsf{L1}}} \wedge \mathsf{CheckMultiSig}_{\subnetConf^A_{e-1}}\vspace{14pt})}
 {?((\mathsf{tx}', 0), \mathsf{CheckMultiSig}_{\subnetConf^A_e}\vspace{11pt})}
 {(*, \mathsf{CheckMultiSig}_{\subnetConf^A_e})}}
{{?(\amountShort_0, \colS(\pk_{V_0}^{\mathsf{L1}}, \pk_{V_0}^{\mathsf{eots}}, \colDuration_0, \subnetConf^A_{e+1})\vspace{8pt})}
 {?(\amountShort_1, \colS(\pk_{V_1}^{\mathsf{L1}}, \pk_{V_1}^{\mathsf{eots}}, \colDuration_1, \subnetConf^A_{e+1})\vspace{8pt})}
 {?(\amountShort_2, \unbS(\pk_{V_2}^{\mathsf{L1}}, \pk_{V_2}^{\mathsf{eots}}, \epochLength, \subnetConf^A_e)\vspace{14pt})}
 {?(\amountShort_B, \mathsf{CheckMultiSig}_{\subnetConf^B_{e'}})}
 {?(\amountShort_C, \mathsf{CheckMultiSig}_{\subnetConf^C_{e''}})}
 {?(f', \mathsf{dataS}(\transferData \concat \slashData)\vspace{7pt})}
 {?(\amountShort_u, \mathsf{CheckSig}_{\pk_u}\vspace{11pt})}
 {(0, \opReturn(\checkpointKeyword \concat \subnetHeight \concat \comBlock)\vspace{8pt})}
 {?(\amountShort', \mathsf{CheckMultiSig}_{\subnetConf^A_{e+1}})}}
{{?\{\sigma_{V}\}^{> 2/3}_{V \in \subnetConf^A_e}, \sigma_{V_0}^{\mathsf{L1}}}
 {?\{\sigma_{V}\}^{> 2/3}_{V \in \subnetConf^A_{e-1}}, \sigma_{V_1}^{\mathsf{L1}}}
 {?\{\sigma_{V}\}^{> 2/3}_{V \in \subnetConf^A_{e-1}}, \sigma_{V_2}^{\mathsf{L1}}}
 {?\{\sigma_{V}\}^{> 2/3}_{V \in \subnetConf^A_e}}
 {\{\sigma_{V}\}^{> 2/3}_{V \in \subnetConf^A_e}}}
\begin{tikzpicture}[remember picture, overlay]
  \txBox[name=col0]{in-0,out-0}[out-1,out-2]
  \txBox[name=col1]{in-1,out-1}[out-0,out-2]
  \txBox[name=col2]{in-2,out-2}[out-0,out-1]
  \txBox[name=fund]{in-3}
  \txBox[name=tfr]{out-3,out-4,out-5}
  \txBox[name=wdr]{out-6}
  \txBox[name=anc]{out-7}
  \node[font=\tiny, anchor=west, align=left, inner sep=1pt, xshift=2pt] at (col0.east) {\sighashSingle\unskip\texttt{|}\\\sighashACP};
  \node[font=\scriptsize, anchor=west, xshift=2pt] at (col1.east) {(i)};
  \node[font=\scriptsize, anchor=west, xshift=2pt] at (tfr.east) {(ii)};
  \node[font=\scriptsize, anchor=west, xshift=2pt] at (wdr.east) {(iii)};
  \node[font=\scriptsize, anchor=west, xshift=2pt] at (anc.east) {(iv)};
  \draw[dashed, thin] ($(fund.north east)!0.3!(fund.south east)$) -- ($(tfr.north west)!0.15!(tfr.south west)$);
  \draw[dashed, thin] ($(fund.north east)!0.7!(fund.south east)$) -- (wdr.west);
\end{tikzpicture}

\txMarks{}
\txLabelsInline
\txTable{\checkpointTxTwo}
{{((\checkpointTxOne, 5), \mathsf{dataS}(\transferData \concat \slashData))}}
{{(*)}}
{{\transferData \concat \slashData}}
\end{txfigure*}

\iflncs\vspace{-3pt}\fi
\dotparagraph{Signing the checkpoint transactions}\label{sec:sign-checkpoints}
An honest validator only signs \checkpointTxOne for subnet height \subnetHeight if it contains an \opReturn output with values $(\subnetHeight,\comBlock)$, where \comBlock is the hash of the subnet block at height \subnetHeight.
It further requires that the inputs and outputs of \checkpointTxOne match the subnet's state at height \subnetHeight: the transfers in \transferData and the withdrawals are exactly those finalized on the subnet since the previous checkpoint; for every destination subnet, \checkpointTxOne carries one output with value the sum of the transfers to that subnet in \transferData, locked by the multisig of that subnet's current configuration; every withdrawal has one output with its value and address; every input is a UTXO of the subnet; and the change output returns the remaining value to the subnet's multisig.

The collateral-refresh inputs of \checkpointTxOne (ind.~0 and~1 in \Cref{fig:tx-checkpoint}) require a signature by the owner (key $\pkExit$).
The owners sign \checkpointTxOne with sighash type \sighashSingleACP: such a signature commits only to \emph{one input and one output with the same index}.
This allows validators to produce a valid, signed \checkpointTxOne, even if some owners do not sign it, while at the same time, \checkpointTxOne cannot spend a collateral UTXO without including the output signed by its owner.
The validators sign \checkpointTxOne with \sighashAll, after the set of pairs is fixed. An honest validator signs a \checkpointTxOne only after it sees, in the subnet contract that collects the signatures, valid pre-signatures on the \slashTx of every collateral or unbonding UTXO that \checkpointTxOne creates. As with the other commit-reveal transaction pairs, \checkpointTxTwo needs no signature.

\iflncs\vspace{-3pt}\fi
\dotparagraph{Validating checkpoints on Bitcoin}\label{sec:validating-checkpoints}
\AnIpcAwareNode accepts a finalized transaction as a checkpoint of a subnet if it spends a UTXO of that subnet, locked under the multisig of its current or previous configuration, and carries an \opReturn output $(\checkpointKeyword, \subnetHeight, \comBlock)$. For every height \subnetHeight, it records the tuple $(\subnetHeight, \comBlock)$ of the first such transaction that is finalized, which it returns in \retrieveStateCom.


\subsection{User operations: \depositCommand, \withdrawCommand, and \transferCommand}\label{sec:subnet-interact}
 \label{sec:deposit}
  \label{sec:transfer}
   \label{sec:withdraw}

The \depositCommand operation is implemented as a single Bitcoin transaction \depositTx (shown in \Cref{fig:tx-deposit}), signed and submitted by the user on Bitcoin.
It spends UTXOs of the user's wallet and creates two outputs: one that locks the deposited amount under the subnet's current configuration, and one \opReturn that carries the \depositCommand keyword and address \userAddress. Once \depositTx is finalized, the validators mint the deposited amount of \wBTC to \userAddress. A \depositTx whose \opReturn output carries no address \userAddress mints nothing and only adds funds to the subnet: this is one possible way for validators to fund the fees of checkpoint transactions.

The \transferCommand operation 
allows users to transfer \wBTC to different subnets (intra-subnet transfers are handled by standard EVM functionality). 
The operation starts with the user submitting an \emph{L2 transaction} on the source subnet, naming the destination subnet, recipient, and amount.
The subnet handles all outgoing transfers in the next \checkpointCommand: it decreases the balances of the senders and creates a string \transferData, containing tag \transferKeyword and the metadata of the transfers, which is committed to in \checkpointTxOne and revealed in \checkpointTxTwo.
Transaction \checkpointTxOne also carries one output per target subnet, with value the sum of all transfers to that subnet, locked by the multisig of that subnet's current configuration. Incoming transfers are detected by a \btcMonitor through the tag \checkpointKeyword and validated by extracting \transferData from the witness and verifying that the outputs of \checkpointTxOne match the sums it specifies. The validators of the destination subnet then mint the transferred amounts to the recipients.
Hence, every cross-subnet transfer is routed through Bitcoin, rather than an external bridge.

In \Cref{fig:tx-checkpoint}, outputs $3$ and $4$ pay two target subnets, output $5$ commits to \transferData, and \checkpointTxTwo reveals it. As an example, assume the subnet has four outgoing transfers: amounts $\amountShort_c$ to address $c$ and $\amountShort_d$ to address $d$ on subnet $B$; amounts $\amountShort_j$ to address $j$ and $\amountShort_k$ to address $k$ on subnet $C$.
Then \transferData would contain $(\subnetId_B, (c,\amountShort_c), (d,\amountShort_d)))$ and $(\subnetId_C, (k, \amountShort_k), (j, \amountShort_j))$, and outputs $3$ and $4$ would carry values $\amountShort_B = \amountShort_c + \amountShort_d$ and $\amountShort_C = \amountShort_j + \amountShort_k$, respectively.

The \withdrawCommand operation lets a user withdraw \wBTC from a subnet to a Bitcoin address. Similar to \transferCommand, it starts with the user submitting an L2 transaction, specifying the withdrawal amount and a Bitcoin address. The subnet burns the \wBTC and handles all withdrawals in the next checkpoint. To that goal, \checkpointTxOne carries one output per withdrawal with the withdrawn value and locked by the public key the user named. The funds come from the subnet's fund UTXOs, so a withdrawal is the exact inverse of a deposit. In \Cref{fig:tx-checkpoint}, output $6$ pays a withdrawal of $\amountShort_u$ to a user $u$.

\subsection{Formalization}
\Cref{sec:model-full} formalizes anchoring, the bridge protocol, and operations \depositCommand, \withdrawCommand, \transferCommand. \Cref{sec:proofs} proves invariants for the create and join operations, then demonstrates that subnet validators alone securely operate the anchoring and bridge protocols, establishing properties P1 - P4 from \Cref{sec:preliminaries}.
Concretely, assuming that a Bitcoin block is finalized after $\depth = 6$ confirming blocks, that a transaction is included within \inclDelay blocks, and that a subnet signs and submits a checkpoint within $\Delta_{\mathsf{sign}}$ blocks,
then anchoring, \depositCommand, \withdrawCommand, and \transferCommand are secure for every $1/3$-bounded subnet with epoch length $\epochLength \geq \depth + \Delta_{\mathsf{sign}} + \inclDelay$, while the firewall property holds for every subnet.
Moreover, assuming that a configuration stops finalizing blocks within \adoptDelay blocks after the cutoff that ends its epoch is finalized, and that honest validators observe an event and react to it within \obsDelay blocks,
then equivocation is slashed in every subnet with $\epochLength \geq 2\depth + \adoptDelay + 2(\obsDelay + \inclDelay)$ in which the Byzantine validators, including those that have withdrawn their collateral, hold at most $2/3$ of the stake of each configuration.

\iflncs\else
  \section{Staking incentivization}
\label{app:staking}

\bitcoinIpc does not require a staking reward token for safety.
Properties P1--P4 hold for any subnet whose seated collateral and scripts are as specified in \Cref{sec:operations}, whether or not validators receive additional issuance.
This section records the \emph{optional} incentive layer that the implementation uses to pay validators for locking BTC and remaining available, including during the two epochs after they leave a committee (\Cref{sec:checkpoint}).
Exact emission parameters are a deployment choice; the figures below are the values used in the current design and are not a security assumption of \Cref{thm:state-anchoring,thm:deposit,thm:withdrawal,thm:transfer}.

\subsection{What is being paid for?}

A \bitcoinIpc validator already earns wBTC on its subnet, because the subnet uses wBTC as execution-layer fees.
That income is local and depends on that subnet's traffic. It does not compensate the opportunity cost of BTC locked on L1 or the requirement that a departing validator keep signing for one more epoch.

The optional layer therefore pays a separate reward token, denoted by \IPCBTC below, to validators who opt in.
A validator opts in by locking collateral as in \Cref{sec:create-subnet} and remaining seated for a full \emph{snapshot} (defined below).
Issuance is independent of which subnet the stake is used in: two validators with the same seated BTC over the same snapshot receive the same issuance, whether they serve a busy subnet or an empty one.
The goal is to pay for \emph{L1-auditable} collateral and availability, not to steer stake toward particular subnets.

\subsection{Emission schedule}

Inspired by Bitcoin tokenomics, \IPCBTC is capped at $21{,}000{,}000$ units.
Issuance is partitioned into \emph{epochs} of $210{,}000$ Bitcoin blocks (about four years at the target block interval).
Epoch $i = 0,1,2,\ldots$ mints $10{,}500{,}000 / 2^{i}$ tokens; after a finite number of epochs the residual issuance is negligible and the cap is exhausted.

Each epoch is split into snapshots of $1{,}000$ Bitcoin blocks (about one week).
There are therefore $210$ snapshots per epoch. In epoch $0$ every snapshot has a budget of $\frac{10{,}500{,}000}{210} = 50{,}000$ \IPCBTC . 
Later epochs scale that budget by the same halving factor as the epoch mint.
A validator receives a share of a snapshot budget only if some collateral of that validator is seated for every cutoff that falls inside the snapshot (\Cref{tab:stake-lifecycle}).
The share is proportional to the BTC-weighted seated stake recorded by $\mathsf{getSubnet}$ over that interval.

A configurable fraction $\alpha \in [0,1]$ of each snapshot budget is reserved for users who originated cross-subnet transfers that were checkpointed during the snapshot, split in proportion to transferred value as recorded on L1. 
The remaining $1-\alpha$ goes to validators.
The current implementation uses $\alpha = 0$; a non-zero $\alpha$ is a policy parameter, not a protocol requirement.

\subsection{Accounting on Bitcoin, minting on subnets}

Which validator is owed how much for snapshot $s$ is a function of finalized Bitcoin data alone: accepted $\mathsf{join}$/$\mathsf{stake}$/refresh outputs, checkpointed slash pre-signatures (so that only slashable seated collateral counts), and, if $\alpha > 0$, the transfer lists revealed by $\mathsf{tx}^{2}_{\mathrm{cpt}}$.
Any \ipcAwareNode can independently recompute the vector of entitlements.

Minting is not performed by Bitcoin, but accounting data is on Bitcoin. We distinguish two deployments.

\begin{itemize}
\item \emph{Target decentralized design.}
After snapshot $s$ closes, each subnet $S$ that has opted into issuance may mint, on $S$, the \IPCBTC owed to \emph{its} validators (and, if $\alpha > 0$, to users whose transfers left $S$).
The mint is an ordinary subnet transaction.
Other subnets accept incoming \textsc{Ipcbtc} only up to the amount the sender is entitled to under the publicly recomputed vector.
The excess is rejected.
This is the firewall property (\Cref{def:bridge-subnet}) applied to the reward token rather than to wBTC: a compromised subnet can inflate its own \textsc{Ipcbtc} book, but it cannot force a $1/3$-bounded destination to credit more than L1 data justifies.

\item \emph{Current MVP.}
Our current implementation does not yet have every subnet mint.
A dedicated \emph{Emission Subnet} watches Bitcoin, computes the snapshot vector, and mints \IPCBTC to validator and user addresses.
Until issuance is moved to decentralized design as above, holders of MVP \IPCBTC trust that Emission Subnet's committee in the same three-regime sense as any other subnet (\Cref{sec:preliminaries}).
The Emission Subnet is a current state of development, not an additional trust assumption of the bridge theorems.
\end{itemize}

In both deployments, \IPCBTC that a validator has already received can be  transferred cross-subnet under the same rules as wBTC. 
Slashing a validator's BTC collateral does not burn tokens already paid; it only removes that collateral from future snapshot tallies and, if the slash occurs during a snapshot, disqualifies that validator for the current snapshot.

Together with subnet wBTC fees, this incentivization gives a validator a reason to (i)~lock \BTC, (ii)~remain seated through a snapshot rather than oscillating around cutoffs, and (iii)~stay online through the bonded window of \Cref{tab:stake-lifecycle} if that window overlaps a snapshot for which it wants a share. Equivocation that leads to an L1 slash forfeits future reward shares tied to the burned collateral.

\fi

\newcommand{\experimentHeading}[1]{%
    \iflncs
        \dotparagraph{#1}%
    \else
        \subsection{#1}%
    \fi
}

\newcommand{\throughputPlot}{\includegraphics[width=\linewidth]{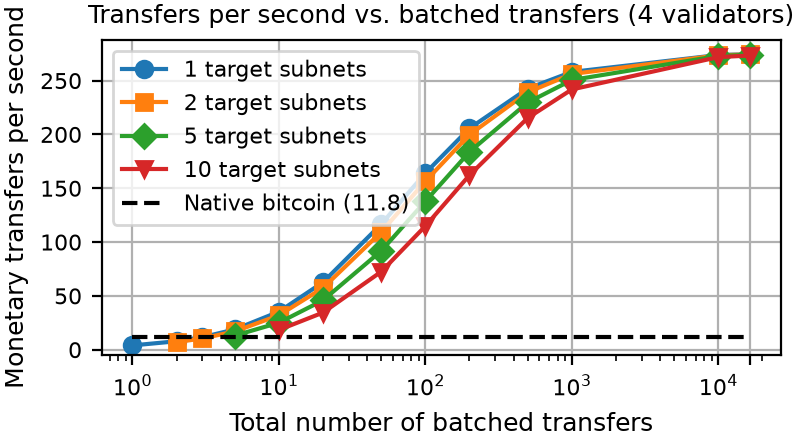}}
\newcommand{\throughputCaption}{Monetary transfers per second if blocks are filled with batched transfers, compared to blocks filled with native payments.}

\newcommand{\checkpointSizesTable}{%
\begin{table*}[t]
\centering
\begin{tabular}{|c|c|c|c|c|c|}
\hline
$n$ & regular & $t$ transfers & $w$ withdrawals & $r$ refreshes & $s$ pre-signed \slashTx \\
\hline
$4$   & $260$   & $+R + 51 + 6t$ & $+43w$ & $+217r$   & $+R + 60s$ \\
$25$  & $705$   & $+R + 51 + 6t$ & $+43w$ & $+663r$   & $+R + 293s$ \\
$100$ & $2{,}293$ & $+R + 51 + 6t$ & $+43w$ & $+2{,}250r$ & $+R + 1{,}125s$ \\
\hline
\end{tabular}
\vspace{1.5ex}

\setlength{\tabcolsep}{2pt}%
\adjustbox{max width=\linewidth}{%
\begin{tabular}{|c|c|c|c|c|c|}
\hline
 & reg., $t{=}0$ & reg., $t{=}100$ & reg., $t{=}10^4 $ & reg., $n$ joins & $100\%$ refresh \\
$n$ & $w{=}r{=}s{=}0$ & $w{=}r{=}s{=}0$ & $w{=}r{=}s{=}0$ & $t{=}w{=}r{=}0$, $s{=}n$ & $t{=}w{=}0$, $r{=}s{=}n$ \\
\hline
$4$ & $260$ (\$0.8) & $1{,}063$ (\$3.2) & $60{,}921$ (\$183) & $648$ (\$1.9) & $1{,}516$ (\$4.5) \\
$25$ & $705$ (\$2.1) & $1{,}508$ (\$4.5) & $61{,}366$ (\$184) & $8{,}178$ (\$25) & $24{,}753$ (\$74) \\
$100$ & $2{,}293$ (\$6.9) & $3{,}096$ (\$9.3) & $62{,}954$ (\$189) & $114{,}941$ (\$345) & $339{,}941$ (\$1{,}020) \\
\hline
\end{tabular}}
\vspace{6pt}
\setlength{\belowcaptionskip}{0pt}%
\caption{Size in vB of a regular checkpoint (\checkpointTxOne) and its operational additions for subnets with $n$ validators: $t$ transfers to a target subnet, $w$ withdrawals, $r$ refreshed collateral UTXOs, and $s$ pre-signed \slashTx. Overhead $R = 148$~vB is added once (if transfers or pre-signed \slashTx exist) for the commitment output of \checkpointTxOne and its spending \checkpointTxTwo. Output paying the target subnet adds $51$~vB ($43$~vB output + $8$~vB \transferData header), plus $6$~vB per transfer.\\
The second table reports total vB written to Bitcoin by \checkpointCommand (and USD fees) across five cases: regular checkpoints with $0$, $100$, or $10,000$ transfers; a post-join checkpoint publishing $n$ validators' pre-signed \slashTx; and a checkpoint where every validator refreshes its collateral (requiring a new pre-signed \slashTx).}

\label{tab:checkpoint-sizes}
\label{tab:rotation-total}
\end{table*}}

\section{Evaluation}\label{sec:benchmarks}
\iflncs\else
    \checkpointSizesTable
\fi
We benchmark the \bitcoinIpc protocol by measuring the amount of data submitted to Bitcoin (and the corresponding Bitcoin fees) for its various operations.


\begin{description}[leftmargin=0pt, font=\normalfont\bfseries]
    \item[Setup:]
    We assume the $n$ validators of a subnet hold equal stake. A checkpoint is signed by $\lfloor 2n/3 \rfloor + 1$ validators. We also assume the subnet keeps its funds consolidated in a single UTXO, which pays fees and funds all withdrawal and transfer outputs. Fees are computed at $3$~sat/vB, a typical fee rate at the time of writing, and converted at 1~BTC = \$100,000~USD.
    The numbers reported are obtained from the \emph{actual} signed transactions that \bitcoinIpc creates and submits to Bitcoin, verified against the actual Bitcoin consensus rules. We provide details on implementation in \Cref{app:implementation-notes}%
    \iflncs
        { }and reproducibility in \Cref{sec:moreevaluation}%
    \fi.

    \iflncs\else
        \item[Reproducibility:]
We obtain the plots of this section by running the \textsl{test\_transfer\_size()} and \textsl{test\_withdraw\_size()} functions in the \textsl{src/} directory of the \BslBitcoinIpcRepo repository at version \versionNumber.\footnote{\anonurl{https://github.com/bitcoinscalinglabs/bitcoin-ipc}}
These functions create the Bitcoin transactions for all defined parameters and write the results to \textsl{bench-plots/bench-transfer-sizes.csv} and \textsl{bench-plots/bench-withdraw-sizes.csv}, respectively.
The script \textsl{bench-plots/plot.py} is then used to generate the plots.
The results are independent of the hardware used for the benchmarks and the actual consensus protocol used in the subnets.

    \fi
    \item[The \emph{vbytes} metric:]
    Sizes are reported in \emph{vbytes}: each transaction's weight equals $\textsl{base\_size} \times 3 + \textsl{total\_size}$, where $\textsl{base\_size}$ excludes witness data and $\textsl{total\_size}$ includes it; dividing by 4 yields vbytes, giving witness data a 75\% discount over non-witness data.
\end{description}

\iflncs
    \vspace{-5pt}
\fi
\experimentHeading{The cost of joining, exiting, and slashing}
A validator pays $360$~vB to join (\joinTx) and $146$~vB to exit unilaterally (the \uniExit leaf is spent with one signature and does not depend on $n$). This costs $1.08$ and $0.44$~USD, respectively. Submitting a pre-signed \slashTx costs $213$~vB at $n = 4$ and $2{,}246$~vB at $n = 100$, as it reveals the multisig script of the configuration and carries the signatures of its quorum; that is $0.64$ and $6.74$~USD, respectively.

\iflncs
    \vspace{-4pt}
\fi
\experimentHeading{The cost of \checkpointCommand}\label{sec:exp-checkpoint}
A \emph{regular checkpoint} happens often (e.g., after every Bitcoin block, max. every few hours): it commits to the subnet's latest block and handles outgoing \transferCommand and \withdrawCommand operations; a \emph{rotation checkpoint} happens less often (e.g., every week) and updates the configuration (\Cref{sec:checkpoint}).
\Cref{tab:checkpoint-sizes} builds up the size of a checkpoint incrementally: the first column is a regular checkpoint that carries nothing but the checkpoint metadata, and each further column reports what one kind of operation adds to it. Only refreshes and pre-signed \slashTx grow with $n$, as each reveals the multisig script of the configuration and carries the signatures. Based on these entries, we give the size of some typical checkpoints, from a regular checkpoint with nothing to batch to a checkpoint in which every validator refreshes its collateral.
Refreshes that exceed the standard transaction size of 100K~vB are spread over several consecutive checkpoints of the epoch, and the pre-signed \slashTx follow in the next checkpoint in any case.
\ifshorten\else
In numbers, for one target subnet, as the table assumes, a batch of transfers costs $199$~vB ($51$~vB per additional target subnet) and then only $6$~vB per transfer, at any number of validators. A withdrawal has constant size, $43$~vB.
\fi

\iflncs
    \checkpointSizesTable

    \vspace{-3pt}
\fi
\experimentHeading{Taking batching to the limits}\label{sec:exp-limits}
A transfer adds only $6$~vB on top of the fixed cost of a batch. For large batches, this fixed cost amortizes away: a batch is only limited by the standard transaction size of 100K~vB, which \checkpointTxTwo reaches at approximately \emph{16,500} transfers, where the amortized size per transfer is \emph{6.06}~vB (\emph{6.1}~vB for 10 target subnets): \checkpointTxTwo is then $99{,}875$~vB, and the batch adds $99{,}961$~vB to the checkpoint. A transfer is settled once the block that contains the checkpoint transactions is finalized on Bitcoin. 
This amounts to amortized \emph{23x fewer data} than a typical native payment (141~vB / 6.1~vB), that is, \emph{23x} more transfers in the same block space or, equivalently, \emph{23x} lower Bitcoin fees.

\iflncs
    \par\bigskip\noindent
    \begin{minipage}{\linewidth}
\else
    \begin{figure}[t]
        \centering
        \throughputPlot
        \caption{\throughputCaption}
        \label{fig:throughput-vs-n-transfers}
    \end{figure}
\fi

\experimentHeading{Improving Bitcoin throughput by batching \bitcoinIpc transfers}\label{sec:throughput-vs-n-transfers}
\iflncs
    \mbox{}

    \vspace{2pt}
    \noindent
    \begin{minipage}[t]{0.44\linewidth}
\fi
We now translate the \emph{23x} improvement of \Cref{sec:exp-limits} into monetary throughput, assuming that blocks contain only monetary transfers, as we did for our baseline throughput of 11.8 monetary transfers per second (\Cref{sec:intro}). 
\bitcoinIpc essentially compresses Bitcoin transfers, so that one monetary transfer fits in significantly less than 141~vBytes. Blocks filled with batches of \bitcoinIpc transfers, at 6.1~vB each, carry \emph{273} monetary transfers per second. We remark that native payments can also be batched, but only by a single sender, whereas a checkpoint batches the transfers of all users of a subnet.
In \Cref{fig:throughput-vs-n-transfers} we report the monetary transfers per second that correspond to the amortized transfer size -- one line per number of target subnets. Each
\iflncs
    \end{minipage}\hfill
    \begin{minipage}[t]{0.53\linewidth}
        \vspace{-3pt}
        \centering
        \throughputPlot
        \captionof{figure}{\throughputCaption}
        \label{fig:throughput-vs-n-transfers}
    \end{minipage}
    \end{minipage}

    \smallskip\noindent
\fi
additional target subnet adds 51~vB, shared by all transfers in the batch: the lines differ for small batches and converge to about 273 monetary transfers per second as the batch grows.

\iflncs\else
\iflncs
    \section{The complete benchmarks}
    \label{sec:moreevaluation}

    \dotparagraph{Reproducibility}
We obtain the plots of this section by running the \textsl{test\_transfer\_size()} and \textsl{test\_withdraw\_size()} functions in the \textsl{src/} directory of the \BslBitcoinIpcRepo repository at version \versionNumber.\footnote{\anonurl{https://github.com/bitcoinscalinglabs/bitcoin-ipc}}
These functions create the Bitcoin transactions for all defined parameters and write the results to \textsl{bench-plots/bench-transfer-sizes.csv} and \textsl{bench-plots/bench-withdraw-sizes.csv}, respectively.
The script \textsl{bench-plots/plot.py} is then used to generate the plots.
The results are independent of the hardware used for the benchmarks and the actual consensus protocol used in the subnets.

\fi

\subsection{The size of the \transferCommand transactions}\label{sec:transfer-size-vs-n-transfers}
This experiment asks how many vbytes a cross-subnet \transferCommand writes on Bitcoin L1, relative to a native Bitcoin transfer. \Cref{fig:amortized-size-vs-n-transfers} plots amortized size per transfer against the total number of batched transfers (log scale), with one curve per number of target subnets and a dashed line for a native Bitcoin transfer (\emph{141~vB}). \bitcoinIpc undercuts the native cost once four transfers are batched. Further batching lowers the amortized cost in all cases. More targets raise the cost because \checkpointTxOne has one output UTXO per target and \checkpointTxTwo stores each target address; that overhead vanishes once enough transfers are batched.



\begin{figure}[t]
    \centering
    \iflncs
        \includegraphics[width=0.9\linewidth]{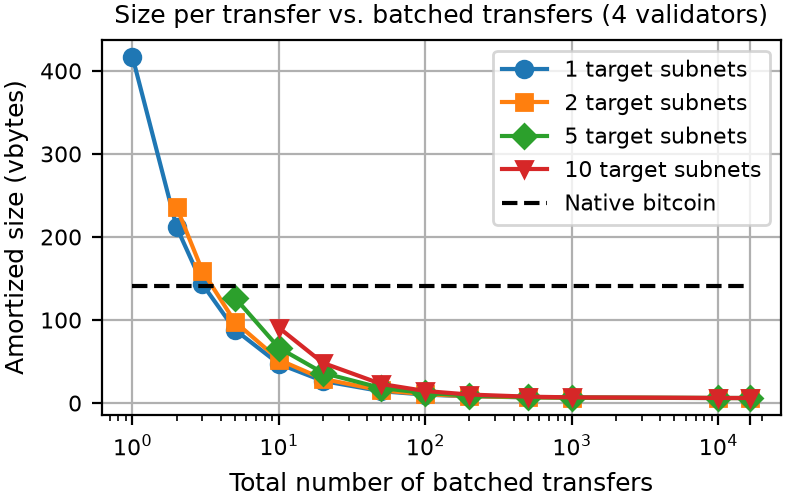}
    \else
        \includegraphics[width=\linewidth]{images/3D-amortized_vsize_vs_n_transfers_vary_n_subnets_log_small.png}
    \fi
    \caption{Amortized size per transfer vs total number of batched transfers, for varying numbers of target subnets.}
    \label{fig:amortized-size-vs-n-transfers}
\end{figure}


\begin{remark}
The experiments of this section measure the total size of the checkpoint transactions, base checkpoint included, divided by the number of transfers. \Cref{sec:exp-checkpoint,sec:exp-limits,sec:throughput-vs-n-transfers} instead measure additively what each transfer adds on top of the base checkpoint.
\end{remark}

\subsection{Different validator sizes}\label{sec:transfer-size-vs-n-validators}
So far we have assumed that the source L2 subnet has four validators. We now explore what happens when this number changes.
\Cref{fig:transfer-size-vs-validators} shows the amortized size for a transfer against the total number of batched transfers, this time fixing the number of target subnets and varying the number of validators in the source subnet.

\begin{figure}[h!]
    \centering
    \includegraphics[width=\linewidth]{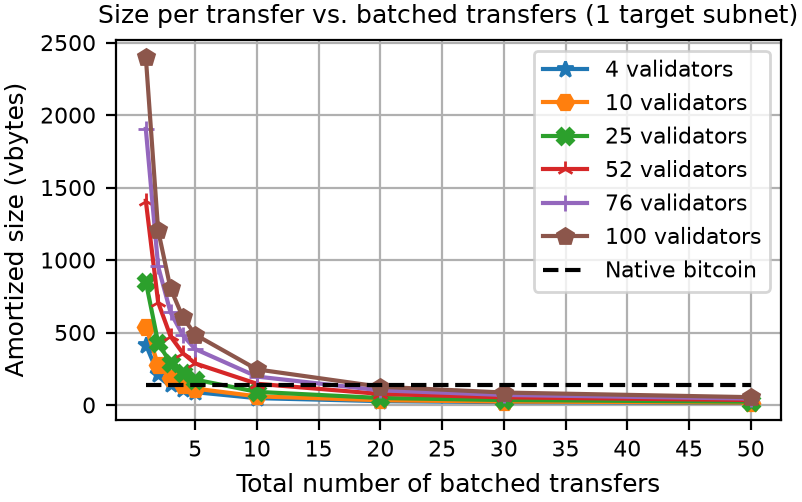}
    \caption{Amortized size per transfer vs total number of batched transfers, for varying numbers of validators in the source subnet.}
    \label{fig:transfer-size-vs-validators}
\end{figure}

We observe that the amortized size increases with the number of validators. This is due to the additional signatures that must be included in the Bitcoin transaction: the current instantiation uses Bitcoin multisig addresses (\Cref{sec:relwork} discusses threshold signatures as a direction for future work). As before, this cost is smaller if we batch many transactions. Despite the overhead of Bitcoin multisigs, \bitcoinIpc still leads to lower transaction costs than native Bitcoin. The break-even point now changes with the number of validators -- for example, when the source subnet has 37 validators the batching pays off after at least 9 transfers.

\subsection{Throughput and latency in a  subnet}\label{sec:throughput-and-latency-within-l2-subnet}

\bitcoinIpc allows each L2 subnet to run its own consensus protocol, as long as it implements certain interfaces and functionalities defined by \bitcoinIpc. We have maintained this feature in the design of \bitcoinIpc, allowing subnets to use any efficient consensus protocol.

The current implementation uses Fendermint as the consensus protocol in all L2 subnets. Hence, the throughput and latency within an L2 subnet are determined by Fendermint, which is a fork of CometBFT\footnote{\url{https://github.com/cometbft/cometbft}}. We do not repeat benchmarks on Fendermint or CometBFT at this point. Previous results report the throughput of implementations of CometBFT at about 500~tps. For more details we refer to previous works~\cite{DBLP:conf/srds/CasonFM0BP21,cometbft-bench}.

\subsection{Improving the transfer size using threshold signatures}\label{sec:threshold-signatures}

We saw in \Cref{sec:exp-checkpoint,sec:transfer-size-vs-n-validators} that the amortized transfer cost increases with the number of validators in the source subnet. Even though it becomes lower than the Bitcoin transaction size when we batch many transfers, ideally we would like that cost to be \emph{independent} of the number of validators.
\Cref{fig:threshold-signatures} shows the amortized size of a transfer for different numbers of validators in the source subnet. The plot contains a line for \emph{one validator} in the source subnet, which corresponds to an implementation of \bitcoinIpc with threshold signatures, where only one signature is included in the Bitcoin transaction.

\begin{figure}[h!]
    \centering
    \includegraphics[width=\linewidth]{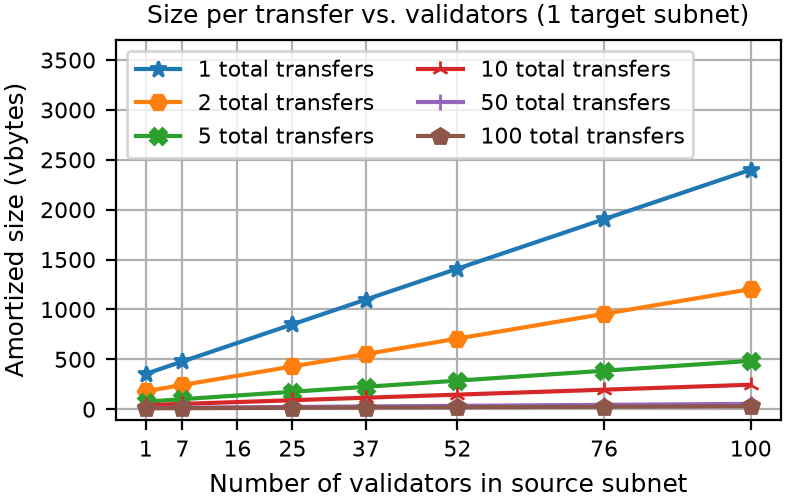}
    \caption{Amortized size per transfer vs number of validators, for varying numbers of batched transfers.}
    \label{fig:threshold-signatures}
\end{figure}

The benefit of threshold signatures (which we plan to introduce in a future release of \bitcoinIpc) is significant if we batch a small number of transfers, such as 1, 2, 5, or 10 transfers. For larger batches, consisting for example of 50 or 100 transfers, the benefit is smaller. For example, for batches of 50 transfers, the amortized cost per transfer is \emph{13}~vB with threshold signatures and \emph{54}~vB with a multisig implementation among 100 validators. For batches of 100 transfers, the corresponding sizes are \emph{9.5}~vB with threshold signatures and \emph{30}~vB with multisigs.


\subsection{Benchmarks showing fiat fees}
\label{sec:benchmarks-fees}
In this section we repeat previous plots, showing the fees instead of the transaction size in the vertical axis.


We now show how, by batching transfers together, the fee per transfer becomes smaller than sending the transfer directly as a Bitcoin L1 transaction.
\Cref{fig:fees-vs-n-transfers} shows the fee (in satoshis) per transfer against the total number of batched transfers, for a varying number of target L2 subnets (1, 2, 5, 10 target subnets). It also shows the fee for sending the transfer natively over Bitcoin, which is constant at 423~satoshis, or 0.42~USD (3~sat/vB for an average-sized transfer of 141~vB). Notice that the horizontal axis is logarithmic.
As we can see, the amortized fee per transfer converges to approx.~\emph{18} satoshis (0.02~USD), for one and for 10 target subnets alike: a saving of approx.~\emph{0.41~USD per transfer}.
Fee rates vary widely, and the saving grows with them: at the median fee rate of \emph{200~sat/vB} observed at the last halving block,\footnote{\url{https://mempool.space/block/0000000000000000000320283a032748cef8227873ff4872689bf23f1cda83a5}} a native transfer costs 28,200~satoshis (28.2~USD) and a batched one 1,214~satoshis (1.21~USD), a saving of \emph{27~USD per transfer}.

\begin{figure}[h]
    \centering
    \includegraphics[width=\linewidth]{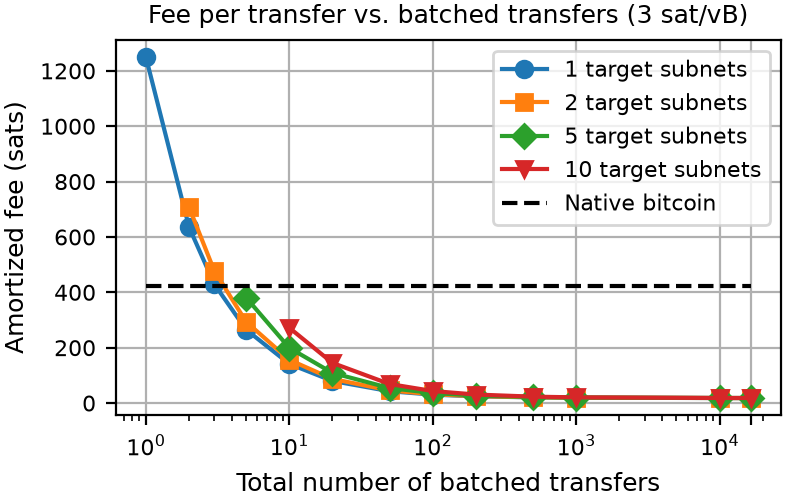}
    \caption{Fee per transfer vs total number of batched transfers, for varying numbers of target L2 subnets.}
    \label{fig:fees-vs-n-transfers}
\end{figure}



\fi

\iflncs\else
  \section{Discussion and future work}\label{sec:discussion}

If validators holding more than $1/3$ of the stake of configuration $\subnetConf_e$ stop signing, no checkpoint is produced, and \withdrawCommand and outgoing \transferCommand operations stall, even with $F_e < 2/3$. This is a liveness violation, as in case \textbf{(ii)} of our security model (\Cref{sec:preliminaries}). Unlike equivocation, withholding signatures is not attributable and cannot be slashed. Funds remain safe, as spending them requires more than $2/3$ of the stake. Recent results on accountable liveness~\cite{DBLP:conf/ccs/Lewis-PyeNRZ25} identify and penalize validators that stall a protocol. Integrating them is outside the scope of this work.

A related question is what happens to fund UTXOs if such a stall is permanent. In this work, fund UTXOs bear no unilateral-exit script, so recovering them requires more than $2/3$ of the stake, or an external process, e.g., governance, based on the history anchored on Bitcoin. In separate ongoing work, we design a construction that returns user funds on Bitcoin when a subnet stops checkpointing, without the cooperation of its validators.

Checkpoints may also stall if the subnet cannot pay for them. Checkpoint fees are paid from fee UTXOs, which are locked under the multisig of the configuration and replenished by the validators. Separate mechanisms can also fund them, e.g., a fee on deposited amounts.
We leave such mechanisms out of scope.
Validators are rewarded for staking their \BTC (\Cref{app:staking}), and the reward token can be exchanged for \BTC.

One might also ask whether many subnets can spam Bitcoin with checkpoints. Every \bitcoinIpc operation is a standard Bitcoin transaction and pays the prevailing fee rate from the subnet's own fee UTXOs, so such spam costs market fees per vB, as for any other transaction. Moreover, a subnet is activated only once enough validators have locked collateral of at least \minCollateral each.

Finally, batching affects latency. A cross-subnet transfer is batched in the next checkpoint of the source subnet, settled once the Bitcoin block containing it is finalized, and processed by the next checkpoint of the destination subnet. Its latency is thus roughly one checkpoint period of each subnet plus Bitcoin finality. The checkpoint period \checkpointInterval, fixed at subnet creation, trades latency for cost: a longer period amortizes the fixed cost of a batch over more transfers, approaching 6~vB per transfer, while a shorter period lowers latency. As argued in \Cref{sec:benchmarks}, the interval for \emph{regular} checkpoints can be low.

\dotparagraph{Future work} \label{sec:relwork}
Previous works~\cite{DBLP:conf/consensusday/AzouviV22,FMT24} envision threshold signatures for checkpointing PoS chains onto Bitcoin. While \bitcoinIpc currently relies on multisigs for certain operations, we plan to integrate threshold signatures as future work.
Moreover, future work will allow custom tokens, natively deployed on any \bitcoinIpc subnet, to benefit from the interoperability of \bitcoinIpc -- relying on the firewall property to minimize the impact of malicious subnets.
We also plan to explore integration of LN with \bitcoinIpc for even better scalability.



\fi
\clearpage

\iflncs\bibliographystyle{splncs04}\else\bibliographystyle{plain}\fi
\bibliography{references,blockchain}

\iflncs\clearpage\else\bigskip\fi

\appendix

\iflncs 
    \dotparagraph{Appendix structure}
    The structure of the Appendix is as follows.
    \Cref{sec:protocol-details} presents construction elements omitted from the main body due to space limitation, and \Cref{sec:pseudocode-all} the pseudocode for all \bitcoinIpc operations.
    \Cref{sec:model-full} contains the model and formal definitions for the protocols used in \bitcoinIpc,
    while \Cref{sec:proofs} contains the corresponding proofs.
    \Cref{app:implementation-notes} provides more details on the implementation used in our benchmarks.
    \Cref{sec:moreevaluation} provides additional benchmarks and reproducibility details.
    We discuss incentives in \Cref{app:staking} and conclude with general discussion and future work in \Cref{sec:discussion}.

    \input{sections/construction-details}
    \section{Pseudocode}\label{sec:pseudocode-all}
The following algorithms show the code of \anIpcAwareNode, in this order: its state and how it processes Bitcoin blocks; the \createCommand, \joinCommand, and \stakeCommand operations, that is, how it submits the corresponding transactions to Bitcoin and how it verifies them once they are finalized; how it determines the configuration of an epoch and processes checkpoints; and how the validators of a subnet create the checkpoint transactions.

\dotparagraph{Notation}
Notation `\textbf{require} $P$' causes a function or operation to terminate immediately and return \false if $P$ evaluates to \false; in that case the state of the node is not modified.
Notation `\textbf{upon} $e$' causes a block of code to be executed when event $e$ occurs.
Finally, $x: a \in A \rightarrow b \in B$ denotes that variable $x$ is a map from elements of type $A$ to elements of type $B$.
With $\emptyset$ we denote an empty map, and $x[a] = \bot$ denotes that map $x$ has no entry for key $a$.
For a sequence $X$, $X.\mathsf{push}(x)$ appends $x$ and $X.\mathsf{pop}()$ removes and returns the last element. Assigning $x[a] \gets \bot$ removes the entry. A script built from an undefined configuration $\subnetConf = \bot$ is $\bot$ and locks no output. A UTXO is denoted by its outpoint $(\mathsf{tx}, n)$.
Functions \opSign{\sk, m} and \opVerify{\pk, m, \sigma} denote Schnorr signing and verification, and \opMerkleRoot{x_1, \ldots, x_n} the root of a Merkle tree with leaves $x_1, \ldots, x_n$.

\begin{pseudocodefloat}
    \caption{Protocol \bitcoinIpc: Code for \anIpcAwareNode $V$, part 1: state and processing of Bitcoin blocks.}
    \label{alg:node-state}
    \begin{algorithmic}[1]
        \small
        \small
        \State{\textbf{General \bitcoinIpc state:}}
            \State{$\nodeParams : \subnetId \rightarrow \subnetParams$}   \Comment{Parameters \subnetParams of each created subnet}\label{line:node-state-begin}
            \State{$\nodeStatus : \subnetId \rightarrow \{\statusInitialized, \statusActive, \statusToBeKilled, \statusKilled\}$}
            \State{$\nodeEpoch : \subnetId \rightarrow \BN \cup \{0\}$}   \Comment{Current epoch of each subnet}
            \State{$\nodeActHeight : \subnetId \rightarrow \BN$}   \Comment{Activation height \heightActive of each subnet}
            \State{$\nodeConf : \subnetId \rightarrow (e \rightarrow \subnetConf_e)$}   \Comment{Configuration $\subnetConf_e$ of each epoch $e \in \BN \cup \{0\}$}
            \State{$\nodeCollaterals : \subnetId \rightarrow (\mathit{col} \rightarrow (\pkCons[], \pkExit[], \colDuration))$}   \Comment{Accepted collaterals, by outpoint}
            \State{$\nodeValidators : \subnetId \rightarrow (\pkCons[] \rightarrow (\pkExit[], \pkEots[]))$}   \Comment{Registered validators}
            \State{$\nodeNonces : \pkEots[] \rightarrow \{\nonceCom\}$}   \Comment{Nonce commitments per EOTS key, $\nonceCom = (\nonceStart, \nonceCount, \nonceRoot)$}
            \State{$\nodeCheckpoints : \subnetId \rightarrow [(\subnetHeight, \comBlock, \height)]$}   \Comment{Anchored tuples and their Bitcoin height \height}
            \State{$\nodeSlashable : \subnetId \rightarrow \{\mathit{col}\}$}   \Comment{UTXOs with a valid pre-signed \slashTx}\label{line:node-state-end}
        \Statex
        \vspace{-2mm}
        \State{\textbf{Validator state:}} \Comment{State of $V$ for subnets it joins}
            \State{$\nodeEotsSk : \subnetId \rightarrow \skEots[]$}   \Comment{EOTS secret key}
            \State{$\nodeSecNonces : \subnetId \rightarrow (h \rightarrow \secNonce{h})$}   \Comment{Secret nonce \secNonce{h} for subnet height $h$}
        \Statex
        \vspace{-2mm}
        \On{\opInit{}} \Comment{Called once when the \ipcAwareNode is initialized}
            \State{$\nodeParams \gets \emptyset ;\; \nodeStatus \gets \emptyset ;\; \nodeEpoch \gets \emptyset ;\; \nodeActHeight \gets \emptyset ;\; \nodeConf \gets \emptyset$}
            \State{$\nodeCollaterals \gets \emptyset ;\; \nodeValidators \gets \emptyset ;\; \nodeNonces \gets \emptyset ;\; \nodeCheckpoints \gets \emptyset$}
            \State{$\nodeSlashable \gets \emptyset ;\; \nodeEotsSk \gets \emptyset ;\; \nodeSecNonces \gets \emptyset$}
        \EndOn
        \Statex
        \vspace{-2mm}
        \On{Bitcoin block $B$ at height $\height$ is finalized} \label{line:node-block-begin} \Comment{In increasing order of $\height$}
            \For{$\tx \in B$ \textbf{in the order of} $B$}
                \Let{\data}{\text{data in the witness of input $0$ or in \opReturn output of \tx, or $\bot$}} \label{line:node-canonical-data}
                \State{\textbf{if} \data not starts with one of $\{\createKeyword, \joinKeyword, \stakeKeyword, \checkpointKeyword, \depositKeyword, \transferKeyword\}$ \textbf{then continue}} \label{line:node-tag-barrier}
                \State{\textbf{if} \data starts with \createKeyword \textbf{then} \opProcessCreate{\tx}} \label{line:node-call-create} \Comment{\Cref{alg:node-create}}
                \State{\textbf{if} \data starts with \joinKeyword \textbf{then} \opProcessJoin{\tx}} \label{line:node-call-join} \Comment{\Cref{alg:node-join}}
                \State{\textbf{if} \data starts with \stakeKeyword \textbf{then} \opProcessStake{\tx}} \label{line:node-call-stake}
                \State{\textbf{if} \data starts with \checkpointKeyword \textbf{then} \opProcessCheckpoint{\tx, \height}} \label{line:node-call-checkpoint} \Comment{\Cref{alg:node-process}}
            \EndFor
            \LineComment{At the cutoff of an epoch, start the next one and fix its configuration}
            \For{\subnetId\ \textbf{s.t.} $\nodeStatus[\subnetId] \in \{\statusActive, \statusToBeKilled\}$} \label{line:node-cutoff}
                \If{$\height = \nodeActHeight[\subnetId] + \nodeEpoch[\subnetId] \cdot \nodeParams[\subnetId].\epochLength$}
                    \Let{\nodeEpoch[\subnetId]}{\nodeEpoch[\subnetId] + 1}
                    \Let{\nodeConf[\subnetId][\nodeEpoch[\subnetId]]}{\opGetConfiguration{\subnetId}} \label{line:node-next-conf}
                \EndIf
            \EndFor
        \EndOn\label{line:node-block-end}
    \end{algorithmic}
\end{pseudocodefloat}

\begin{pseudocodefloat}
    \caption{Protocol \bitcoinIpc: Code for \anIpcAwareNode $V$, part 2: the \createCommand operation.}
    \label{alg:node-create}
    \begin{algorithmic}[1]
        \small
        \Operation{\createCommand(\subnetParams)} \label{line:create-op-begin} \Comment{Part of \bitcoinIpc interface}
            \State{\textbf{require} \opValidCreateParams{\subnetParams}} \label{line:create-op-params} \Comment{\subnetParams: subnet parameters (\Cref{sec:create-subnet})}
            \Let{\subnetData}{\createKeyword \concat \subnetParams}
            \Let{\createTx}{\text{transactions from \Cref{fig:tx-create} for data \subnetData}} \label{line:create-op-txs}
            \State{submit \createTx to Bitcoin}
        \EndOperation\label{line:create-op-end}
        \Statex
        \vspace{-2mm}
        \Function{\opProcessCreate{\createTxTwo}}{} \label{line:create-verify-begin}
            \Let{\createKeyword \concat \subnetParams}{\text{data in the witness of input $0$ of } \createTxTwo}
            \State{\textbf{require} \opValidCreateParams{\subnetParams}} \Comment{Otherwise, ignore \createTxTwo}
            \Let{\subnetId}{\opDeriveSubnetId{\opTxid{\createTxTwo}}} \label{line:node-derive-id}
            \State{\textbf{require} $\nodeParams[\subnetId] = \bot$} \label{line:create-verify-unique} \Comment{A subnet is created only once}
            \State{$\nodeParams[\subnetId] \gets \subnetParams ;\; \nodeStatus[\subnetId] \gets \statusInitialized$}
            \State{$\nodeEpoch[\subnetId] \gets 0 ;\; \nodeCollaterals[\subnetId] \gets \emptyset ;\; \nodeValidators[\subnetId] \gets \emptyset$}
            \State{$\nodeCheckpoints[\subnetId] \gets [\,] ;\; \nodeSlashable[\subnetId] \gets \emptyset$}
            \State{$\nodeConf[\subnetId][0] \gets \{(\pk, 1) : \pk \in \subnetParams.\whitelist\}$} \Comment{\whitelist, equal weights}
        \EndFunction\label{line:create-verify-end}
        \Statex
        \vspace{-2mm}
        \Function{\opValidCreateParams{\subnetParams}}{} \label{line:valid-params-begin}
            \Let{(\minCollateral, \whitelistThreshold, \checkpointInterval, \epochLength, \minTimeout, \maxTimeout, \permissionPolicy, \killVotePeriod, \killDelayPeriod, \whitelist)}{\subnetParams}
            \State{\textbf{require} $\minCollateral > 0$ \andl $\whitelistThreshold \geq 1$ \andl $\checkpointInterval > 0$ \andl $\killVotePeriod > 0$}
            \State{\textbf{require} $0 < \epochLength \leq 21845$ \andl $0 < \minTimeout \leq \maxTimeout$} \Comment{So timelock $3\epochLength$ is encodable}
            \State{\textbf{require} $\whitelist \neq \emptyset$} \label{line:valid-params-bootstrap} \Comment{Needed to build $\colS$}
            \State{\textbf{require} the public keys in \whitelist are valid and distinct} \label{line:valid-params-keys}
            \State{\textbf{require} $|\whitelist| \geq \whitelistThreshold$} \label{line:valid-params-whitelist-size}
        \EndFunction\label{line:valid-params-end}
        \Statex
        \vspace{-2mm}
        \Function{\opDeriveSubnetId{s}}{}
            \State{\textbf{return} the first $20$ bytes of $s$}
        \EndFunction
        \Statex
        \vspace{-2mm}
        \Function{\opTxid{\tx}}{}
            \State{\textbf{return} the hash of \tx, computed without its witnesses}
        \EndFunction
    \end{algorithmic}
\end{pseudocodefloat}

\begin{pseudocodefloat}
    \caption{Protocol \bitcoinIpc: Code for \anIpcAwareNode $V$, part 3: the \joinCommand operation.}
    \label{alg:node-join}
    \begin{algorithmic}[1]
        \small
        \Operation{\joinCommand($\subnetId, \pkCons, \pkExit, \amountShort, \colDuration$)} \label{line:join-op-begin}\Comment{Part of \bitcoinIpc interface}
            \State{\textbf{require} $\nodeValidators[\subnetId][\pkCons] = \bot$} \Comment{Join once per subnet}
            \State{\textbf{require} \opValidateJoinParams{\subnetId, \pkCons, \pkExit, \amountShort, \colDuration}} \label{line:join-op-params} \Comment{\Cref{alg:node-scripts}}
            \Let{\joinData}{\joinKeyword \concat \subnetId \concat \pkCons \concat \pkExit \concat \colDuration}
            \State{$(\skEots, \pkEots) \gets \opInitEotsKeys{} ;\; \nodeEotsSk[\subnetId] \gets \skEots$} \label{line:join-op-eots} \Comment{\Cref{alg:node-stake}}
            \State{choose $\nonceStart, \nonceCount > 0$} \label{line:join-op-range}
            \For{$h \in [\nonceStart, \nonceStart + \nonceCount)$} \label{line:join-op-sample-nonces}
                \Let{(\secNonce{h}, \nonce{h})}{\opGenEotsNonce{}} \Comment{\Cref{alg:node-stake}} 
                \Let{\nodeSecNonces[\subnetId][h]}{\secNonce{h}}
            \EndFor
            \Let{\nonceRoot}{\opMerkleRoot{\nonce{\nonceStart}, \ldots, \nonce{\nonceStart + \nonceCount - 1}}}
            \State{$\nonceCom \gets (\nonceStart, \nonceCount, \nonceRoot) ;\; \joinData \gets \joinData \concat \pkEots \concat \nonceCom$}
            \State{$\nonceComSig \gets \opSign{\skEots, \joinData} ;\; \joinData \gets \joinData \concat \nonceComSig$} \label{line:join-op-sign}
            \State{$e \gets \nodeEpoch[\subnetId] ;\; \subnetConf \gets \nodeConf[\subnetId][e]$} 
            \Let{S}{\opBuildCollateralScript{\pkExit, \pkEots, \colDuration, \subnetConf}} \label{line:join-op-script} \Comment{\Cref{alg:node-scripts}}
            \Let{\joinTx}{\text{transactions from \Cref{fig:tx-join} using \joinData, \amountShort, and $S$}} \label{line:join-op-txs}
            \State{submit \joinTx to Bitcoin}
        \EndOperation\label{line:join-op-end}
        \Statex
        \vspace{-2mm}
        \Function{\opProcessJoin{\joinTxTwo}}{} \label{line:join-verify-begin} \Comment{$W$: the validator that joins}
            \Let{\joinData}{\text{data in the witness of input $0$ of } \joinTxTwo}
            \Let{\joinKeyword \concat \subnetId \concat \pkCons[W] \concat \pkExit[W] \concat \colDuration \concat \pkEots[W] \concat \nonceCom \concat \nonceComSig}{\joinData} \label{line:join-verify-parse}
            \Let{\mathit{col}}{(\joinTxOne, 1), \text{ for the } \joinTxOne \text{ spent by input $0$ of } \joinTxTwo} \label{line:join-verify-output}
            \State{\textbf{require} $\mathit{col}$ exists \andl is unspent} \label{line:join-verify-parse-output}
            \Let{(\amountShort, S')}{\text{output } 1 \text{ of } \joinTxOne}
            \State{\textbf{require} $\nodeValidators[\subnetId][\pkCons[W]] = \bot$} \label{line:join-verify-once} \Comment{Join once per subnet}
            \State{\textbf{require} \opValidateJoinParams{\subnetId, \pkCons[W], \pkExit[W], \amountShort, \colDuration}} \label{line:join-verify-params} 
            \State{$e \gets \nodeEpoch[\subnetId]$} \label{line:join-verify-conf}
            \Let{S_e}{\opBuildCollateralScript{\pkExit[W], \pkEots[W], \colDuration, \nodeConf[\subnetId][e]}} \label{line:join-verify-script}
            \Let{S_{e-1}}{\opBuildCollateralScript{\pkExit[W], \pkEots[W], \colDuration, \nodeConf[\subnetId][e-1]}} \label{line:join-verify-script-prev}
            \State{\textbf{require} $S' \in \{S_e, S_{e-1}\}$} \label{line:join-verify-script-match} \Comment{Script must be correct, either for $\subnetConf_e$ or $\subnetConf_{e-1}$}
            \State{\textbf{require} $\nodeNonces[\pkEots[W]] = \bot$} \label{line:join-nonce-fresh}
            \Let{m}{\joinKeyword \concat \subnetId \concat \pkCons[W] \concat \pkExit[W] \concat \colDuration \concat \pkEots[W] \concat \nonceCom}
            \State{\textbf{require} \opCheckNonceCom{\pkEots[W], m, \nonceCom, \nonceComSig}} \Comment{\Cref{alg:node-scripts}}
            \Let{\nodeValidators[\subnetId][\pkCons[W]]}{(\pkExit[W], \pkEots[W])} \label{line:join-verify-register}
            \Let{\nodeCollaterals[\subnetId][\mathit{col}]}{(\pkCons[W], \pkExit[W], \colDuration)} \label{line:join-verify-pending}
        \EndFunction\label{line:join-verify-end}
    \end{algorithmic}
\end{pseudocodefloat}

\begin{pseudocodefloat}
    \caption{Protocol \bitcoinIpc: Code for \anIpcAwareNode $V$, part 4: auxiliary functions of the \joinCommand operation.}
    \label{alg:node-scripts}
    \begin{algorithmic}[1]
        \small
        \Function{\opValidateJoinParams{\subnetId, \pkCons[], \pkExit[], \amountShort, \colDuration}}{} \label{line:join-params-begin}
            \State{\textbf{require} $\nodeParams[\subnetId] \neq \bot$ \textbf{and} $\nodeStatus[\subnetId] \neq \statusKilled$} \label{line:join-verify-status}
            \Let{(\minCollateral, \cdot, \cdot, \epochLength, \minTimeout, \maxTimeout, \permissionPolicy, \cdot, \cdot, \whitelist)}{\nodeParams[\subnetId]}
            \State{\textbf{require} $\amountShort \geq \minCollateral$} \label{line:join-verify-value} \Comment{Collateral value must be sufficient}
            \State{\textbf{require} \permissionPolicy allows $\pkCons[]$} \label{line:join-verify-policy}
            \If{$\nodeStatus[\subnetId] = \statusInitialized$} \label{line:join-verify-whitelist}
                \State{\textbf{require} $\pkExit[] \in \whitelist$} \Comment{Before activation, only whitelisted validators join}
                \Let{\heightActive}{\text{latest finalized Bitcoin height}} \label{line:join-verify-h0}
            \Else
                \Let{\heightActive}{\nodeActHeight[\subnetId]}
            \EndIf
            \Let{e}{\nodeEpoch[\subnetId]} \Comment{Current epoch}        
            \Let{E_{\colDuration}}{\lceil (\colDuration - \heightActive) \,/\, \epochLength \rceil} \label{line:join-verify-epoch-of-timeout} \Comment{Epoch that contains height \colDuration$^{\dagger}$}
            \State{\textbf{require} $e + \minTimeout + 3 \leq E_{\colDuration} \leq e + \maxTimeout + 3$} \label{line:join-verify-timeout} \Comment{Min \minTimeout, max \maxTimeout epochs}
        \EndFunction\label{line:join-params-end}
        \Statex
        \vspace{-2mm}
        \Function{\opCheckNonceCom{\pkEots[W], m, \nonceCom, \nonceComSig}}{} \Comment{$m$: the \joinData fields before \nonceComSig}
            \State{\textbf{require} \opVerify{\pkEots[W], m, \nonceComSig}} \label{line:join-nonce-verify}
            \Let{(\nonceStart, \nonceCount, \nonceRoot)}{\nonceCom}
            \State{\textbf{require} $\nonceStart, \nonceCount > 0$}
            \For{$(\nonceStart', \nonceCount', \cdot) \in \nodeNonces[\pkEots[W]]$}
                \State{\textbf{require} $\nonceStart \geq \nonceStart' + \nonceCount'$} \label{line:join-nonce-overlap}
            \EndFor
            \Let{\nodeNonces[\pkEots[W]]}{\nodeNonces[\pkEots[W]] \cup \{\nonceCom\}} \label{line:join-nonce-store}
        \EndFunction
        \Statex
        \vspace{-2mm}
        \Function{\opBuildCollateralScript{\pkExit[], \pkEots[], \colDuration, \subnetConf}}{} \label{line:build-cols-begin}
            \State{\textbf{if} $\subnetConf = \bot$ \textbf{then return} $\bot$} 
            \Let{\mathcal{S}}{\emptyset} \Comment{Leaf scripts of the tree}
            \State{$S_{\mathsf{UE}} \gets \mathsf{CheckSig}_{\pkExit[]} \wedge \mathsf{AbsTimelock}(\colDuration) ;\; \mathcal{S} \gets \mathcal{S} \cup \{S_{\mathsf{UE}}\}$} \Comment{Unilateral-exit leaf}
            \State{\textbf{require} $\pkEots[] \neq \bot$} \label{line:build-cols-eots} 
            \State{$S_{\mathsf{slash}} \gets \mathsf{CheckSig}_{\pkEots[]} \wedge \mathsf{CheckMultiSig}_{\subnetConf} ;\; \mathcal{S} \gets \mathcal{S} \cup \{S_{\mathsf{slash}}\}$} \Comment{Slash leaf}
            \State{$S_{\mathsf{CE}} \gets \mathsf{CheckSig}_{\pkExit[]} \wedge \mathsf{CheckMultiSig}_{\subnetConf} ;\; \mathcal{S} \gets \mathcal{S} \cup \{S_{\mathsf{CE}}\}$} \Comment{Cooperative-exit leaf}
            \Let{P_{\mathit{int}}}{\text{unspendable NUMS point (\Cref{sec:tx-notation})}}
            \return $\langle P_{\mathit{int}}, \mathcal{S} \rangle$ \Comment{Script tree with internal key $P_{\mathit{int}}$ and leaves $\mathcal{S}$}
        \EndFunction\label{line:build-cols-end}
    \end{algorithmic}
    \vspace{1mm}{\footnotesize $^{\dagger}$ Each epoch is \epochLength Bitcoin blocks and epoch 1 starts at activation height $\heightActive=\nodeActHeight[\subnetId]$. For this check only, if the subnet is still \statusInitialized (hence \heightActive not known), we assume it activates at the current height.\par}
\end{pseudocodefloat}

\begin{pseudocodefloat}
    \caption{Protocol \bitcoinIpc: Code for \anIpcAwareNode $V$, part 5: the \stakeCommand operation, without nonce commitments. A \stakeCommand carrying nonce commitments is a commit-reveal pair, and they are processed as in \joinCommand.}
    \label{alg:node-stake}
    \begin{algorithmic}[1]
        \small
        \Operation{\stakeCommand($\subnetId, \pkCons, \amountShort, \colDuration, \mathit{col}'$)} \label{line:stake-op-begin} \Comment{Part of \bitcoinIpc interface}
            \State{\textbf{require} $\nodeValidators[\subnetId][\pkCons] \neq \bot$} \Comment{where $\mathit{col}'$: matured or $\bot$}
            \Let{(\pkExit, \pkEots)}{\nodeValidators[\subnetId][\pkCons]}
            \State{\textbf{require} \opValidateJoinParams{\subnetId, \pkCons, \pkExit, \amountShort, \colDuration}}
            \Let{\stakeData}{\stakeKeyword \concat \subnetId \concat \pkCons \concat \colDuration}
            \State{$e \gets \nodeEpoch[\subnetId] ;\; S \gets \opBuildCollateralScript{\pkExit, \pkEots, \colDuration, \nodeConf[\subnetId][e]}$}
            \State{$I \gets [\,]$ ;\; \textbf{if} $\mathit{col}' \neq \bot$ \textbf{then} $I.\mathsf{push}((\mathit{col}', S_{\mathsf{UE}}))$} \Comment{$S_{\mathsf{UE}}$: unilateral-exit leaf of $\mathit{col}'$}
            \State{$I.\mathsf{push}((*))$} \Comment{Funds of $V$}
            \Let{\stakeTx}{(I, [(0, \opReturn(\stakeData)), (\amountShort, S)])}
            \State{submit \stakeTx to Bitcoin}
        \EndOperation \label{line:stake-op-end}
        \Statex
        \vspace{-2mm}
        \Function{\opProcessStake{\stakeTx}}{} \label{line:stake-verify-begin} 
            \Let{\stakeKeyword \concat \subnetId \concat \pkCons[W] \concat \colDuration}{\text{data in the \opReturn output of } \stakeTx}
            \State{\textbf{require} $\nodeValidators[\subnetId][\pkCons[W]] \neq \bot$}
            \Let{(\pkExit[W], \pkEots[W])}{\nodeValidators[\subnetId][\pkCons[W]]}
            \Let{(\amountShort, S')}{\text{output } 1 \text{ of } \stakeTx}
            \State{\textbf{require} \opValidateJoinParams{\subnetId, \pkCons[W], \pkExit[W], \amountShort, \colDuration}}
            \Let{e}{\nodeEpoch[\subnetId]}
            \Let{S_e}{\opBuildCollateralScript{\pkExit[W], \pkEots[W], \colDuration, \nodeConf[\subnetId][e]}}
            \Let{S_{e-1}}{\opBuildCollateralScript{\pkExit[W], \pkEots[W], \colDuration, \nodeConf[\subnetId][e-1]}}
            \State{\textbf{require} $S' \in \{S_e, S_{e-1}\}$} \label{line:stake-verify-script}
            \Let{\nodeCollaterals[\subnetId][(\stakeTx, 1)]}{(\pkCons[W], \pkExit[W], \colDuration)} \label{line:stake-verify-register}
        \EndFunction \label{line:stake-verify-end}
        \Statex
        \vspace{-2mm}
        \Function{\opInitEotsKeys{}}{}
            \State{\textbf{return} a fresh EOTS key pair} \label{line:init-eots}
        \EndFunction
        \Statex
        \vspace{-2mm}
        \Function{\opGenEotsNonce{}}{} \label{line:gen-eots-nonce}
            \State{$k \getsrandomly \BZ_q$ ;\; \textbf{return} $(k, k \cdot G)$} \Comment{$G$: generator of the EOTS group, of order $q$}
        \EndFunction
        \Statex
        \vspace{-2mm}
        \Function{\opEotsSign{\skEots[], \secNonce{h}, m}}{} \label{line:eots-sign} \Comment{$\secNonce{h}$: secret nonce for height $h$, used once}
            \State{$e \gets H(\nonce{h} \concat \pkEots[] \concat m)$ ;\; \textbf{return} $\secNonce{h} + e \cdot \skEots[] \bmod q$} \Comment{$\nonce{h} = \secNonce{h} \cdot G$}
        \EndFunction
        \Statex
        \vspace{-2mm}
        \Function{\opEotsVerify{\pkEots[], \nonce{h}, m, s}}{} \label{line:eots-verify} \Comment{$\nonce{h}$: committed public nonce}
            \State{$e \gets H(\nonce{h} \concat \pkEots[] \concat m)$ ;\; \textbf{return} $s \cdot G = \nonce{h} + e \cdot \pkEots[]$}
        \EndFunction
        \Statex
        \vspace{-2mm}
        \Function{\opEotsExtract{\pkEots[], \nonce{h}, (m, s), (m', s')}}{} \label{line:eots-extract}
            \State{\textbf{require} $m \neq m'$ \andl \opEotsVerify{\pkEots[], \nonce{h}, m, s}}
            \State{\textbf{require} \opEotsVerify{\pkEots[], \nonce{h}, m', s'}}
            \State{$e \gets H(\nonce{h} \concat \pkEots[] \concat m) ;\; e' \gets H(\nonce{h} \concat \pkEots[] \concat m')$}
            \State{\textbf{return} $(s - s') \cdot (e - e')^{-1} \bmod q$} \Comment{The secret key $\skEots[]$}
        \EndFunction
    \end{algorithmic}
\end{pseudocodefloat}

\begin{pseudocodefloat}
    \caption{Protocol \bitcoinIpc: Code for \anIpcAwareNode $V$, part 6: auxiliary functions and the configuration of an epoch.}
    \label{alg:node-config}
    \begin{algorithmic}[1]
        \small
        \Function{\opUnspent{S}}{} \label{line:unspent}
            \State{\textbf{return} the outpoints of the unspent outputs locked under script $S$}
        \EndFunction
        \Statex
        \vspace{-2mm}
        \Function{\opBuildUnbondingScript{\pkExit[], \pkEots[], \epochLength, \subnetConf}}{} \label{line:build-unbs}
            \State{$S_{\mathsf{UU}} \gets \mathsf{CheckSig}_{\pkExit[]} \wedge \mathsf{RelTimelock}(3\epochLength)$} \Comment{Unilateral-unbonding leaf}
            \State{$S_{\mathsf{slash}} \gets \mathsf{CheckSig}_{\pkEots[]} \wedge \mathsf{CheckMultiSig}_{\subnetConf}$} \Comment{Slash leaf}
            \State{\textbf{return} $\langle P_{\mathit{int}}, \{S_{\mathsf{UU}}, S_{\mathsf{slash}}\} \rangle$} \Comment{$P_{\mathit{int}}$: unspendable NUMS point}
        \EndFunction
        \Statex
        \vspace{-2mm}
        \Function{\opBuildSlashTx{\mathit{col}}}{} \label{line:build-slash}
            \State{\textbf{return} the slashing transaction \slashTx of \Cref{fig:tx-slash} spending $\mathit{col}$}
        \EndFunction
        \Statex
        \vspace{-2mm}
        \Function{\opSubnetOf{\tx^{\ddagger}}}{} \label{line:subnet-of}
            \For{\subnetId\ \textbf{s.t.} $\nodeParams[\subnetId] \neq \bot$}
                \State{$e \gets \nodeEpoch[\subnetId] ;\; \mathcal{M} \gets \{\mathsf{CheckMultiSig}_{\nodeConf[\subnetId][e]}, \mathsf{CheckMultiSig}_{\nodeConf[\subnetId][e-1]}\}$}
                \State{\textbf{if} an input of \tx is locked under a script in $\mathcal{M}$ \textbf{then return} \subnetId}
            \EndFor
            \State{\textbf{return} $\bot$}
        \EndFunction
        \Statex
        \vspace{-2mm}
        \Statex $\triangleright$ $\subnetConf$: a configuration, map $\pkExit[] \rightarrow$ stake, the total value of its seated collaterals
        \Function{\opGetConfiguration{\subnetId}}{} \label{line:conf-begin} \Comment{Returns $\subnetConf_e$ for the current epoch $e$}
            \State{$e \gets \nodeEpoch[\subnetId] ;\; \subnetConf \gets \emptyset$}
            \For{$((\mathsf{tx}, n), (\pkCons[W], \pkExit[W], \colDuration)) \in \nodeCollaterals[\subnetId]$} \label{line:conf-loop} \Comment{$W$: owner}
                \State{$(\amountShort, S) \gets \text{output } n \text{ of } \mathsf{tx} ;\; (\cdot, \pkEots[W]) \gets \nodeValidators[\subnetId][\pkCons[W]]$}
                \If{$e = 0$}
                    \Let{\mathit{isSeated}}{(\mathsf{tx}, n) \text{ is among the first } \nodeParams[\subnetId].\whitelistThreshold \text{ accepted}} \label{line:conf-first}
                    \State{$\mathit{isSeated} \gets \mathit{isSeated}$ \andl $\colDuration > \nodeActHeight[\subnetId] + 3 \cdot \nodeParams[\subnetId].\epochLength$} \label{line:conf-first-timeout}
                    \Let{\subnetConf'}{\nodeConf[\subnetId][0]}
                \Else \Comment{Matures after epoch $e+2$}
                    \Let{\mathit{isSeated}}{(\colDuration > \nodeActHeight[\subnetId] + (e+2) \cdot \nodeParams[\subnetId].\epochLength)} \label{line:conf-cutoff}
                    \Let{\subnetConf'}{\nodeConf[\subnetId][e-1]}
                \EndIf
                \State{$\mathit{isSeated} \gets \mathit{isSeated}$ \andl $S = \opBuildCollateralScript{\pkExit[W], \pkEots[W], \colDuration, \subnetConf'}$} \label{line:conf-locked}
                \If{$(\mathsf{tx}, n)$ is unspent \andl $(\mathsf{tx}, n) \in \nodeSlashable[\subnetId]$ \andl $\mathit{isSeated}$} \label{line:conf-seat}
                    \State{\textbf{if} $\subnetConf[\pkExit[W]] = \bot$ \textbf{then} $\subnetConf[\pkExit[W]] \gets 0$}
                    \Let{\subnetConf[\pkExit[W]]}{\subnetConf[\pkExit[W]] + \amountShort} \label{line:conf-add}
                \EndIf
            \EndFor
            \State{\textbf{return} \subnetConf} \label{line:conf-end}
        \EndFunction
    \end{algorithmic}
    \vspace{1mm}{\footnotesize $^{\ddagger}$ The multisig script $\mathsf{CheckMultiSig}_{\subnetConf}$ of a subnet is placed in a script tree with an unspendable leaf carrying \subnetId, so it differs per subnet even for equal configurations.\par}
\end{pseudocodefloat}

\begin{pseudocodefloat}
    \caption{Protocol \bitcoinIpc: Code for \anIpcAwareNode $V$, part 7: processing checkpoints.}
    \label{alg:node-process}
    \begin{algorithmic}[1]
        \small
        \Function{\opProcessCheckpoint{\checkpointTxOne, \height}}{} \label{line:pcpt-begin} \Comment{\height: Bitcoin height of \checkpointTxOne}
            \Let{\checkpointKeyword \concat \subnetHeight \concat \comBlock}{\text{data in the \opReturn output of } \checkpointTxOne}
            \State{$\subnetId \gets \opSubnetOf{\checkpointTxOne}$ ;\; \textbf{require} $\subnetId \neq \bot$} \label{line:pcpt-subnet}
            \State{\textbf{require} $(\subnetHeight, \cdot, \cdot) \notin \nodeCheckpoints[\subnetId]$} \label{line:pcpt-first} \Comment{One checkpoint per subnet height}
            \State{$\nodeCheckpoints[\subnetId].\mathsf{push}((\subnetHeight, \comBlock, \height))$} \label{line:pcpt-record}
            \LineComment{The first checkpoint activates a subnet}
            \If{$\nodeStatus[\subnetId] = \statusInitialized$} \label{line:pcpt-activate}
                \State{\textbf{require} $|\nodeValidators[\subnetId]| \geq \nodeParams[\subnetId].\whitelistThreshold$}
                \State{$\nodeActHeight[\subnetId] \gets \height$} \label{line:pcpt-act-height}
                \Let{\nodeConf[\subnetId][1]}{\opGetConfiguration{\subnetId}}
                \State{$\nodeStatus[\subnetId] \gets \statusActive ;\; \nodeEpoch[\subnetId] \gets 1$}
            \EndIf
            \Let{e}{\nodeEpoch[\subnetId]}
            \LineComment{Spent collaterals leave; a refreshed one stays under its new outpoint}
            \For{outpoint $(\mathsf{tx}, n)$ spent by \checkpointTxOne\ \textbf{s.t.} $\nodeCollaterals[\subnetId][(\mathsf{tx}, n)] \neq \bot$} \label{line:pcpt-pairs}
                \Let{k}{\text{index of the input spending } (\mathsf{tx}, n) \text{ in } \checkpointTxOne}
                \State{$(\pkCons[], \pkExit[], \colDuration) \gets \nodeCollaterals[\subnetId][(\mathsf{tx}, n)] ;\; \nodeCollaterals[\subnetId][(\mathsf{tx}, n)] \gets \bot$}
                \Let{(\cdot, \pkEots[])}{\nodeValidators[\subnetId][\pkCons[]]}
                \State{$(\amountShort, \cdot) \gets \text{output } n \text{ of } \mathsf{tx} ;\; (\amountShort', S') \gets \text{output } k \text{ of } \checkpointTxOne$}
                \State{$S_e \gets \opBuildCollateralScript{\pkExit[], \pkEots[], \colDuration, \nodeConf[\subnetId][e]}$}
                \State{$S_{e-1} \gets \opBuildCollateralScript{\pkExit[], \pkEots[], \colDuration, \nodeConf[\subnetId][e-1]}$}
                \If{$\amountShort' = \amountShort$ \andl $S' \in \{S_e, S_{e-1}\}$} \Comment{Store refreshed collateral}
                    \Let{\nodeCollaterals[\subnetId][(\checkpointTxOne, k)]}{(\pkCons[], \pkExit[], \colDuration)} \label{line:pcpt-refresh}
                \EndIf
            \EndFor
            \LineComment{Mark the collaterals for which a pre-signed \slashTx was posted}
            \On{\checkpointTxTwo spending the $\mathsf{dataS}$ output of \checkpointTxOne is finalized} \label{line:prv-begin}
                \Let{\transferData \concat \slashData}{\text{data in the witness of input } 0 \text{ of } \checkpointTxTwo}
                \For{$(\mathit{col}, \Sigma) \in \slashData$ \textbf{s.t.} $\nodeCollaterals[\subnetId][\mathit{col}] \neq \bot$} \label{line:prv-loop}
                    \Let{\slashTx}{\opBuildSlashTx{\mathit{col}}}
                    \Let{\subnetConf}{\text{the configuration in the slash leaf of } \mathit{col}}
                    \If{$\Sigma$ holds valid signatures on \slashTx from more than $2/3$ of \subnetConf} \label{line:prv-slashable}
                        \Let{\nodeSlashable[\subnetId]}{\nodeSlashable[\subnetId] \cup \{\mathit{col}\}}
                    \EndIf
                \EndFor
            \EndOn \label{line:prv-end}
        \EndFunction \label{line:pcpt-end}
    \end{algorithmic}
\end{pseudocodefloat}

\begin{pseudocodefloat}
    \caption{Protocol \bitcoinIpc: Code for \anIpcAwareNode $V$, part 8: creating the checkpoint transactions.}
    \label{alg:node-checkpoint}
    \begin{algorithmic}[1]
        \small
        \Statex $\triangleright$ $T$: outgoing transfers $(\subnetId_B, a, \amountShort)$ of value \amountShort to address $a$ on subnet $B$
        \Statex $\triangleright$ $W$: withdrawals $(\pk, \amountShort)$; $f, f'$: fees of \checkpointTxOne and \checkpointTxTwo
        \Statex $\triangleright$ $P$: owner-signed collaterals $((\mathsf{tx}, n), \mathit{action}, \sigma)$, $\mathit{action} \in \{\mathsf{refresh}, \mathsf{exit}\}$
        \Statex $\triangleright$ $\Sigma$: pre-signatures $(\mathit{col}, \{\sigma\})$ on the \slashTx of newly created collateral UTXOs
        \Function{\opCreateCheckpointTxs{\subnetId, \subnetHeight, \comBlock, T, W, P, \Sigma, f, f'}}{} \label{line:cpt-begin}
            \Let{\mathit{currentEpoch}}{\nodeEpoch[\subnetId]}
            \Let{\mathit{isRotation}}{\text{no checkpoint of } \subnetId \text{ finalized in } \mathit{currentEpoch}} \label{line:cpt-rot}
            \State{\textbf{if} $\mathit{isRotation}$ \textbf{then} $e \gets \mathit{currentEpoch} - 1$ \textbf{else} $e \gets \mathit{currentEpoch}$}
            \Let{\subnetConf_{\mathit{next}}}{\nodeConf[\subnetId][\mathit{currentEpoch}]} \Comment{Receives the change}
            \State{$\subnetConf_e \gets \nodeConf[\subnetId][e] ;\; \subnetConf_{e-1} \gets \nodeConf[\subnetId][e-1]$} \label{line:cpt-conf}
            \State{$I \gets [\,] ;\; O \gets [\,]$} \Comment{Inputs and outputs of \checkpointTxOne, ordered as in \Cref{fig:tx-checkpoint}}
            \LineComment{Process the refreshes and exits of collaterals}
            \For{$((\mathsf{tx}, n), \mathit{action}, \sigma) \in P$} \label{line:cpt-pairs-begin}
                \State{$(\amountShort, S) \gets \text{output } n \text{ of } \mathsf{tx} ;\; (\pkCons[], \pkExit[], \colDuration) \gets \nodeCollaterals[\subnetId][(\mathsf{tx}, n)]$}
                \Let{(\cdot, \pkEots[])}{\nodeValidators[\subnetId][\pkCons[]]}
                \Let{\subnetConf_{\mathit{col}}}{\text{the } \subnetConf \in \{\subnetConf_e, \subnetConf_{e-1}\} \textbf{ s.t. } S = \opBuildCollateralScript{\pkExit[], \pkEots[], \colDuration, \subnetConf}} \label{line:cpt-pair-locked}
                \State{\textbf{if} $\subnetConf_{\mathit{col}} = \bot$ \textbf{then continue}} \Comment{Locked under an older configuration}
                \State{\textbf{if} $V$ saw $\pkCons[]$ equivocate \textbf{then continue}} \label{line:cpt-no-equivocator} 
                \State{\textbf{require} pre-signed \slashTx for each pair output} \label{line:cpt-presig}
                \If{$\mathit{action} = \mathsf{refresh}$}
                    \Let{S_{\mathit{new}}}{\opBuildCollateralScript{\pkExit[], \pkEots[], \colDuration, \subnetConf_{\mathit{next}}}} \label{line:cpt-refresh}
                \Else
                    \Let{S_{\mathit{new}}}{\opBuildUnbondingScript{\pkExit[], \pkEots[], \nodeParams[\subnetId].\epochLength, \subnetConf_e}} \label{line:cpt-exit}
                \EndIf
                \State{$I.\mathsf{push}(((\mathsf{tx}, n), S) \text{ with witness } \sigma) ;\; O.\mathsf{push}((\amountShort, S_{\mathit{new}}))$} \Comment{Equal index}
            \EndFor \label{line:cpt-pairs-end}
            \LineComment{Process the transfers and the pre-signatures}
            \For{subnet $B$ \textbf{s.t.} $T_B = \{(a, \amountShort) : (\subnetId_B, a, \amountShort) \in T\} \neq \emptyset$} \label{line:cpt-transfers}
                \State{$O.\mathsf{push}((\sum_{(a, \amountShort) \in T_B} \amountShort,\; \mathsf{CheckMultiSig}_{\nodeConf[\subnetId_B][\nodeEpoch[\subnetId_B]]}))$}
            \EndFor
            \State{$\transferData \gets \transferKeyword \concat (\subnetId_B, T_B)_B ;\; \slashData \gets \slashKeyword \concat \Sigma$} \label{line:cpt-data}
            \If{$T \neq \emptyset$ \textbf{or} $\Sigma \neq \emptyset$}
                \State{$D \gets \transferData \concat \slashData ;\; j \gets |O| ;\; O.\mathsf{push}((f', \mathsf{dataS}(D)))$} \Comment{Spent by \checkpointTxTwo}
            \EndIf
            \LineComment{Process the withdrawals}
            \State{\textbf{for} $(\pk, \amountShort) \in W$ \textbf{do} $O.\mathsf{push}((\amountShort, \mathsf{CheckSig}_{\pk}))$} \label{line:cpt-withdrawals}
            \LineComment{Write the block commitment}
            \State{$O.\mathsf{push}((0, \opReturn(\checkpointKeyword \concat \subnetHeight \concat \comBlock)))$} \label{line:cpt-anchor}
            \Let{\amountShort_{\mathit{out}}}{\text{total value of the transfer, commitment, and withdrawal outputs in } O}
            \LineComment{Fund the outputs and the fees from the subnet's UTXOs}
            \Let{U}{\opUnspent{\mathsf{CheckMultiSig}_{\subnetConf_e}}} \label{line:cpt-funds}
            \If{$\mathit{isRotation}$}
                \Let{U}{U \cup \opUnspent{\mathsf{CheckMultiSig}_{\subnetConf_{e-1}}}} \Comment{Include delayed arrivals too}
            \Else
                \Let{U}{\text{a subset of } U \text{ \textbf{s.t.} its total value is at least } \amountShort_{\mathit{out}} + f}
            \EndIf
            \State{\textbf{for} $(\mathsf{tx}, n) \in U$ \textbf{do} $I.\mathsf{push}(((\mathsf{tx}, n), S))$}, \Comment{$S$: the script of output $n$ of \textsf{tx}}
            \State{$\amountShort_{\mathit{change}} \gets \text{total value of } U - \amountShort_{\mathit{out}} - f ;\; O.\mathsf{push}((\amountShort_{\mathit{change}}, \mathsf{CheckMultiSig}_{\subnetConf_{\mathit{next}}}))$} \label{line:cpt-change-value}\label{line:cpt-change}
            \Let{\checkpointTxOne}{(I, O)} \label{line:cpt-txs}
            \State{\textbf{if} $T = \emptyset$ \andl $\Sigma = \emptyset$ \textbf{then return} \checkpointTxOne} \Comment{No data to post}
            \Let{\checkpointTxTwo}{[\text{inputs:}((\checkpointTxOne, j), \mathsf{dataS}(D)), \text{  witness: } D]}
            \State{\textbf{return} $(\checkpointTxOne, \checkpointTxTwo)$} \label{line:cpt-end}
        \EndFunction
    \end{algorithmic}
\end{pseudocodefloat}

\FloatBarrier

\section{Model and definitions}\label{sec:model-full}
This section specifies the assumptions we make on Bitcoin and subnets and formalizes what it means for a subnet to inherit security from Bitcoin.
We also formulate, for our setting, and prove the \emph{firewall property}~\cite{DBLP:conf/sp/GaziKZ19, DBLP:conf/icdcs/RochaKSV22}, which ensures that compromised subnets do not affect the security of other subnets. In particular, since subnets may transfer \wBTC among each other, this property ensures that a malicious subnet cannot send out more \wBTC than what has been sent into it.

\subsection{Cryptographic primitives}\label{sec:crypto-primitives}
We use the following standard primitives, which we describe informally. We denote by $\lambda$ the security parameter, and an adversary is a probabilistic polynomial-time algorithm.

\begin{definition}[Cryptographic hash function~\cite{rogaway2004cryptographic}]\label{def:hash}
A cryptographic hash function $H: \{0,1\}^* \rightarrow \{0,1\}^\lambda$ satisfies the following properties:
\begin{enumerate}
    \item \textbf{Preimage resistance}: Given $y = H(x)$ for a random $x$, no adversary finds $x'$ with $H(x') = y$, except with negligible probability.
    \item \textbf{Collision resistance}: No adversary finds $x \neq x'$ with $H(x) = H(x')$, except with negligible probability.
\end{enumerate}
\end{definition}
Bitcoin transaction identifiers are computed with such a function. In the security analysis of the signature schemes below, $H$ is modeled as a random oracle.

\begin{definition}[Merkle tree~\cite{merkle1989certified}]\label{def:merkle}
A Merkle tree over a sequence $(x_1, \ldots, x_n)$ is a binary tree whose leaves are the hashes of $x_1, \ldots, x_n$ and whose inner nodes are the hashes of their two children, computed with $H$. Its root is $\rho = \opMerkleRoot{x_1, \ldots, x_n}$. A Merkle proof for position $i$ consists of the $\lceil \log_2 n \rceil$ siblings on the path from the $i$-th leaf to the root, and verifies that $x_i$ is the $i$-th element of the sequence committed by $\rho$. It satisfies the following properties:
\begin{enumerate}
    \item \textbf{Correctness}: A proof computed for $x_i$ from the tree verifies against $\rho$.
    \item \textbf{Binding}: No adversary finds a root $\rho$, a position $i$, and Merkle proofs that verify against $\rho$ for two values $x_i \neq x_i'$ at position $i$, except with negligible probability. This follows from the collision resistance property of $H$.
\end{enumerate}
\end{definition}

\begin{definition}[Schnorr signatures~\cite{Schnorr91}]\label{def:schnorr}
Bitcoin uses Schnorr signatures over the group of the secp256k1 curve, of prime order $q$ and with generator $G$\cite{bip340,bip341}. A key pair consists of a secret key $\sk \in \{1, \ldots, q-1\}$ and a public key $\pk = \sk \cdot G$. Function $\opSign{\sk, m}$ returns a signature $\sigma$ on a message $m$, and $\opVerify{\pk, m, \sigma}$ returns whether $\sigma$ is a valid signature on $m$ under $\pk$. It satisfies the following properties:
\begin{enumerate}
    \item \textbf{Correctness}: $\opVerify{\pk, m, \opSign{\sk, m}}$ returns true.
    \item \textbf{Unforgeability}: No adversary that does not know $\sk$ produces a valid signature under $\pk$ on a message that was not signed with $\sk$, except with negligible probability, even if it obtains signatures on messages of its choice. This holds under the discrete-logarithm assumption, with $H$ modeled as a random oracle.
\end{enumerate}
\end{definition}

\begin{definition}[Extractable one-time signatures~\cite{DBLP:journals/corr/abs-2408-01896}]\label{def:eots}
An extractable one-time signature (EOTS) scheme is a Schnorr signature scheme in which the nonce of every signature is fixed in advance. Its functions are given in \Cref{alg:node-stake}. Function $\opInitEotsKeys{}$ returns a key pair $(\skEots, \pkEots)$. For every subnet height $h$, $\opGenEotsNonce{}$ returns a secret nonce $\secNonce{h}$ and a public nonce $\nonce{h} = \secNonce{h} \cdot G$, and the public nonces of a range of heights are committed in advance by their Merkle root. Function $\opEotsSign{\skEots, \secNonce{h}, m}$ returns a signature $s$ on message $m$, $\opEotsVerify{\pkEots, \nonce{h}, m, s}$ returns whether $s$ is valid for $m$ under $\pkEots$ and $\nonce{h}$, and $\opEotsExtract{\pkEots, \nonce{h}, (m, s), (m', s')}$ returns a secret key. In \bitcoinIpc, a validator signs message $m = (\subnetId, h, b)$ for block $b$ at height $h$ of subnet \subnetId. The scheme satisfies the following properties:
\begin{enumerate}
    \item \textbf{Correctness}: The following returns true: $\opEotsVerify{\pkEots, \nonce{h}, m, \opEotsSign{\skEots, \secNonce{h}, m}}$.
    \item \textbf{Unforgeability}: If every secret nonce is sampled freshly, kept secret, and used for at most one message, no adversary that does not know $\skEots$ produces a valid signature under $\pkEots$ and a public nonce $\nonce{h}$ on a message that was not signed with $\skEots$ and $\secNonce{h}$, except with negligible probability.
    \item \textbf{Extractability}: If $s$ and $s'$ are valid signatures under $\pkEots$ and the same public nonce $\nonce{h}$, on two messages $m \neq m'$, then function $\opEotsExtract{\pkEots, \nonce{h}, (m, s), (m', s')}$ returns $\skEots$, except with negligible probability.
\end{enumerate}
\end{definition}

\subsection{Bitcoin ledger model}\label{sec:bitcoin-model}
We do not model Bitcoin itself. Instead, we state the properties that \bitcoinIpc requires from it. These properties are established for the Bitcoin backbone protocol~\cite{DBLP:conf/crypto/GarayKL17}, under the assumption that honest parties control the majority of the hashing power. Time is measured in Bitcoin blocks.
\begin{enumerate}
    \item \textbf{Common Prefix}, with parameter \depth: If a block has at least \depth blocks on top of it in the chain of an honest party, then it remains in the chain of every honest party from then on. We call such a block, and the transactions it contains, \emph{finalized}.
    \item \textbf{Chain Growth}: The chain of every honest party keeps growing; in particular, a block at every height is eventually finalized.
    \item \textbf{Liveness}, with parameter \inclDelay: If an honest party submits a valid transaction that pays a sufficient fee when the chain of honest parties has height \height, then by height $\height + \inclDelay$ either this transaction or a transaction that spends one of the same outputs is included in a block. In particular, if no other transaction can spend these outputs, the transaction is included within \inclDelay blocks and finalized within $\inclDelay + \depth$ blocks. A fee is sufficient if it leads to inclusion within \inclDelay blocks.
\end{enumerate}

\subsection{Anchoring Protocol}\label{sec:state-anchoring}
A \bitcoinIpc subnet periodically anchors its latest block on Bitcoin.
We model this as a function $\computeStateCom{\subnetId; \subnetHeight} \rightarrow (\subnetHeight, \comBlock, \{\sigma_V\}^{> 2/3}_{V \in \subnetConf})$, which is parametrized by a subnet identifier \subnetId and gets as input a subnet block height \subnetHeight. It returns that height \subnetHeight, a commitment \comBlock to the block of the subnet finalized at height \subnetHeight, and signatures on $(\subnetHeight, \comBlock)$ from validators of the current configuration \subnetConf of the subnet that collectively hold more than $2/3$ of its stake. The exact format of the commitment can be specified by the subnet. Function $\computeStateCom$ is run by the validators every time the subnet creates a checkpoint.
The block commitment \comBlock is written on the Bitcoin transaction \checkpointTxOne, and the signatures form the witness of \checkpointTxOne. We model this as an \emph{anchoring protocol}.

\begin{definition}[Anchoring protocol]\label{def:state-anchoring}
    An anchoring protocol of a subnet $S$ is run by the validators of $S$ and by \ipcAwareNodes. It exposes an operation $\retrieveStateCom{\subnetId}$, which gets the identifier \subnetId of $S$ and returns an array of tuples $[(\subnetHeight_1, {\comBlock}_1), \ldots, (\subnetHeight_m, {\comBlock}_m)]$, for some $m \geq 0$. The tuples are ordered by height in ascending order.
    It is \emph{secure} if it satisfies the following properties:
    \begin{itemize}
        \item \textbf{Ever-growing (liveness)}: New tuples $(\subnetHeight, \comBlock)$ are eventually appended to the output of $\retrieveStateCom{\subnetId}$.
        \item \textbf{Append-only (safety)}:
        If array  $A = [(\subnetHeight_1, {\comBlock}_1), \ldots, (\subnetHeight_m, {\comBlock}_m)]$ is returned by a call to $\retrieveStateCom{}$, then any subsequent output of $\retrieveStateCom{}$ will contain $A$ as a prefix.
        \item \textbf{Binding (safety)}: Calls to $\retrieveStateCom{}$ do not return tuples $(\subnetHeight, \comBlock)$ and $(\subnetHeight, {\comBlock}')$, such that $\comBlock \neq {\comBlock}'$.
        \item \textbf{No forging (safety)}: If tuple $(\subnetHeight, \comBlock)$ is included in the output of $\retrieveStateCom{\subnetId}$, then $(\subnetHeight, \comBlock, \cdot) = \computeStateCom{\subnetId; \subnetHeight}$.
        \item \textbf{Anchored finality (safety)}: If tuple $(\subnetHeight, \comBlock)$ is included in the output of $\retrieveStateCom{\subnetId}$, then no honest party accepts a chain of $S$ whose block at height \subnetHeight is not the one committed by \comBlock. This holds even if every validator of $S$ that has withdrawn its collateral becomes Byzantine.
    \end{itemize}
\end{definition}

The \emph{ever-growing} property ensures that a subnet can always anchor its blocks on Bitcoin.
The \emph{append-only} and \emph{binding} properties ensure that, once a block has been anchored, it cannot be removed or modified, even if that subnet is later compromised.
The \emph{no forging} property ensures that an anchored commitment is a commitment to the subnet's actual block. In other words, an adversary cannot forge a commitment.
The \emph{anchored finality} property captures protection against long-range attacks (LRA)~\cite{DBLP:conf/consensusday/AzouviV22}. Anchored finality states that no honest party accepts such a chain if it conflicts with an anchored block, even if every validator that has withdrawn its collateral is Byzantine.
Notice that the anchoring protocol only involves commitments to blocks, while the blocks and the state of the subnet are abstracted away.
The protocol does not detect a compromise of a subnet on-chain, and does not revert a compromised subnet. Detecting a compromise and any recovery from the anchored history happen outside the protocol, e.g., through governance.

\subsection{Bridge protocols}\label{sec:bridge-protocols}
\label{sec:bridge-protocols-for-transfer}

Informally, a \emph{bridge protocol} runs between two chains, a \emph{source} chain and a \emph{destination} chain, and lets a user move value from the source to the destination. The user invokes a bridge function which locks or burns the amount on the source chain; in response, the protocol credits the equivalent amount to the user on the destination chain. A bridge protocol is correct if it neither creates nor destroys value: every credit on the destination chain is backed by a matching debit on the source chain (\emph{safety}), and every transfer honestly initiated on the source chain eventually completes on the destination chain (\emph{liveness}).

In \bitcoinIpc, the \depositCommand operation is an instance of a bridge protocol with Bitcoin as the source chain and a subnet as the destination chain; the \withdrawCommand operation is the instance with the roles reversed. Similarly, a cross-subnet \transferCommand is a bridge protocol between a source subnet and a destination subnet. Every such bridge is operated by the validators of the subnets involved and by no other party, hence it requires trust in no additional entity.

Traditional bridge definitions~\cite{DBLP:conf/fc/ZamyatinAZKMKK21} assume that both chains are secure, and guarantee nothing otherwise: a compromised source chain can fake bridge transactions and have arbitrary amounts minted on the destination chain. Ga\v{z}i et al.~\cite{DBLP:conf/sp/GaziKZ19} address this for pegged ledgers with the \emph{firewall} property, which gives guarantees to a ledger whose security assumption holds even when other ledgers are compromised; Rocha et al.~\cite{DBLP:conf/icdcs/RochaKSV22} adopt it for hierarchical subnets. We follow this approach and formulate it for our setting. Subnets are permissionless and dynamic, their configurations change every epoch, and any subnet may exceed its corruption bound at any time, so every definition states what a subnet obtains when it is $1/3$-bounded, regardless of subnets.

In the definitions below we make use of three kinds of balances: the \emph{Bitcoin balance} of a subnet $S$, which represents the amount of \BTC attributed to $S$ on Bitcoin, and the \emph{user balance in subnet}, which represents the amount of \wBTC held by the user on the subnet, and the \emph{Bitcoin balance} of a user $u$, which represents the amount of \BTC held by $u$ on Bitcoin.

The \bitcoinIpc protocol allows users to lock \BTC on the Bitcoin network and obtain the same amount of \wBTC in a subnet (an operation we call \emph{deposit}), and vice-versa to burn \wBTC in a subnet and unlock the same amount of \BTC on Bitcoin (an operation we call \emph{withdrawal}).
We first formalize these two protocols.

\begin{definition}[Deposit protocol]\label{def:deposit-bitcoin}
    A deposit protocol of a subnet $S$ is run by the validators of $S$ and exposes an operation $\depositCommand(\amountShort)$ to its users. It is \emph{secure} if it satisfies the following properties:
    \begin{itemize}
        \item \textbf{Liveness}: If an honest user $u$ calls $\depositCommand(\amountShort)$, for some $\amountShort \in \CR$, then the protocol pays \amountShort \wBTC to $u$ on $S$.
        \item \textbf{Safety}: \begin{enumerate*}[label=(\arabic*)]
            \item If the protocol pays \amountShort \wBTC to a user $u$ on $S$, for $\amountShort \in \CR$, then it decreases the Bitcoin balance of $u$ by \amountShort \BTC.
            \item If the protocol pays \amountShort \wBTC to a user $u$ on $S$, for $\amountShort \in \CR$, then it increases the Bitcoin balance of $S$ by \amountShort \BTC.
        \end{enumerate*}
    \end{itemize}
\end{definition}

\begin{definition}[Withdrawal protocol]\label{def:withdrawal-bitcoin}
    A withdrawal protocol of a subnet $S$ is run by the validators of $S$ and exposes an operation $\withdrawCommand(\amountShort)$ to its users. It is \emph{secure} if it satisfies the following properties:
    \begin{itemize}
        \item \textbf{Liveness}: If an honest user $u$ holding at least \amountShort \wBTC on $S$ calls $\withdrawCommand(\amountShort)$, for some $\amountShort \in \CR$, then the protocol pays \amountShort \BTC to $u$ on Bitcoin.
        \item \textbf{Safety}: \begin{enumerate*}[label=(\arabic*)]
            \item If the protocol pays \amountShort \BTC to a user $u$ on Bitcoin, for $\amountShort \in \CR$, then it decreases the balance of $u$ on $S$ by \amountShort \wBTC.
            \item If the protocol decreases the Bitcoin balance of $S$ by \amountShort \BTC, for $\amountShort \in \CR$, then it decreases the balance of some user $u$ on $S$ by \amountShort \wBTC.
        \end{enumerate*}
        \item \textbf{Firewall}: \begin{enumerate*}[label=(\arabic*)]
            \item If the protocol pays \amountShort \BTC to a user $u$ on Bitcoin, for $\amountShort \in \CR$, then it decreases the Bitcoin balance of $S$ by \amountShort \BTC.
            \item The Bitcoin balance of $S$ is always non-negative.
        \end{enumerate*}
    \end{itemize}
\end{definition}

Finally, we define a protocol for transferring \wBTC between two subnets.
Its properties cover four cases. If both subnets are $1/3$-bounded, safety and liveness guarantee \wBTC conservation: user $u_2$ is credited on $B$ exactly what user $u_1$ is debited on $A$. If only $B$ is $1/3$-bounded, the \emph{Firewall~(3)} property guarantees to its users that every \wBTC minted on $B$ is backed by \BTC in the Bitcoin balance of $B$, and the \emph{Firewall~(1)} and \emph{(2)} properties guarantee that this \BTC came out of $A$, whose Bitcoin balance cannot go negative; together these imply that a compromised $A$ cannot inflate $B$ with more \wBTC it ever received. If only $A$ is $1/3$-bounded, the \emph{Firewall~(4)} property guarantees to its users that \BTC leaves $A$ only against \wBTC burnt on $A$. The \emph{Firewall~(1)} and \emph{(2)} properties hold regardless of $A$ and $B$, as they are enforced by Bitcoin.

\begin{definition}[Cross-subnet transfer protocol]\label{def:bridge-subnet}
    A transfer protocol between a \emph{source} subnet $A$ and a \emph{destination} subnet $B$ is run by the validators of $A$ and $B$ and exposes an operation $\transferCommand(\amountShort)$ to the users of $A$. It is \emph{secure} if it satisfies the following properties:
    \begin{itemize}
        \item \textbf{Liveness}: If an honest user $u_1$ holding at least \amountShort \wBTC on $A$ calls $\transferCommand(\amountShort)$ with recipient $u_2$ on $B$, for $\amountShort \in \CR$, then the protocol pays \amountShort \wBTC to $u_2$ on $B$.
        \item \textbf{Safety}: If the protocol pays \amountShort \wBTC to a user $u_2$ on $B$, for $\amountShort \in \CR$, then some user $u_1$ on $A$ called $\transferCommand(\amountShort)$ with recipient $u_2$, and the protocol decreases the balance of $u_1$ on $A$ by \amountShort \wBTC.
        \item \textbf{Firewall}: \begin{enumerate*}[label=(\arabic*)]
            \item If the protocol increases the Bitcoin balance of $B$ by \amountShort \BTC, for $\amountShort \in \CR$, then it decreases the Bitcoin balance of $A$ by \amountShort \BTC.
            \item The Bitcoin balance of $A$ is always non-negative.
            \item If the protocol pays \amountShort \wBTC to a user $u_2$ on $B$, for $\amountShort \in \CR$, then it increases the Bitcoin balance of $B$ by \amountShort \BTC.
            \item If the protocol increases the Bitcoin balance of $B$ by \amountShort \BTC, for $\amountShort \in \CR$, then it decreases the balance of some user $u_1$ on $A$ by \amountShort \wBTC.
        \end{enumerate*}
    \end{itemize}
\end{definition}

The firewall property in the last two definitions protects the destination subnet from adversarial source subnets: it limits the amount of \BTC that can be withdrawn and the amount of \wBTC that can be transferred through our protocols, preventing a malicious source subnet from sending more \wBTC than it has received.
Note that the firewall is a \emph{per-subnet} bound by design: it protects each subnet's aggregate balance, while the correctness of individual users' balances \emph{within} a subnet is that subnet's own responsibility, guaranteed when the subnet is $1/3$-bounded.
    \subsection{Theorems}\label{sec:theorems}

The following theorems establish the security of the \bitcoinIpc protocol.
The proofs are shown in \Cref{sec:proofs}.

\begin{restatable}[Security of anchoring]{theorem}{thmStateAnchoring}
  \label{thm:state-anchoring}
  Let $\Pi$ be the anchoring protocol of \bitcoinIpc for a subnet $S$, described in \Cref{sec:persist-state,sec:checkpoint}. Regardless of $S$, $\Pi$ satisfies the append-only, binding, and anchored finality properties of \Cref{def:state-anchoring}. If $S$ is $1/3$-bounded, $\Pi$ is secure.
\end{restatable}

\begin{restatable}[Security of \depositCommand operation]{theorem}{thmDeposit}
  \label{thm:deposit}
  Let $\Pi$ be the \depositCommand protocol of \bitcoinIpc for a subnet $S$, described in \Cref{sec:deposit,sec:checkpoint}. If $S$ is $1/3$-bounded and its epoch length satisfies $\epochLength \geq \depth + \inclDelay$, $\Pi$ is a secure deposit protocol (\Cref{def:deposit-bitcoin}).
\end{restatable}

\begin{restatable}[Security of \withdrawCommand operation]{theorem}{thmWithdrawal}
  \label{thm:withdrawal}
 Let $\Pi$ be the \withdrawCommand protocol of \bitcoinIpc for a subnet $S$, described in \Cref{sec:withdraw,sec:checkpoint}. Regardless of $S$, $\Pi$ satisfies the firewall property of \Cref{def:withdrawal-bitcoin}. If $S$ is $1/3$-bounded, $\Pi$ is secure.
\end{restatable}

\begin{restatable}[Security of \transferCommand operation]{theorem}{thmTransfer}
  \label{thm:transfer}
  Let $\Pi$ be the \transferCommand protocol of \bitcoinIpc between a source subnet $A$ and a destination subnet $B$, described in \Cref{sec:transfer,sec:checkpoint}. Regardless of $A$ and $B$, $\Pi$ satisfies the \emph{Firewall~(1)} and \emph{(2)} properties of \Cref{def:bridge-subnet}. If $B$ is $1/3$-bounded, $\Pi$ satisfies the \emph{Firewall~(3)} property. If $A$ is $1/3$-bounded, $\Pi$ satisfies the \emph{Firewall~(4)} property. If $A$ and $B$ are $1/3$-bounded and the epoch length of $B$ satisfies $\epochLength_B \geq \depth + \Delta_{\mathsf{sign}} + \inclDelay$, $\Pi$ is secure. Here $\Delta_{\mathsf{sign}}$ bounds the number of Bitcoin blocks from the finalization of the latest Bitcoin block that a checkpoint of $A$ takes into account until its signed \checkpointTxOne is submitted to Bitcoin.
\end{restatable}
    \section{Proofs for the \bitcoinIpc protocol}\label{sec:proofs}

\iflncs

\else
    We prove the four theorems of \Cref{sec:theorems}.
\fi
Throughout this section we assume that Bitcoin is correct.

\subsection{Invariants of the \createCommand operation}\label{sec:create-invariants}
We first state what \anIpcAwareNode is guaranteed about every created subnet. We say that a node \emph{accepts} a \createCommand when function \opProcessCreate{\createTxTwo} (\Cref{alg:node-create}) stores the parameters of the new subnet.

\begin{lemma}[Create invariants]\label{lem:create}
Assume an honest \ipcAwareNode accepts a \createCommand of a subnet. Then the following hold:
\begin{description}[leftmargin=0pt, font=\normalfont\bfseries]
    \item[C1 (Parameters).] The parameters of the subnet are the ones revealed by a finalized Bitcoin transaction, and they are well-formed.
    \item[C2 (Unique identifier).] No other subnet has the same identifier, and the parameters of the subnet never change.
    \item[C3 (Bootstrap).] The subnet has an initial configuration with enough keys for it to become \activeSubnet.
    \item[C4 (Agreement).] All honest nodes accept the same \createCommand operations, with the same identifiers and parameters, in the same order, and a \createCommand that is not accepted does not change the state of a node.
\end{description}
\end{lemma}

\begin{proof}
We refer to the lines of \Cref{alg:node-create}. A \createCommand is accepted only if every \textbf{require} of \opProcessCreate{\createTxTwo} holds.
\textbf{C1:} The parameters are parsed from the data revealed by \createTxTwo, which is finalized, and function \opValidCreateParams{\subnetParams} requires that they are well-formed: all values are present, the collateral, the number of validators, and the periods are positive, the epoch length can be encoded in a relative timelock of Bitcoin, and $\minTimeout \leq \maxTimeout$.
\textbf{C2:} \Cref{line:create-verify-unique} requires that no subnet with the derived identifier exists, and the parameters of a subnet are written only after this line; hence they are written once. Moreover, the identifier consists of the first $20$ bytes of the txid of \createTxTwo; txid identifiers are unique on Bitcoin (except with negligible probability), as they are the output of a cryptographic hash function, so two correctly created subnets obtain the same identifier only with negligible probability, in which case the second one is not accepted.
\textbf{C3:} \Cref{line:valid-params-bootstrap} requires a non-empty initial configuration, and \cref{line:valid-params-whitelist-size} requires that it has at least \whitelistThreshold keys.
\textbf{C4:} Acceptance depends only on the finalized transaction \createTxTwo (\anIpcAwareNode only parses Bitcoin blocks after 6 additional blocks are mined~\cite{DBLP:conf/crypto/GarayKL17}) and on the state of the node, which results from processing finalized blocks in order; by the \emph{Common Prefix} property of Bitcoin~\cite{DBLP:conf/crypto/GarayKL17}, honest nodes process the same blocks in the same order and hence reach the same decisions. All writes to the state occur after the last \textbf{require}, so a \createCommand that is not accepted writes nothing.
\end{proof}

\subsection{Invariants of the \joinCommand operation}\label{sec:join-invariants}
We next state what \anIpcAwareNode is guaranteed about every \joinCommand it accepts. We say that a node \emph{accepts} a \joinCommand of a validator $W$ in a subnet with identifier \subnetId when function \opProcessJoin{\joinTxTwo} (\Cref{alg:node-join}) adds the entry $\nodeCollaterals[\subnetId][\mathit{col}] = (\pkCons[W], \pkExit[W], \colDuration)$.

\begin{lemma}[Join invariants]\label{lem:join}
Assume an honest \ipcAwareNode accepts a \joinCommand of a validator $W$ in a subnet. Then the following hold:
\begin{description}[leftmargin=0pt, font=\normalfont\bfseries]
    \item[J1 (Collateral).] The collateral of $W$ is a finalized, unspent Bitcoin output that holds at least the minimum collateral of the subnet and is locked under the collateral script tree of \Cref{tab:collateral-script}, built from the keys and timeout that $W$ declared and from the current or the previous configuration of the subnet.
    \item[J2 (Timeout).] The timeout of the collateral gives $W$ a seat for at least $\minTimeout$ and at most $\maxTimeout$ epochs, the bounds declared by the subnet. For a \joinCommand accepted before the subnet is active, this holds assuming that the subnet activates at the height of the \joinCommand.
    \item[J3 (Admission).] The \joinCommand is allowed by the \permissionPolicy of the subnet; if the subnet is not yet active, the Bitcoin key of $W$ is in its initial configuration; and the subnet is not being killed.
    \item[J4 (Authentic).] All values declared in the \joinCommand were chosen by the holder of the declared EOTS key; no other party can join with the EOTS key or the nonce commitments of $W$; in particular, $(\nonceCom, \nonceComSig)$ of $W$ cannot be replayed in another \joinCommand, in any subnet.
    \item[J5 (One key per validator).] The consensus key of $W$ joins the subnet at most once, and its EOTS key is registered for no other consensus key, in any subnet.
    \item[J6 (Unique nonces).] Every EOTS key commits to at most one nonce per subnet height.
    \item[J7 (Agreement).] All honest nodes accept the same \joinCommand operations, in the same order, and a \joinCommand that is not accepted does not change the state of a node.
\end{description}
\end{lemma}

\begin{proof}
We refer to the lines of \opProcessJoin{\joinTxTwo} in \Cref{alg:node-join}. A \joinCommand is accepted only in \cref{line:join-verify-pending}, which is reached only if every \textbf{require} before it holds.
\textbf{J1:} \Cref{line:join-verify-script,line:join-verify-script-prev} build the expected script trees from the declared fields, using the current and the previous configuration, and accept the transaction only if the collateral UTXO is locked with one of the two scripts (\cref{line:join-verify-script-match}). \Cref{line:join-verify-output} fixes the collateral as output~$1$ of \joinTxOne, \cref{line:join-verify-parse-output} requires that it is unspent, and function \opValidateJoinParams{\subnetId, \pkCons[W], \pkExit[W], \amountShort, \colDuration} requires that its value \amountShort is at least \minCollateral. Transaction \joinTxOne is finalized because \joinTxTwo, which spends it, is finalized.
\textbf{J2}, \textbf{J3:} \Cref{line:join-verify-params} calls function \opValidateJoinParams{\subnetId, \pkCons[W], \pkExit[W], \amountShort, \colDuration}, which requires J2 in \cref{line:join-verify-timeout} and J3 in lines~\ref{line:join-verify-policy}, \ref{line:join-verify-whitelist}, and~\ref{line:join-verify-status}; for J2, a collateral accepted in epoch $e$ with timeout in epoch $E_{\colDuration}$ has a seat in epochs $e+1, \ldots, E_{\colDuration}-3$ (\Cref{tab:stake-lifecycle}), that is, for $E_{\colDuration} - e - 3$ epochs, which \cref{line:join-verify-timeout} bounds by \minTimeout and \maxTimeout. Before the subnet is active its cutoff heights are not yet known, and $E_{\colDuration}$ is computed assuming that the subnet activates at the current height (\cref{line:join-verify-h0}).
\textbf{J4:} Function \opCheckNonceCom{\pkEots[W], m, \nonceCom, \nonceComSig} requires a signature under the declared EOTS key on $m$ (\cref{line:join-nonce-verify}), which \opProcessJoin{\joinTxTwo} sets to all other declared values; by unforgeability, its holder signed them, and a signature of $W$ does not verify for a \joinCommand that declares any other value. Hence a replay of $(\nonceCom, \nonceComSig)$ must declare the same values, including \subnetId and \pkEots[W], and is rejected by \cref{line:join-nonce-fresh} (see J5).
\textbf{J5:} \Cref{line:join-verify-once} requires that no validator is registered under the consensus key, and \cref{line:join-verify-register} registers it, together with its declared keys, for every accepted \joinCommand. \Cref{line:join-nonce-fresh} requires that no nonce commitment is stored for the declared EOTS key, and every accepted \joinCommand stores one (\cref{line:join-nonce-store}); hence no later \joinCommand with the same EOTS key is accepted, under any consensus key and in any subnet.
\textbf{J6:} Nonce commitments are stored only by function \opCheckNonceCom{\pkEots[W], m, \nonceCom, \nonceComSig}, which stores a range for an EOTS key only if it does not overlap any range already stored for that key. Hence every subnet height lies in at most one committed range of the key, and a range commits to one nonce per height, as its Merkle root \nonceRoot opens to at most one nonce per height by the collision resistance property of the hash function.
\textbf{J7:} Acceptance depends only on finalized Bitcoin transactions and on the state of the node, which itself results from processing finalized blocks in order (\Cref{alg:node-create}); by the Common Prefix property of Bitcoin~\cite{DBLP:conf/crypto/GarayKL17}, honest nodes process the same blocks in the same order and hence reach the same decisions. All writes to the state occur after the last \textbf{require}, so a \joinCommand that is not accepted writes nothing.
\end{proof}

\subsection{Invariants of computing the configuration}\label{sec:conf-invariants}
We next state what \anIpcAwareNode is guaranteed about the configuration it determines for an epoch.

\begin{lemma}[Configuration invariants]\label{lem:conf}
Assume an honest \ipcAwareNode determines the configuration $\subnetConf_e$ of a subnet $S$ with identifier \subnetId for an epoch $e$, through function \opGetConfiguration{\subnetId}. Then the following hold:
\begin{description}[leftmargin=0pt, font=\normalfont\bfseries]
    \item[G1 (Backed power).] The power of every Bitcoin key in $\subnetConf_e$ equals the total value of the collaterals of that key that are unspent at the cutoff, slashable, locked under $\subnetConf_{e-1}$, and mature after epoch $e+2$ (for $\subnetConf_1$: that are also among the first \whitelistThreshold accepted). Each of these collaterals has a slash leaf under the EOTS key registered for the consensus key of its owner, and that EOTS key is registered for no other consensus key.
\end{description}
\end{lemma}

\begin{proof}
We refer to the lines of \Cref{alg:node-config}.
\textbf{G1:} \Cref{line:conf-add} adds to the power of a Bitcoin key exactly the value of each of its entries whose collateral is unspent, slashable, locked under $\subnetConf_{e-1}$ (\cref{line:conf-locked}), and whose timeout exceeds the cutoff $\height_{e+2}$ of epoch $e+2$ (lines~\ref{line:conf-first} to~\ref{line:conf-seat}; for $\subnetConf_1$, also among the first \whitelistThreshold accepted). Every entry was added in one of three ways: by an accepted \joinCommand\ (\cref{line:join-verify-pending}), by an accepted \stakeCommand\ (\cref{line:stake-verify-register}), or by a refresh in a checkpoint (\cref{line:pcpt-refresh}). In the first case, by J1 of \Cref{lem:join}, its collateral is locked under the expected script tree, whose slash leaf carries the declared EOTS key (\cref{line:build-cols-eots} requires $\pkEots[W] \neq \bot$), and \cref{line:join-verify-register} registers this key for the consensus key of $W$; by J5, that key is registered for no other consensus key. In the other two cases, the new collateral is required to be locked under the collateral script tree built from the keys registered for the consensus key of its owner (\cref{line:stake-verify-script} and \cref{line:pcpt-refresh}), so its slash leaf carries the same registered EOTS key.
\end{proof}

\begin{lemma}[Unilateral exit]\label{lem:unilateral-exit}
Let $V$ be an honest validator with keys $\pkExit[V]$ and $\pkEots[V]$, and $\mathit{col}$ a collateral UTXO of $V$ with timeout \colDuration. UTXO $\mathit{col}$, and any UTXO created from it by a refresh or a cooperative exit, can be spent only with a signature of $V$, and $V$ can spend it alone through its unilateral-exit leaf once height \colDuration is finalized, or through its unilateral-unbonding leaf $3\epochLength$ blocks after it is created.
\end{lemma}

\begin{proof}
A collateral UTXO is locked under $\colS(\pkExit[V], \pkEots[V], \colDuration, \subnetConf)$ and an unbonding UTXO under $\unbS(\pkExit[V], \pkEots[V], \epochLength, \subnetConf)$, both with an unspendable internal key. Their leaves are $\mathsf{CheckSig}_{\pkExit[V]} \wedge \mathsf{AbsTimelock}(\colDuration)$ or $\mathsf{CheckSig}_{\pkExit[V]} \wedge \mathsf{RelTimelock}(3\epochLength)$, $\mathsf{CheckSig}_{\pkEots[V]} \wedge \mathsf{CheckMultiSig}_{\subnetConf}$, and $\mathsf{CheckSig}_{\pkExit[V]} \wedge \mathsf{CheckMultiSig}_{\subnetConf}$. Every leaf requires a signature under $\pkExit[V]$ or $\pkEots[V]$, so no configuration, compromised or not, spends the UTXO alone.

\emph{EOTS key.} The honest $V$ signs at most one block per subnet height, as its EOTS signature is a vote in the last round of consensus, on the block already decided at that height; its secret nonces are sampled freshly and kept secret, with the nonce committed for that height, and by J6 every height has at most one committed nonce. Hence no nonce is used for two messages, the secret EOTS key of $V$ cannot be extracted, and, by the unforgeability of Schnorr signatures, no other party produces a signature under $\pkEots[V]$. The slash leaf is therefore never satisfied.

\emph{Exit key.} The honest $V$ signs with the secret key of $\pkExit[V]$ only its own spends through the unilateral-exit or unilateral-unbonding leaf, and the pairs of checkpoints, with \sighashSingleACP. The output of such a pair has the same value and is locked under $\colS$ with the same keys and timeout \colDuration (a refresh) or under $\unbS$ with the same keys (a cooperative exit). As \sighashSingleACP binds the signed input to the output at the same index, any transaction that uses this signature creates that output. By induction over these spends, every UTXO created from $\mathit{col}$ is locked under $\colS$ or $\unbS$ of $V$, and the claim follows. Since Bitcoin is correct, a spend by $V$ through the unilateral-exit or unilateral-unbonding leaf is eventually finalized.
\end{proof}

This establishes property \textbf{P3: Unilateral exit} from the main body.

We call a UTXO a \emph{descendant} of a collateral UTXO $\mathit{col}$ if it is $\mathit{col}$, or if a checkpoint creates it by spending a descendant of $\mathit{col}$ through its \coopExit leaf.

\begin{lemma}[Slashing]\label{lem:slashing}
Let $W$ be a validator seated in configuration $\subnetConf_e$ of a subnet $S$ through a collateral UTXO $\mathit{col}$ with timeout \colDuration, and let $\height_e$ be the cutoff height that ends epoch $e$. Let \adoptDelay bound the number of Bitcoin blocks from the finalization of the block at height $\height_e$ until $\subnetConf_e$ stops finalizing blocks of $S$. Let \obsDelay bound the number of Bitcoin blocks until every honest validator of $S$ observes an event, namely two conflicting blocks signed by $W$ or a transaction that spends a descendant of $\mathit{col}$, and one of them submits a \slashTx in response. Let $\slashDelay = 2(\obsDelay + \inclDelay) + \depth$, and assume that
\textbf{(a)} the epoch length of $S$ satisfies $\epochLength \geq \depth + \adoptDelay + \slashDelay$, and
\textbf{(b)} the Byzantine validators hold at most $2/3$ of the stake of each of $\subnetConf_{e-1}$, $\subnetConf_e$, and $\subnetConf_{e+1}$.
If $\subnetConf_e$ finalizes two conflicting blocks of $S$ at the same subnet height, both signed by $W$, the second at Bitcoin height $t_0$, then, except with negligible probability, a \slashTx that burns a descendant of $\mathit{col}$ is finalized by height $t_0 + \slashDelay$, and no descendant of $\mathit{col}$ is spent through its unilateral-exit or unilateral-unbonding leaf.
\end{lemma}

\begin{proof}
\emph{Timing.} The block at height $\height_e$ is finalized at height $\height_e + \depth$, so $t_0 \leq \height_e + \depth + \adoptDelay$, and by (a), $t_0 + \slashDelay \leq \height_e + \epochLength$. As $\mathit{col}$ is seated in $\subnetConf_e$, its timeout satisfies $\colDuration > \height_e + 2\epochLength$ (by the seating condition of \opGetConfiguration{\subnetId}, which for $e = 1$ reads $\colDuration > \nodeActHeight[\subnetId] + 3\epochLength$). Honest validators build every checkpoint, and each requires their signature (see below): a refresh keeps the keys and the timeout of the collateral, and a cooperative exit creates an unbonding UTXO with the same keys. Hence, the unilateral-exit leaf of a descendant can be spent only at a height greater than $\colDuration$, and the unilateral-unbonding leaf only $3\epochLength$ blocks after the unbonding UTXO is created, which happens after epoch $e$ starts, at height $\height_e - \epochLength$. So no unilateral leaf of a descendant can be spent up to height $\height_e + \epochLength$.

\emph{Key extraction.} $W$ signed both blocks at the same height $h$ with its EOTS key. Honest validators accept such a signature only with the nonce committed for $h$, which is unique by J6. By the extractability property of EOTS, every honest validator obtains $\skEots[W]$ by height $t_0 + \obsDelay$, except with negligible probability.

\emph{Spending paths.} Let $u$ be a descendant of $\mathit{col}$, locked under the multisig of $\subnetConf_j$. Up to height $\height_e + \epochLength$, we have $j \in \{e-1, e, e+1\}$: $\mathit{col}$ is locked under $\subnetConf_{e-1}$ (G1), and a checkpoint created in epoch $e'$ locks the descendant it creates under a configuration that is not older than the one of the spent descendant and not newer than $\subnetConf_{e'}$ (lines~\ref{line:cpt-refresh} and~\ref{line:cpt-exit}). Before its unilateral leaves can be spent, $u$ can be spent only through its slash leaf or its \coopExit leaf, and both need signatures from more than $2/3$ of the stake of $\subnetConf_j$, so, by (b), the signature of an honest validator. An honest validator signs a spend of the slash leaf only as the pre-signature of the \slashTx of $u$. It signs a spend of the \coopExit leaf only in a checkpoint, with \sighashAll, only after it sees the pre-signatures of the \slashTx of the new descendant, and never after it has observed the equivocation of $W$. Hence, honest validators hold a pre-signed \slashTx for every descendant: for $\mathit{col}$, it is published on Bitcoin, as $\mathit{col}$ is slashable (G1).

\emph{Slashing.} A \slashTx pays a sufficient fee. Let $\tau = t_0 + \obsDelay$. By height $\tau$, every honest validator has observed the equivocation of $W$, and one of them submits the \slashTx of the descendant $u_0$ that is unspent at height $\tau$. By the spending paths, only a checkpoint can spend $u_0$ instead. At most one checkpoint that spends a descendant is included after height $\tau$: a second one would spend the output of the first, which an honest validator includes in a checkpoint only after it records this output at line~\ref{line:pcpt-refresh}, once the first is finalized, hence after $\tau$, and no honest validator signs it. If such a checkpoint spends $u_0$, it is included by height $\tau + \inclDelay$ by the Liveness property of Bitcoin, and within \obsDelay further blocks an honest validator submits the \slashTx of the new descendant, which nothing else can spend. Hence, by the Liveness property of Bitcoin, a \slashTx of a descendant of $\mathit{col}$ is included at a height $x \leq t_0 + 2(\obsDelay + \inclDelay) = t_0 + \slashDelay - \depth$.

At height $x + \depth \leq t_0 + \slashDelay \leq \height_e + \epochLength$, this \slashTx is finalized, before any unilateral leaf of a descendant can be spent. By the Common Prefix property of Bitcoin, it remains in the chain of every honest party, so no descendant of $\mathit{col}$ is spent through a unilateral leaf.
\end{proof}
Under the conditions of \Cref{lem:slashing}, this establishes that equivocation is slashed, as stated in case \textbf{(ii)} of the security model (\Cref{sec:preliminaries}).

\subsection{Proofs of the main theorems}

\thmStateAnchoring*
\begin{proof}
Let $S$ be a \bitcoinIpc subnet with identifier \subnetId. We prove the append-only, binding, and anchored finality properties of \Cref{def:state-anchoring} for arbitrary $S$, and the ever-growing and no forging properties assuming that $S$ is $1/3$-bounded. Recall that a $1/3$-bounded subnet is safe and live, and that its honest validators hold more than $2/3$ of the stake of every configuration $\subnetConf_e$.

\medskip\noindent\textit{Append-only.}
The output of $\retrieveStateCom{\subnetId}$ is assembled from the finalized Bitcoin chain, which is itself append-only.
By the \emph{Common Prefix} property of Bitcoin~\cite{DBLP:conf/crypto/GarayKL17}, once a block is finalized it is present in every honest node's view of the chain.
Therefore, any tuple $(\subnetHeight, \comBlock)$ that has been appended to the output of $\retrieveStateCom{\subnetId}$ will remain there in all future calls, so every subsequent output contains the current array as a prefix.

\medskip\noindent\textit{Binding.}
Suppose for contradiction that two calls to $\retrieveStateCom{\subnetId}$ return tuples $(\subnetHeight, \comBlock)$ and $(\subnetHeight, {\comBlock}')$ with $\comBlock \neq {\comBlock}'$.
Each such tuple corresponds to a distinct finalized \checkpointTxOne on Bitcoin whose \opReturn encodes height \subnetHeight.
By the Common Prefix property of Bitcoin, all honest nodes observe the same finalized chain and therefore agree on which \opReturn output for height \subnetHeight was finalized first.
\AnIpcAwareNode records at most one commitment per height (the first finalized occurrence), so no two distinct commitments for the same height can both appear in the output of $\retrieveStateCom{\subnetId}$, a contradiction.

\medskip\noindent\textit{Anchored finality.}
By \emph{append-only} and \emph{binding}, the anchored tuples of $S$ form a unique history, and by the consensus rule of the subnets, an honest party accepts a chain of $S$ only if its block at every anchored height is the committed one. This holds also if every validator that has withdrawn its collateral becomes Byzantine.

\medskip\noindent\textit{Ever-growing.}
Assume $S$ is $1/3$-bounded.
By the construction of \Cref{sec:persist-state,sec:checkpoint}, $S$ periodically produces a \checkpointTxOne containing an \opReturn output that encodes the \checkpointKeyword, a block height \subnetHeight, and the commitment $\comBlock$ returned by $\computeStateCom{\subnetId; \subnetHeight}$.
Since $S$ is live, new checkpoints are eventually produced, and since the honest validators hold more than $2/3$ of the stake of the current configuration, they sign \checkpointTxOne and submit it to Bitcoin.
By correctness of Bitcoin, these transactions are eventually finalized, at which point every honest \ipcAwareNode detects the \opReturn output and appends the tuple $(\subnetHeight, \comBlock)$ to the output of $\retrieveStateCom{\subnetId}$.
Hence, new tuples are eventually appended.

\medskip\noindent\textit{No forging.}
Assume $S$ is $1/3$-bounded, and suppose tuple $(\subnetHeight, \comBlock)$ is included in the output of $\retrieveStateCom{\subnetId}$.
By the validation rule for checkpoints, this tuple corresponds to a \checkpointTxOne finalized on Bitcoin that spends a UTXO of $S$ locked under the multisig of a configuration $\subnetConf_e$ of $S$; hence \checkpointTxOne carries signatures from validators of $\subnetConf_e$ holding more than $2/3$ of its stake.
Since $S$ is $1/3$-bounded, at least one honest validator signed \checkpointTxOne, and an honest validator signs only if $(\subnetHeight, \comBlock, \cdot) = \computeStateCom{\subnetId; \subnetHeight}$, and since $S$ is safe, the block of $S$ finalized at height \subnetHeight is unique.
Therefore $(\subnetHeight, \comBlock, \cdot) = \computeStateCom{\subnetId; \subnetHeight}$.
Moreover, even if validators in $\subnetConf_e$ become Byzantine after they retrieve their collateral, they cannot forge such a checkpoint: a validator of $\subnetConf_e$ has a timeout in epoch $E_{\colDuration} \geq e+3$ and retrieves its collateral at the earliest in epoch $e+3$, while every UTXO of $S$ locked under $\subnetConf_e$ is spent by the rotation checkpoint at the end of epoch $e+1$, so no UTXO remains for the forged \checkpointTxOne to spend.

\end{proof}
The anchored finality property establishes property \textbf{P2: Long-Range Attack (LRA) protection} from the main body (\Cref{sec:preliminaries}).

\thmDeposit*

\begin{proof}
Let $S$ be a $1/3$-bounded \bitcoinIpc subnet, so that $S$ is safe and live and its honest validators hold more than $2/3$ of the stake of every configuration. We prove each property of \Cref{def:deposit-bitcoin}.

\medskip\noindent\textit{Liveness.}
Assume an honest user $u$ calls $\depositCommand(\amountShort)$.
By the construction of \Cref{sec:deposit}, $u$ submits to Bitcoin a transaction that (i)~creates a UTXO of value $\amountShort$ locked with $S$'s multisig address, and (ii)~attaches an \texttt{\opReturn} output encoding $u$'s address on $S$.
Since $u$ is honest, this transaction is valid, pays a sufficient fee, and locks the UTXO under the latest configuration $\subnetConf_e$ determined by the finalized blocks. Let $t$ be the height of the chain when $u$ submits it; then $\height_{e-1} \leq t - \depth < \height_e$. By the \emph{Liveness} property of Bitcoin, the transaction is contained in a block at a height $\height'$ with
\[ t \;<\; \height' \;\leq\; t + \inclDelay \;<\; \height_e + \depth + \inclDelay \;\leq\; \height_{e+1} . \]
Hence $\height'$ lies in epoch $e$ or $e+1$, and the rotation checkpoint that ends this epoch spends the UTXO. By the \emph{Chain Growth} property of Bitcoin~\cite{DBLP:conf/crypto/GarayKL17}, the block is eventually finalized.
Every honest validator of $S$ is \anIpcAwareNode that continuously scans the finalized Bitcoin chain, and by the \emph{Common Prefix} property of Bitcoin~\cite{DBLP:conf/crypto/GarayKL17}, they all observe the same deposit transaction and extract $u$'s address from the \texttt{\opReturn} output.
Since $S$ is live and its honest validators hold more than $2/3$ of the stake, $S$ eventually produces a checkpoint within $\checkpointInterval$ subnet blocks, at which point the validators process the deposit and mint $\amountShort$~$\wBTC$ to $u$'s address on $S$.
Since $S$ is live, this update is eventually finalized on $S$, so the protocol pays $\amountShort$~$\wBTC$ to $u$ on $S$.

\medskip\noindent\textit{Safety~(1) and~(2).}
Assume the protocol pays $\amountShort$~$\wBTC$ to a user $u$ on $S$.
By the construction of \Cref{sec:deposit,sec:checkpoint}, $\wBTC$ is minted for $u$ only during checkpoint processing, which requires a supermajority (more than $2/3$ of the stake) of $S$'s validators to sign the checkpoint transaction.
Since $S$ is $1/3$-bounded, at least one honest validator signed it.
Each honest validator independently verifies the existence of a deposit transaction on Bitcoin before signing. By the Common Prefix property of Bitcoin, they all observe the same finalized chain, which must contain a transaction creating a UTXO of value $\amountShort$ locked with $S$'s multisig address.
By Bitcoin's UTXO model, the inputs of this transaction are UTXOs of the user who submitted it, and its \opReturn output names $u$ as the recipient; by the construction of \Cref{sec:deposit}, that user is $u$. Hence, the protocol decreases the Bitcoin balance of $u$ by $\amountShort$~$\BTC$ (\emph{Safety~(1)}) and increases the Bitcoin balance of $S$ by $\amountShort$~$\BTC$ (\emph{Safety~(2)}).
\end{proof}

\thmWithdrawal*

\begin{proof}
Let $S$ be a \bitcoinIpc subnet. We prove the firewall property of \Cref{def:withdrawal-bitcoin} for arbitrary $S$, and the liveness and safety properties assuming that $S$ is $1/3$-bounded, hence safe and live, with honest validators holding more than $2/3$ of the stake of every configuration.

\medskip\noindent\textit{Firewall~(1).}
Assume the protocol pays $\amountShort$~$\BTC$ to a user $u$ on Bitcoin.
By construction (\Cref{sec:withdraw,sec:checkpoint}), this corresponds to a UTXO of value $\amountShort$ locked under $\mathsf{CheckSig}_{\pk_u}$ appearing as an output of a \checkpointTxOne.
The \checkpointTxOne spends input UTXOs drawn from $S$'s multisig; by Bitcoin's UTXO model, the sum of outputs (including the $\amountShort$ UTXO to $u$ and any change returned to $S$) cannot exceed the sum of inputs.
Hence, the protocol decreases the Bitcoin balance of $S$ by $\amountShort$~$\BTC$.

\medskip\noindent\textit{Firewall~(2).}
The Bitcoin balance of $S$ is non-negative by Bitcoin's UTXO model: a \checkpointTxOne can only spend UTXOs that $S$ owns, so the balance cannot drop below zero, regardless of whether $S$'s validators are honest. This implies that the subnet cannot withdraw more \BTC than what has been deposited into it.

\medskip\noindent\textit{Liveness.}
Assume $S$ is $1/3$-bounded, and an honest user $u$ holding at least $\amountShort$~$\wBTC$ on $S$ calls $\withdrawCommand(\amountShort)$.
By the construction of \Cref{sec:withdraw}, $u$ submits a withdrawal command on $S$.
Since $u$ is honest and holds at least $\amountShort$~$\wBTC$, this is a valid withdrawal command, and since $S$ is live, it is eventually finalized in $S$'s ledger.
Subnet $S$ eventually produces a checkpoint that processes the withdrawal: the \checkpointTxOne includes an output UTXO of value $\amountShort$ locked under $\mathsf{CheckSig}_{\pk_u}$, spending from $S$'s multisig inputs.
The \checkpointTxOne is signed by the honest validators of $S$, who hold more than $2/3$ of the stake, and submitted to Bitcoin by them.
By security of Bitcoin, the \checkpointTxOne is eventually finalized, so the protocol pays $\amountShort$~$\BTC$ to $u$ on Bitcoin.

\medskip\noindent\textit{Safety~(1).}
Assume $S$ is $1/3$-bounded, and the protocol pays $\amountShort$~$\BTC$ to a user $u$ on Bitcoin.
As in the \emph{Firewall~(1)} property, this corresponds to a UTXO of value $\amountShort$ locked under $\mathsf{CheckSig}_{\pk_u}$ as an output of a \checkpointTxOne, which has been signed by validators holding more than $2/3$ of the stake of $S$; since $S$ is $1/3$-bounded, at least one of them is honest. An honest validator signs only if the withdrawals in \checkpointTxOne are exactly those finalized on $S$'s ledger, which is unique since $S$ is safe.
In particular, the ledger contains a $\withdrawCommand(\amountShort)$ from $u$, whose processing debits $\amountShort$~$\wBTC$ from $u$'s balance on $S$, so the protocol decreases the balance of $u$ on $S$ by $\amountShort$~$\wBTC$.

\medskip\noindent\textit{Safety~(2).}
Assume $S$ is $1/3$-bounded, and the protocol decreases the Bitcoin balance of $S$ by $\amountShort$~$\BTC$.
This decrease corresponds to $S$'s multisig spending $\amountShort$~$\BTC$ in a \checkpointTxOne, which has been constructed from $S$'s finalized ledger state.
The \checkpointTxOne has been signed by validators holding more than $2/3$ of the stake of $S$; since $S$ is $1/3$-bounded, at least one of them is honest. An honest validator signs only if the withdrawals in \checkpointTxOne are exactly those finalized on $S$'s ledger, which is unique since $S$ is safe.
In particular, the ledger contains a $\withdrawCommand(\amountShort)$ from $u$, whose processing debits $\amountShort$~$\wBTC$ from $u$'s balance on $S$, so the protocol decreases the balance of $u$ on $S$ by $\amountShort$~$\wBTC$.
\end{proof}

\thmTransfer*

\begin{proof}
Let $A$ and $B$ be \bitcoinIpc subnets. We prove the \emph{Firewall~(1)} and \emph{(2)} properties of \Cref{def:bridge-subnet} for arbitrary $A$ and $B$, the \emph{Firewall~(3)} property assuming that $B$ is $1/3$-bounded, the \emph{Firewall~(4)} property assuming that $A$ is $1/3$-bounded, and safety and liveness assuming that both are. A $1/3$-bounded subnet is safe and live, and its honest validators hold more than $2/3$ of the stake of every configuration.

\medskip\noindent\textit{Firewall~(1).}
Assume the protocol increases the Bitcoin balance of $B$ by $\amountShort$~$\BTC$.
By construction (\Cref{sec:transfer,sec:checkpoint}), a UTXO locked with $B$'s multisig is an output of $A$'s \checkpointTxOne, with value equal to the sum of all transfers from $A$ to $B$ batched in this checkpoint (at least $\amountShort$~$\BTC$); the \checkpointTxOne also consumes input UTXOs drawn from $A$'s multisig.
By Bitcoin's UTXO model, the total output value (including the UTXO for $B$) cannot exceed the total input value drawn from $A$.
Hence, the protocol decreases the Bitcoin balance of $A$ by $\amountShort$~$\BTC$.

\medskip\noindent\textit{Firewall~(2).}
The Bitcoin balance of $A$ is always non-negative by the same argument as the withdrawal firewall.

\medskip\noindent\textit{Firewall~(3).}
Assume $B$ is $1/3$-bounded, and the protocol pays $\amountShort$~$\wBTC$ to a user $u_2$ on $B$.
By construction (\Cref{sec:transfer,sec:checkpoint}), $\wBTC$ is minted for $u_2$ on $B$ only upon detection of a UTXO locked with $B$'s multisig address on Bitcoin, whose commit-reveal script indicates $u_2$ as the recipient of $\amountShort$~$\wBTC$, during checkpoint processing, which requires validators holding more than $2/3$ of the stake of $B$ to sign the checkpoint transaction.
Since $B$ is $1/3$-bounded, at least one of them is honest, and an honest validator independently verifies the UTXO by the Common Prefix property of Bitcoin.
Therefore, such a UTXO exists on Bitcoin, so the protocol increases the Bitcoin balance of $B$ by $\amountShort$~$\BTC$.

\medskip\noindent\textit{Firewall~(4).}
Assume $A$ is $1/3$-bounded, and the protocol increases the Bitcoin balance of $B$ by $\amountShort$~$\BTC$.
The \checkpointTxOne has been signed by validators holding more than $2/3$ of the stake of $A$; since $A$ is $1/3$-bounded, at least one of them is honest. An honest validator signs only if the transfers in \checkpointTxOne are exactly those finalized on $A$'s ledger, which is unique since $A$ is safe.
In particular, the ledger contains a $\transferCommand(\amountShort)$ from $u_1$, whose processing debits $\amountShort$~$\wBTC$ from $u_1$'s balance on $A$, so the protocol decreases the balance of $u_1$ on $A$ by $\amountShort$~$\wBTC$.

\medskip\noindent\textit{Safety.}
Assume $A$ and $B$ are $1/3$-bounded, and the protocol pays $\amountShort$~$\wBTC$ to a user $u_2$ on $B$.
By the \emph{Firewall~(3)} property and its proof, an honest validator of $B$ verified on Bitcoin a UTXO locked with $B$'s multisig, output of a \checkpointTxOne of $A$, whose commit-reveal script names $u_2$ as the recipient of $\amountShort$ and $u_1$ as the sender.
By the argument of the \emph{Firewall~(4)} property, an honest validator of $A$ signed that \checkpointTxOne, hence $A$'s ledger contains a $\transferCommand(\amountShort)$ from $u_1$ with recipient $u_2$, and the protocol decreases the balance of $u_1$ on $A$ by $\amountShort$~$\wBTC$.

\medskip\noindent\textit{Liveness.}
Assume $A$ and $B$ are $1/3$-bounded, and an honest user $u_1$ holding at least $\amountShort$~$\wBTC$ on $A$ calls $\transferCommand(\amountShort)$ with recipient $u_2$ on $B$.
Since $u_1$ is honest and holds at least $\amountShort$~$\wBTC$, this is a valid transfer command, and since $A$ is live, it is eventually finalized on $A$.
Since $A$ is live, $A$ eventually produces a checkpoint that batches all outgoing transfers: the \checkpointTxOne includes an output UTXO locked with $B$'s multisig address whose value equals the sum of all transfers from $A$ to $B$ batched in this checkpoint (hence at least $\amountShort$~$\BTC$), with each recipient's individual amount encoded in the accompanying commit-reveal script (\Cref{sec:transfer}).
The \checkpointTxOne is signed by the honest validators of $A$, who hold more than $2/3$ of the stake, and submitted to Bitcoin.
Let $\height_{\mathsf{ref}}$ be the height of the latest finalized Bitcoin block that this checkpoint takes into account, and $\subnetConf^B_{e}$ the configuration of $B$ at that height, under which the UTXO is locked; then $\height^B_{e-1} \leq \height_{\mathsf{ref}} < \height^B_{e}$, where $\height^B_{e}$ are the cutoff heights of $B$. The \checkpointTxOne pays a sufficient fee and is submitted when the chain has height at most $\height_{\mathsf{ref}} + \depth + \Delta_{\mathsf{sign}}$. By the \emph{Liveness} property of Bitcoin, it is contained in a block at a height $\height'$ with
\[ \height_{\mathsf{ref}} \;<\; \height' \;\leq\; \height_{\mathsf{ref}} + \depth + \Delta_{\mathsf{sign}} + \inclDelay \;<\; \height^B_{e+1} . \]
Hence $\height'$ lies in epoch $e$ or $e+1$ of $B$, and the rotation checkpoint of $B$ that ends this epoch spends the UTXO.
By the Chain Growth property of Bitcoin, the \checkpointTxOne is eventually finalized, and every honest validator of $B$ detects the UTXO locked with $B$'s multisig.
Since $B$ is live and its honest validators hold more than $2/3$ of the stake, $B$ eventually produces a checkpoint that processes the incoming transfer, minting $\amountShort$~$\wBTC$ to $u_2$'s address on $B$.
Since $B$ is live, this update is eventually finalized, so the protocol pays $\amountShort$~$\wBTC$ to $u_2$ on $B$.
\end{proof}
The firewall properties of the \withdrawCommand and \transferCommand theorems, which hold regardless of the security of the subnets involved, establish property \textbf{P1: Firewall} from the main body (\Cref{sec:preliminaries}).
By construction, the \depositCommand, \withdrawCommand, and \transferCommand protocols of these theorems involve only Bitcoin and the validators of the communicating subnets, which establishes property \textbf{P4: No external bridges} from the main body.

\section{Implementation notes}\label{app:implementation-notes}

This section collects details specific to the implementation used for the benchmarks of this paper.



\ifwhitepaper
  \subsection{Protocol parameters}\label{sec:impl-parameters}

  The following parameters are specific to the current implementation.
  \begin{itemize}
      \item \textbf{\killVotePeriod}: A kill proposal expires after 36 Bitcoin blocks, corresponding to approximately 6 hours on Bitcoin mainnet.
      \item \textbf{\killDelayPeriod}: The kill-delay period is set to five subnet checkpoints, giving users sufficient time to withdraw funds before the subnet terminates.
      \item \textbf{\checkpointInterval}: The checkpoint interval, measured in number of subnet blocks, is configurable at subnet creation. A typical value of 2~hours (in approximate wall-clock time) yields an amortized checkpointing overhead of approx.~1080~vB per 24~hours.
  \end{itemize}
\fi

\subsection{Bitcoin fees and wallet management} \label{sec:fees-and-wallet}
\dotparagraph{Fee calculation model}
The fee-calculation model in our implementation uses the \estimatesmartfee{}\footnote{\url{https://github.com/bitcoin/bitcoin/blob/master/src/rpc/fees.cpp}} rpc call of Bitcoin Core.
This function returns the \feeRate{} per vKB. Function \estimatesmartfee{} takes as parameters $k$, the desired number of blocks after which our transaction should be confirmed, and \textsl{mode}, either \economicalMode or \conservativeMode. Mode \economicalMode is quicker to adapt to fluctuating prices by considering a shorter time frame, and prioritizes cost efficiency. Mode \conservativeMode considers a longer time frame and could potentially return a higher fee.
We use $k=6$ and the \economicalMode mode. The final transaction fee is the transaction weight
multiplied by the \feeRate. As defined in BIP 141\footnote{\url{https://github.com/bitcoin/bips/blob/master/bip-0141.mediawiki}}, the transaction weight is
calculated as $\textsl{weight} = \textsl{base\_size} * 3 + \textsl{total\_size}$, where $\textsl{base\_size}$ excludes
witness data and \textsl{total\_size} includes it. Dividing the weight by 4 yields the equivalent size in virtual bytes (vbytes).


\dotparagraph{Coin selection}
We have implemented a coin selection algorithm, used by the \btcProvider every time it needs to create a Bitcoin transaction that spends a UTXO locked by a \bitcoinIpc subnet. It is used, for example, when a user withdraws funds from the subnet or when a user transfers funds to another \bitcoinIpc subnet, as described in \Cref{sec:transfer}.

We remark here that a \bitcoinIpc subnet might own multiple UTXOs. For example, if ten users have deposited funds to that subnet, there will be at least ten UTXOs locked with the multisig of the subnet. The coin-selection algorithm spends these UTXOs in decreasing order of value.
Future software versions will analyze this coin-selection rule and potentially use a more efficient one.

We have also implemented an algorithm to consolidate all UTXOs owned by a subnet into a single UTXO. This reduces transaction sizes and simplifies wallet management. The current implementation calls this algorithm whenever the validator set changes (rotation checkpoints, \Cref{sec:dynamic-participation}). We remark that the UTXO-consolidation algorithm can also be run on regular intervals (not only on validator-set changes), creating a tradeoff between redundant Bitcoin transactions (when called too often) and fragmented subnet UTXOs (when called too seldom).

\subsection{Script construction and parsing}\label{sec:script-construction}
Bitcoin's scripting language limits each stack push operation to at most 520 bytes.\footnote{\url{https://github.com/bitcoin/bitcoin/blob/master/src/script/script.h}}
To commit data payloads that exceed this limit, \bitcoinIpc splits the payload into chunks of at most 520 bytes and encodes each chunk as a separate push operation.
The opcodes used, depending on chunk size, are:
\begin{itemize}
    \item \opPushBytes\_$n$, for $n \in [1,75]$: push $n$ bytes
    \item \opPushDataOne{} to push 76--255 bytes (the following byte encodes the length)
    \item \opPushDataTwo{} to push 256--520 bytes (the following two bytes encode the length)
\end{itemize}
For a script to be valid, Bitcoin requires the stack to contain exactly one element upon completion; an \opDrop opcode is therefore appended after each push, followed by the value~$1$.
The \btcMonitor reconstructs the committed data by scanning for push opcodes and concatenating the byte sequences that follow them.
As a safety measure, the \btcMonitor rejects any script containing opcodes other than push and \opDrop, preventing adversarial scripts from triggering unintended execution paths.


\ifwhitepaper
  \begin{example}[script-building-example]{Building a script with long binary data}\label{ex:script-building}
      Let's see an example on how binary data is encoded as a script. The following is a structure (simplified for ease of presentation) is used to store outgoing transfers, in this case four transfers with two different target \bitcoinIpc subnets (transfers are explained in detail in \Cref{sec:transfer}).
      \small{
      \begin{verbatim}
  "batched_transfers": [
      { "subnet_id": "b4/f410fb...d6pg6y",
        "transfers": [
          { "address": "0x44d7d6...ec6e",
            "amount_sat": 30000 },
          { "address": "0xdb83d3...4413",
            "amount_sat": 30000 } ]
      },
      { "subnet_id": "b4/f410f...he7kn44a",
        "transfers": [
          { "address": "0x55281b...517f5922",
            "amount_sat": 30000 },
          { "address": "0xc0bccc...aa6822",
            "amount_sat": 30000 } ]
      }
  ]
      \end{verbatim}
      }
      The first step is to encode this data in binary format. We use the postcard\footnote{\url{https://docs.rs/postcard/latest/postcard/}} library for that. In this example the data fits in 144 Bytes, hence we get the following script:
      \small{
      \begin{verbatim}
  OP_PUSHDATA1
  <...144 bytes of data...>
  OP_DROP
  OP_PUSHNUM_1
      \end{verbatim}
      }
      If the data was longer, for example when batching 20 transfers, the script would look like this:
      \small{
      \begin{verbatim}
  OP_PUSHDATA2
  <...520 bytes of data...>
  OP_DROP
  OP_PUSHBYTES_9
  ec7a5f91cec6b0ea01
  OP_DROP
  OP_PUSHNUM_1
      \end{verbatim}
      }
  \end{example}
\fi

\else 
    \dotparagraph{Appendix structure}
    The structure of the Appendix is as follows.
    \Cref{sec:pseudocode-all} contains the pseudocode for all \bitcoinIpc operations.
    \Cref{sec:model-full} provides a formalization of the subprotocols used in \bitcoinIpc and their properties.
    The corresponding theorems are shown in \Cref{sec:theorems},
    while \Cref{sec:proofs} contains the proofs.
    \Cref{app:implementation-notes} provides more details on the implementation used in our benchmarks:
    \Cref{sec:fees-and-wallet} on Bitcoin fee and wallet management, and \Cref{sec:script-construction} on the specific Bitcoin scripts we use.

\fi

\end{document}